\documentclass[preprint]{elsarticle}

\usepackage{lineno}
\usepackage{amsmath}
\usepackage{amsfonts}
\usepackage{amssymb}
\usepackage{graphicx,subcaption,float}
\usepackage{sidecap}
\usepackage{rotating}
\usepackage{epstopdf}
\usepackage{tikz}
\usepackage{wasysym}
\usepackage[mode=text, detect-none]{siunitx}[=v2]  
\usepackage{caption}
\usepackage{varwidth}
\usepackage{textcomp}
\usepackage{hyperref}
\usepackage{booktabs}

\allowdisplaybreaks

\modulolinenumbers[1]

\journal{Some Journal}

\makeatletter
\def\ps@pprintTitle{%
	\let\@oddhead\@empty
	\let\@evenhead\@empty
	\def\@oddfoot{\reset@font\hfil\thepage\hfil}
	\let\@evenfoot\@oddfoot
}
\makeatother

\usepackage[T1]{fontenc}
\usepackage{fouriernc}

\biboptions{sort&compress}  

\usepackage{geometry}
\begin{document}

\begin{frontmatter}

\title{Numerical Simulations of the Nonlinear Quantum Vacuum in the Heisenberg--Euler Weak-Field Expansion}

\author{Andreas Lindner\corref{cor1}}
\ead{and.lindner@physik.uni-muenchen.de}
\author{Baris \"Olmez\corref{cor2}}
\ead{b.oelmez@physik.uni-muenchen.de}
\author{Hartmut Ruhl\corref{cor2}}
\ead{hartmut.ruhl@physik.uni-muenchen.de}
\address{Arnold Sommerfeld Center for Theoretical Physics, Ludwig-Maximilians-Universit\"at M\"unchen\\Theresienstr. 37, D-80333 M\"unchen, Germany}

\begin{abstract}
The Heisenberg--Euler theory of the quantum vacuum supplements Maxwell's theory of electromagnetism with nonlinear light--light interactions.
These originate in vacuum fluctuations, a key prediction of quantum theory, and can be triggered by high-intensity laser pulses, causing a variety of intriguing phenomena.
A highly accurate numerical scheme for solving the nonlinear equations due to the leading orders of the Heisenberg--Euler weak-field expansion is presented.
The algorithm possesses an almost linear vacuum dispersion relation even for comparably small wavelengths and incorporates a nonphysical modes filter.
The implemented solver is tested in one spatial dimension against a set of known analytical results for vacuum birefringence and harmonic generation.
More complex scenarios for harmonic generation are demonstrated in two and three spatial dimensions.
\end{abstract}

\begin{keyword}
vacuum polarization, Heisenberg--Euler, simulations, harmonic generation, birefringence
\end{keyword}

\end{frontmatter}

\begin{figure}[H]
	\centering
	\includegraphics[width=\textwidth]{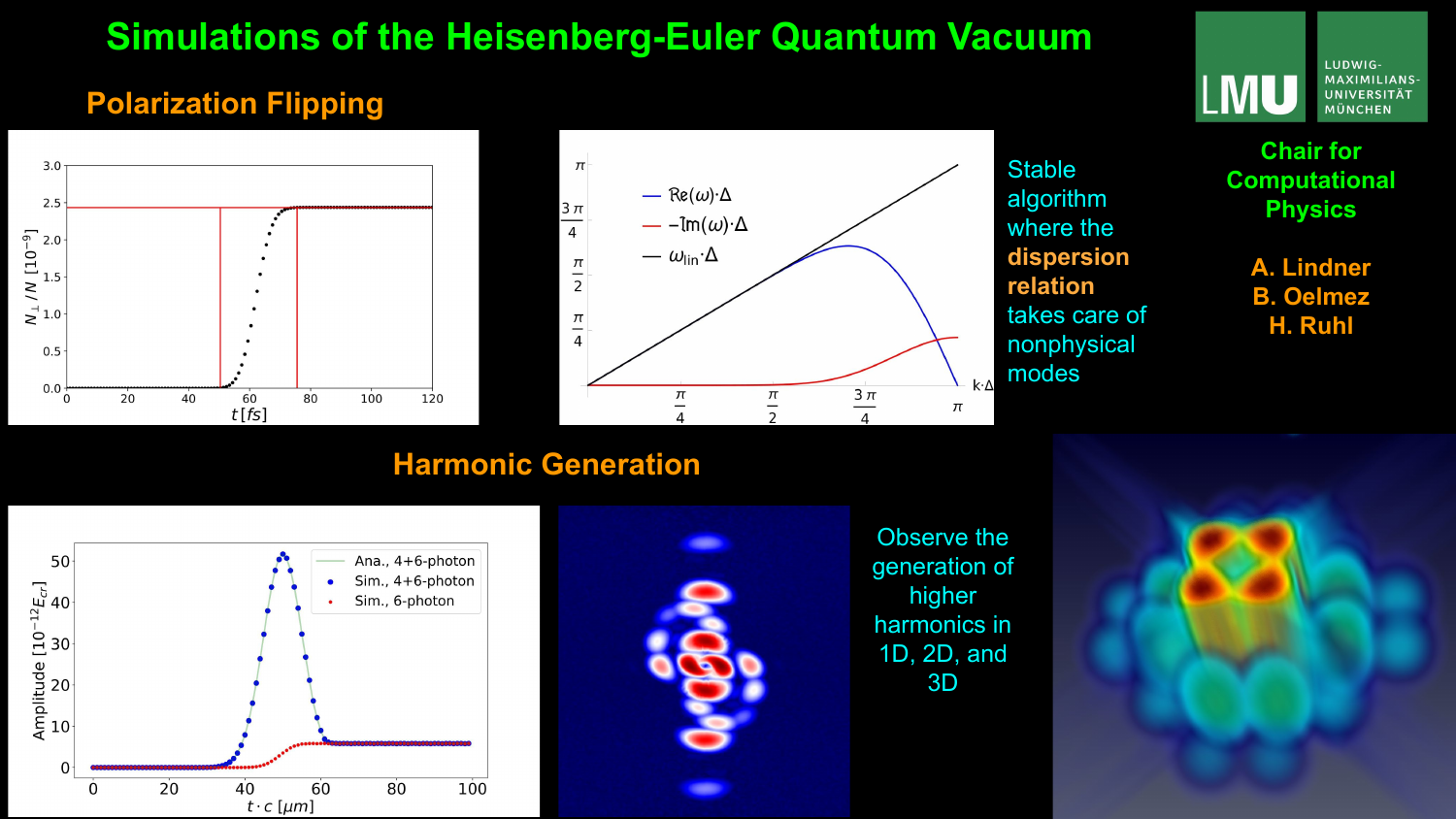}
	\caption*{}
\end{figure}

\section*{Highlights}
\begin{itemize}
	\item A universal 13\textsuperscript{th} order of accuracy numerical scheme with nonphysical modes filter
	\item Universality with respect to pulse configurations in contrast to analytical treatments
	\item Scalability on distributed computing systems
	\item Inclusion of up to six-photon processes in the Heisenberg--Euler weak-field expansion
\end{itemize}


\section{Introduction \label{sec:Introduction}}

Virtual electron--positron pairs are omnipresent in the vacuum of quantum electrodynamics (QED) and can mediate effective photon--photon interactions.
Taking those into account in the low-energy limit below the Compton scale of the electron, nonlinear terms in the electromagnetic field extend Maxwell's theory, invalidating the classical superposition principle in the vacuum.
Radiation on the other hand can influence the behavior of the particle--antiparticle dipoles in the vacuum.
Therefore, the vacuum of QED with its quantum fluctuations acts as a nonlinear, polarizable medium with quantifiable effects.
The corresponding low-energy effective theory was devised by Euler, Kockel, Heisenberg, and Weisskopf \cite{EulerKockel1935,HeisenbergEuler1936,Weisskopf1936}.
Schwinger incorporated the theory into the larger QED picture \cite{Schwinger1951}.
For reviews, see \cite{DittrichReuter1985,DittrichGies2000,MarklundLundin2009,Dunne2009,HeinzlIlderton2009,DiPiazzaetal2012,Dunne2012,KingHeinzl2016,Karbstein2016b,Inadaetal2017}.

There are numerous optical effects that are due to the quantum nature of the vacuum.
These are based on light-by-light scattering phenomena \cite{EulerKockel1935,KarplusNeumann1950,KarplusNeumann1951,McKenna1963,Moulinetal1996,MoulinBernard1999,Bernardetal2000,Lundstrometal2006,Lundinetal2006,Tommasinietal2009,KryuchkyanHatsagortsyan2011,KingKeitel2012,Dinuetal2014b,Giesetal2018,Kingetal2018}.
There is vacuum birefringence caused by a preferred direction for charged particles in a strong electromagnetic field \cite{Toll1952,BaierBreitenlohner1967,BrezinItzykson1971,HeylHernquist1997,LuitenPetersen2004,Heinzletal2006,DiPiazzaetal2006,King2010,Dinuetal2014,Karbsteinetal2015,Schlenvoigtetal2016,KarbsteinSundquist2016,KingElkina2016,Braginetal2017,Karbstein2018,Ataman2018,KarbsteinMosman2019,KarbsteinMosmann2021,Karbsteinetal2021b},
which in inhomogeneous fields is a consequence of broken translational invariance accompanied by vacuum diffraction \cite{DiPiazzaetal2006,TommasiniMichinel2010,Kingetal2010,Kingetal2010b,MondenKodama2011,Karbsteinetal2021}.
A peculiar effect is quantum reflection with photons \cite{Giesetal2013,Giesetal2015}.
Then there are photon merging \cite{Yakovlev1967,DiPiazzaetal2008,Giesetal2014,Giesetal2016}
with the generation of higher harmonics \cite{BhartiaValluri1978,ValluriBhartia1980,Bialynicka-Birula1981,KaplanDing2000,DiPiazzaetal2005,FedotovNarozhny2007,NarozhnyFedotov2007,Kingetal2014,Boehletal2015,Boehl2016,Kadlecovetal2019,Sasorov2021},
and photon splitting \cite{Adleretal1970,Bialynicka-BirulaBialynicki-Birula1970,Adler1971,PapanyanRitus1972,Stoneham1979,Baieretal1996,Adleretal1996,DiPiazzaetal2007,Giesetal2016}.

A promising approach to their detection is the investigation of the asymptotic dynamics of \textit{probe} photons after passing through a strong-field region in the form of a high-intensity laser \textit{pump} pulse.
Probe photons can in this way indirectly sense the applied pump field via the quantum fluctuations which couple to both probe and pump fields \cite{KarbsteinSundquist2016}.
As a result of the nonlinear interaction with the power pulses, signal photons of the probe field scatter in such a way that their dynamics or polarization distinguishes them from the main pulses.
The properties and dynamics of the quantum vacuum, in turn, are hereby encoded into the signal photons.
This class of quantum vacuum experiment, where one electromagnetic field drives the nonlinear effect while the other carries its signature, is called \textit{all-optical}.

Experiments at the high-intensity frontier promise unprecedented detail in the study of the quantum vacuum in the near future.
The hope is that the rapid and steady advancements in ultra-intense laser physics \cite{Dansonetal2019,Scholz2018} combined with theoretically optimized specifications will soon facilitate the experimental discovery \cite{MarklundLundin2009,DiPiazzaetal2012,KingHeinzl2016}.

Almost all analytical approaches to compute nonlinear vacuum effects, as found in the literature cited above, have shortcomings.
As a consequence, these attempts are still limited to simple scenarios, which impedes experimental verification.
Approximations limit the accuracy of predictions and the precision the theory can be tested with.
Precision tests require accurate theoretical predictions for arbitrary laser field configurations \cite{Blinneetal2019a}.

The present paper introduces a numerical algorithm for the solution of the Heisenberg--Euler equations in weak-field approximation, where ``weak'' means below an ultra-high \textit{critical} field strength of more than \SI{e18}{\volt \per \metre}, to be detailed below.
The work is based on the numerical scheme outlined in one spatial dimension in \cite{Kingetal2014,Boehletal2015,Boehl2016} and extended to full three dimensions in \cite{Pons2016,Pons2018}.
The accuracy of the numerical scheme relies on the assumptions that the fields vary on scales larger than the Compton scale of the electron and the field strengths involved are below the critical field for pair creation in the vacuum.
In addition, a viable numerical algorithm is restricted by the computational load it generates, particularly in the realm of high frequencies.
The advantage of a numerical approach, however, is that it can principally solve almost any interaction scenario for nearly any arrangement of laser pulses.

The numerical predictions of the presented solver are benchmarked with analytical predictions for birefringence as calculated in \cite{Karbsteinetal2015,Dinuetal2014} and the generation of higher harmonics as predicted in \cite{Kingetal2014}.

\subsection{Comparison to other approaches}

There exist other approaches to simulate all-optical QED vacuum effects.
The algorithm outlined in \cite{Giesetal2017} and the solver presented in \cite{Blinneetal2019a} are based on the vacuum emission picture \cite{KarbsteinShaisultanov2015}.
In the vacuum emission picture the fields are split into strong background and weak signal fields and the Heisenberg--Euler Lagrangian is expanded to linear order in the signal field strength.
The strong electromagnetic fields driving the nonlinear effects are propagated by the linear Maxwell equations in the absence of vacuum nonlinearities.
This is done by means of a Maxwell solver detailed in \cite{Blinneetal2019b}.

The vacuum subjected to the high-intensity fields constitutes a source term for an outgoing photon via a scattering amplitude using the Heisenberg--Euler interaction Lagrangian.
The picture can be interpreted as describing laser-stimulated signal photon emission from the vacuum.
In certain configurations excellent signal to background separation can be achieved.

However, there is no feedback on the effect-generating electromagnetic fields themselves by the response of the quantum vacuum.
In other words, nonlinearities induced by laser fields, and thus also back-reactions into the latter, are neglected.
Hence, there is no notion of pump and probe laser fields in contrast to the scenarios considered throughout the present work.

The vacuum emission picture is capable of making predictions for asymptotic states of ultra-short emission wavelengths, while the solver outlined in the present paper is limited by the affordable grid resolution.
Numerous analytical works are based on the vacuum emission picture \cite{Karbstein2015,Karbsteinetal2015,KarbsteinSundquist2016,Karbsteinetal2019,Klar2020,Karbstein2020,KarbsteinMosman2020,Karbsteinetal2021b,Karbstein2022,Giesetal2021}.
In addition, the solver based on the vacuum emission picture has already been employed in some studies of optical signatures of the quantum vacuum \cite{Giesetal2018,Blinneetal2019c,Blinneetal2019d}.

Besides vacuum birefringence, the approach has also been employed to photon--photon scattering as well as merging and splitting processes.
The emission process is not restricted to cubic order in the background field \cite{Giesetal2014,Giesetal2016}.
Hence, multi-photon emission processes can also be incorporated in the description.

In \cite{Grismayeretal2021} a generalized Yee scheme of second order accuracy in space and time is devised.
The accuracy order of the numerical scheme discussed in the present paper is arbitrary and implementations for the second to the thirteenth order are given.
The Yee scheme approach in \cite{Grismayeretal2021} requires interpolations in space and time to compute the vacuum nonlinearities.
Also, the dispersion relations of the numerical scheme in \cite{Grismayeretal2021} and in the present paper differ from each other.
The dispersion relation outlined in \cite{Grismayeretal2021} has an imaginary part that can lead to the amplification of nonphysical modes and requires a high grid resolution for a given wavelength.
The dispersion relations outlined in the present paper have imaginary parts that always damp nonphysical modes and at high integration order can afford lower grid resolution.

The solvers in \cite{Grismayeretal2021} and \cite{Blinneetal2019a} include only the four-photon order of the weak-field expansion of the Heisenberg--Euler interaction. 
Some all-optical vacuum nonlinear effects are tiny.
Hence, high-order, high-precision numerical solvers may bear advantages.  
An implementation for distributed computing systems of the scheme discussed in the present paper is available and allows costly simulations in full three spatial dimensions plus time.
High-resolution grids, however, are subject to the curse of dimensionality.
Some applications require very high resolutions in order to model high-frequency waves.
These easily go beyond the limits of any computing system and thus extrapolation techniques have to be employed.

\subsection{Outline}

The paper begins with a recapitulation of the weak-field expansion of the Heisenberg--Euler effective theory in Section \ref{sec:LHE}.
In Section \ref{sec:Maxwell}, the nonlinear modifications to Maxwell's equations due to the Heisenberg--Euler theory are detailed.
In Section \ref{sec:PDEtoODE} the numerical scheme that solves the nonlinear Maxwell equations is outlined.
Dispersion effects on the lattice and the scaling behavior of the code implementation are discussed in Sections \ref{sec:Dispersion} and \ref{sec:scaling}.

In Sections \ref{sec:phasevel}-\ref{sec:2dsims} the capabilities of the Heisenberg--Euler solver is demonstrated by solving various scenarios of nonlinear vacuum phenomena and by successfully benchmarking with analytical results, where they exist.
In Section \ref{sec:phasevel} the phase velocity reduction of a probe pulse propagating through a strong electromagnetic background is successfully compared to analytical predictions for sufficiently large background field strengths.
In Section \ref{sec:flipping} the phenomenon of vacuum birefringence is investigated with the help of simulations for probe--pump setups.
The theoretical prediction of the polarization flipping probability and its scaling with various parameters is verified.
An extrapolation of the results to wavelengths in the x-ray regime is performed.

In Section \ref{sec:harmonics} the capability of simulations to predict higher harmonics generated by vacuum nonlinearities is demonstrated.
It is shown that for specific settings analytical results from \cite{Kingetal2014} are accurately reproduced in the simulations presented in this work.
Examples of harmonic generation in 2D and 3D simulations are presented in Section \ref{sec:2dsims}.
The latter are hard, if not impossible, to obtain by analytical means.
The results are used for comparison with the previous 1D simulations.
Higher-dimensional simulations enable the investigation of complicated scenarios that cannot be solved analytically in the future.

A summary and an outlook to future research employing the Heisenberg--Euler solver are given in Section \ref{sec:Outlook}.

\section{Heisenberg--Euler weak-field expansion} \label{sec:LHE}

The Heisenberg--Euler Lagrangian can be expressed as \cite{Dunne2005}
\begin{align}\label{pureLHE}
	\mathcal{L}_{\text{HE}}=-\frac{c^5m_e^4}{8 \hbar^3 \pi ^2} \int_0^{\infty }& \frac{e^{-s}}{s^3} \left(\frac{s^2}{3}  \left(a^2-b^2\right)-1+a \, b \, s^2 \, \cot (a s) \, \coth (b s)\right) \, ds \ ,
\end{align}
where
\begin{align}
	& a=\sqrt{\sqrt{\mathcal{F}^2+\mathcal{G}^2}+\mathcal{F}}\ , \quad	 b=\sqrt{\sqrt{\mathcal{F}^2+\mathcal{G}^2}-\mathcal{F}} \ , \\
	& \mathcal{F}=-\frac{c^2 F^{\mu\nu}F_{\mu\nu}}{4E_{\textrm{cr}}^2}\ , \quad \mathcal{G}=-\frac{c^2 F^{\mu\nu}F^\ast_{\mu\nu}}{4E_{\textrm{cr}}^2} \ , \\
	& E_{\textrm{cr}}= \frac{m_e^2 \, c^3}{e \, \hbar} =\SI{1.323e18}{\volt \per \metre} \ ,
\end{align}
and $ e $, $ m_e $ are the charge and mass of the electron.
The quantities $F$ and $F^\ast$ denote the electromagnetic field strength tensor and its dual and $ E_{\textrm{cr}} $ is the above mentioned critical field strength, which is the scale where the magnitude of the vacuum nonlinearities becomes dominant.

The Lagrangian for the quantum vacuum is then given by the classical Maxwell Lagrangian and the quantum corrections due to vacuum polarization captured in the Heisenberg--Euler Lagrangian
\begin{equation}\label{LQV}
	\mathcal{L} = \mathcal{L}_{\text{MW}} + \mathcal{L}_{\text{HE}} \ . 
\end{equation}
The Maxwell Lagrangian is given by
\begin{equation}
	\mathcal{L}_{\text{MW}} =-\frac{1}{4\mu_0}F^{\mu\nu}F_{\mu\nu} = \frac{E_{\textrm{cr}}^2}{c^2 \mu_0} \mathcal{F} \ ,
\end{equation}
with the vacuum permeability
\begin{equation}
	\mu_0=
	\SI{1.257e-6}{\newton \per \square \ampere} \ .
\end{equation}
To derive $ \mathcal{L}_{\text{HE}}$ 
it has been assumed that the constant background approximation is valid \cite{DunneHall1999,GiesRoessler2011} as a consequence of the fact that the variations of the field strengths are small on the scale of the Compton wavelength \cite{Tiesingaetal2021}
\begin{equation}
	\lambda_C = \frac{h}{m_e c} = \SI{2.426e-12}{\metre} \ .
\end{equation}
The weak-field expansion of \eqref{pureLHE} up to fourth order in $ \mathcal{F} $ and $ \mathcal{G} $ is given by
\begin{subequations}\label{approxLHE}
	\begin{align}
		\mathcal{L}_{\text{HE}}\approx & \quad 4 \, \epsilon_0 \frac{E_{\textrm{cr}}^2 \alpha}{360 \pi} \left(\,4\mathcal{F}^2+7\mathcal{G}^2\,\right)\label{LHE_T1}
		\\&- 4 \, \epsilon_0 \frac{E_{\textrm{cr}}^2 \alpha}{630 \pi} \left(8 \mathcal{F}^3+13\mathcal{F}\mathcal{G}^2\,\right)\label{LHE_T2}
		\\&+ 4 \, \epsilon_0 \frac{E_{\textrm{cr}}^2 \alpha}{945 \pi} \left(48 \mathcal{F}^4+88 \mathcal{F}^2 \mathcal{G}^2+19 \mathcal{G}^4\,\right)\label{LHE_T3}\ ,
	\end{align}
\end{subequations}
with the vacuum permittivity
\begin{equation}
	\epsilon_0 = \SI{8.854e-12}{ \ampere \second \per \volt \per \meter} 
\end{equation}
and the fine-structure constant
\begin{equation}
 \alpha = \frac{e^2}{4 \pi \epsilon_0 \hbar c} \approx 1/137 \ .
\end{equation}
It should be noted that the Heisenberg--Euler Lagrangian solely depends on the electromagnetic field invariants $\mathcal{F}$ and $\mathcal{G}$, and subsequently (\ref{LHE_T1}) is of the order $\mathcal{O}((E/E_{\textrm{cr}})^4)$.

\begin{figure}
	\centering
	\includegraphics[width=0.8\linewidth]{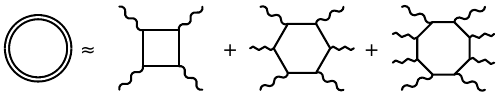}
	\caption[Heisenberg--Euler weak-field expansion]{Graphical representation of the Heisenberg--Euler Lagrangian in weak-field expansion. 
		The double-lined loop to the left represents $ \mathcal{L}_{\text{HE}} $ and corresponds to the irradiated electron--positron loop, denoting the coupling of the external field to the fermion loop to all orders in the field strength.
		The diagrams to the right are the nonlinear four-, six-, and eight-photon contributions due to the weak-field expansion of $ \mathcal{L}_{\text{HE}} $, corresponding to the terms in \eqref{LHE_T1}, \eqref{LHE_T2}, and \eqref{LHE_T3}.}
	\label{fig:Diagramme}
\end{figure}

The effective interaction $ \mathcal{L}_{\text{HE}} $ can be graphically represented by a double-lined loop as in Figure \ref{fig:Diagramme}.
As a consequence, (\ref{LHE_T1}) represents the four-photon contributions, (\ref{LHE_T2}) the six-photon contributions, and (\ref{LHE_T3}) the eight-photon contributions to the closed loop as illustrated in Figure \ref{fig:Diagramme}.
Since the field strengths are constrained to below the critical field strength, $c \vert F_{\mu\nu} \vert < E_{\textrm{cr}}$, higher order terms can be neglected in the low-intensity regime.

It has to be mentioned that pair production is exponentially suppressed in the Heisenberg--Euler Lagrangian.
Therefore, the properties of the nonlinear vacuum can be described exclusively by radiation fields.

\section{Modified Maxwell equations \label{sec:Maxwell}}

To obtain the nonlinear modifications to the linear Maxwell equations an explicit rescaling of the electric and magnetic fields $\vec{E}$ and $\vec{B}$ to the critical field strengths is instructive such that the field strengths are given in units of $E_{\textrm{cr}}$,
\begin{align} 
\vec{E} \rightarrow  \frac{\vec{E}}{E_{\textrm{cr}}} \ , \quad  \vec{B} \rightarrow  \frac{\vec{B}}{E_{\textrm{cr}}} \ .
\end{align}

From the Lagrangian of the quantum vacuum \eqref{LQV} the equations of motion are obtained with the help of the Euler--Lagrange equations with respect to the normalized fields
\begin{equation}\label{eq:MaxPDE}
\partial_t \left(\frac{\vec{E}}{c^2} + \mu\frac{\partial \mathcal{L}_{\text{HE}}}{\partial \vec{E}}\right)=\nabla\times\left(\vec{B}-\mu\frac{\partial \mathcal{L}_{\text{HE}}}{\partial \vec{B}}\right) \ ,
\end{equation}
with the definition
\begin{equation}
	\mu = 
	\frac{\mu_0}{E_{\textrm{cr}}^2} \, .
\end{equation}
The classical Maxwell--Amp\`{e}re circuital law 
\begin{equation}\label{eq:Ampere}
\partial_t \left(\frac{\vec{E}}{c^2}+ \mu_0 \, \vec{P}\right) = \nabla\times\left(\vec{B}- \mu_0 \, \vec{M}\right)
\end{equation}
is compared to \eqref{eq:MaxPDE}
in such a way that photon--photon interactions are included, which implies the appearance of macroscopic nonlinear polarization and magnetization terms given by
\begin{equation}\label{MPtoLHE}
	\vec{P}= \frac{\mu}{\mu_0} \frac{\partial \mathcal{L}_{\text{HE}}}{\partial \vec{E}} = \frac{1}{E_{\textrm{cr}}^2}  \frac{\partial \mathcal{L}_{\text{HE}}}{\partial \vec{E}} 
	\quad \text{and} \quad \vec{M}= \frac{\mu}{\mu_0} \frac{\partial \mathcal{L}_{\text{HE}}}{\partial \vec{B}} =
	\frac{1}{E_{\textrm{cr}}^2}  \frac{\partial \mathcal{L}_{\text{HE}}}{\partial \vec{B}} \ .
\end{equation}
The modified Maxwell--Amp\`{e}re law \eqref{eq:Ampere} with the terms in \eqref{MPtoLHE} is merged with the Maxwell--Faraday law of induction 
\begin{equation}\label{eq:Faraday}
	\partial_t \vec{B} = -\nabla\times\vec{E} \ ,
\end{equation}
which persists in the nonlinear vacuum,
such that a single partial differential equation that describes the whole dynamics of the system is formulated.
The curl of $(\vec{B}-\vec{M})$ can be written as
\begin{equation}
\begin{aligned}
\label{eq:Qs1}
    \nabla \times (\vec{B}-\vec{M}) &= \underbrace{\begin{pmatrix} 
0 & 0 & 0 \\ %
0 & 0 & -1 \\ 
0 & 1 & 0 \\ 
\end{pmatrix}}_{\mathbf{Q}_x} \partial_{x} \, (\vec{B}-\vec{M})
+
\underbrace{\begin{pmatrix}
0 & 0 & 1 \\ 
0 & 0 & 0 \\ 
-1 & 0 & 0 \\ 
\end{pmatrix}}_{\mathbf{Q}_y} \partial_{y} \, (\vec{B}-\vec{M})
\\
& +
\underbrace{\begin{pmatrix} 
0 & -1 & 0 \\ 
1 & 0 & 0 \\ 
0 & 0 & 0 \\ 
\end{pmatrix}}_{\mathbf{Q}_z} \partial_{z} \, (\vec{B}-\vec{M})
 = \sum _ {j \in \lbrace x,y,z\rbrace}\mathbf{Q}_{j} \partial_{j} \, (\vec{B}-\vec{M}) \ .
 \end{aligned}
\end{equation}
With the help of the vector
\begin{equation}
	\vec{f} = \left( 
	\vec{E}/c , \vec{B}
	\right)^{\text{T}} 
\end{equation}
the Maxwell equations \eqref{eq:Ampere} and \eqref{eq:Faraday} can then be expressed as
\begin{equation}\label{eq:NumPDE}
	 \left( \mathbf{1}_{6} + \mathbf{A} \right) \frac{\partial_{t}}{c} \vec{f}= \sum_{j\in\{x,y,z\}} \mathbf{Z}_{j} \partial_{j} \vec{f} \ .
\end{equation}
with the $6 \times 6$ identity matrix $\mathbf{1}_{6}$ and
the matrices $ \mathbf{A} $ and $ \mathbf{Z}_{j} $ are given by
\begin{align} \label{eq:AZ}
    \mathbf{A} = \begin{pmatrix}
    \mathbf{J}_{\vec{P}} \left(\vec{E}\right) & \mathbf{J}_{\vec{P}} \left(\vec{B}\right) \\
\mathbf{0}_{3} & \mathbf{0}_{3} \\ 
\end{pmatrix} \ , \quad
\mathbf{Z}_{j} =  \begin{pmatrix} 
    - \mathbf{Q}_{j} \mathbf{J}_{\vec{M}} \left(\vec{E}\right) &\mathbf{Q}_{j} -\mathbf{Q}_{j}\mathbf{J}_{\vec{M}} \left(\vec{B}\right) \\
-\mathbf{Q}_{j} & \mathbf{0}_{3} \\ 
\end{pmatrix} \ ,
\end{align}
where $\mathbf{0}_{3}$ is the $3 \times 3$ zero matrix and
\begin{align}
	\mathbf{J}_{\vec{P}} (\vec{E}) = \epsilon_0 \frac{\partial \vec{P}}{\partial \vec{E}} \, , \quad \mathbf{J}_{\vec{P}} (\vec{B}) = \mu_0 c \frac{\partial \vec{P}}{\partial \vec{B}} \, , \quad \mathbf{J}_{\vec{M}} (\vec{E}) = \mu_0 c \frac{\partial \vec{M}}{\partial \vec{E}} \, , \quad \mathbf{J}_{\vec{M}} (\vec{B}) = \mu_0 \frac{\partial \vec{M}}{\partial \vec{B}} \ ,
\end{align}
with matrix elements
\begin{equation}
	\left( \mathbf{J}_{\vec{P}} (\vec{E}) \right)_{ij} = \epsilon_0 \frac{\partial P_i}{\partial E_j} \ .
\end{equation}

Equation \eqref{eq:NumPDE} contains the full dynamics of electromagnetic fields in the weak-field approximation of the Heisenberg--Euler model.
For illustrative purposes, first the linear vacuum is considered by setting $\vec{P}=\vec{M}=0$, leading to
\begin{equation} \label{LinPropEq}
\frac{\partial_{t}}{c} \vec{f}= \sum_{j\in\{x,y,z\}} \mathbf{Z}^{\text{lin}}_{j} \partial_j \vec{f} \ , \quad \mathbf{Z}^{\text{lin}}_{j} =  \begin{pmatrix} \mathbf{0}_3 &\mathbf{Q}_j\\-\mathbf{Q}_j& \mathbf{0}_3 \end{pmatrix} \ .
\end{equation}
To deduce modes with specific propagation directions, which becomes important in the construction of the numerical scheme, $\mathbf{Z}_{j}^{\text{lin}}$ is diagonalized using the matrices
\begin{equation}\label{eq:rotations}
	\begin{aligned}
		\mathbf{R}_{x} &=\frac{1}{\sqrt{2}} \begin{pmatrix}
			\sqrt{2} & 0 & 0 & 0 & 0 & 0\\
			0 & 1 & 0 & 0 & 0 & -1\\
			0 & 0 & 1& 0 & 1 & 0\\
			0 & 0 & 0 & \sqrt{2} & 0 & 0\\
			0 & 1 & 0 & 0 & 0 & 1\\
			0 & 0 & 1 & 0 & -1 & 0\\
		\end{pmatrix}  \ , \quad
		\mathbf{R}_{y} = \frac{1}{\sqrt{2}} \begin{pmatrix}
			0 & \sqrt{2} & 0 & 0 & 0 & 0\\
			-1 & 0 & 0 & 0 & 0 & -1\\
			0 & 0 & -1 & 1 & 0 & 0\\
			0 & 0 & 0 & 0& \sqrt{2} &0\\
			-1 & 0 & 0 & 0 & 0 & 1\\
			0 & 0 & -1 & -1 & 0 & 0\\
		\end{pmatrix} \ , \\ 
		\mathbf{R}_{z} &= \frac{1}{\sqrt{2}} \begin{pmatrix}
			0 & 0 & \sqrt{2} & 0 & 0 & 0\\
			1 & 0 & 0 & 0 & -1 & 0\\
			0 & 1 & 0 & 1 & 0 & 0\\
			0 & 0 & 0 & 0 & 0 & \sqrt{2}\\
			1 & 0 & 0 & 0 & 1 & 0\\
			0 & 1 & 0 & -1 & 0 & 0\\
		\end{pmatrix}   \ ,
	\end{aligned}
\end{equation}
so that for $\mathbf{Z}_{j}^{\text{lin}}$ the expressions
\begin{align}
    \mathbf{Z}_{j}^{\text{lin}} &= \mathbf{R}^{\text{T}}_{j} \begin{pmatrix}
    0 & 0 & 0 & 0 & 0 & 0\\
    0 & 1 & 0 & 0 & 0 & 0\\
    0 & 0 & 1 & 0 & 0 & 0\\
    0 & 0 & 0 & 0 & 0 & 0\\
    0 & 0 & 0& 0 & -1 & 0\\
    0 & 0 & 0 & 0 & 0 & -1\\
    \end{pmatrix} \mathbf{R}_{j}
    = \mathbf{R}^{\text{T}}_{j}\text{diag}(0,1,1,0,-1,-1) \, \mathbf{R}_{j} 
\end{align}
are obtained.
Now, Equation \eqref{LinPropEq} becomes
\begin{align}\label{eq:dtrx}
	\frac{\partial_{t}}{c} \, \mathbf{R}_{j}\vec{f}(x,y,z;t) = \text{diag}(0,1,1,0,-1,-1) \, \partial_{j} \, \mathbf{R}_{j}\vec{f}(x,y,z;t) \ ,
\end{align}
where
\begin{align}\label{eq:Rf}
 \mathbf{R}_{x}\vec{f} = \frac{1}{\sqrt{2}} \begin{pmatrix}
	\sqrt{2}E_x\\
	E_y - B_z\\
	E_z + B_y\\ 
	\sqrt{2}B_x\\
	E_y + B_z\\
	E_z - B_y\\
\end{pmatrix}
\ , \quad
\mathbf{R}_{y}\vec{f} = \frac{1}{\sqrt{2}} \begin{pmatrix}
	\sqrt{2}E_y\\
	- B_z - E_x\\
	B_x - E_z\\ 
	\sqrt{2}B_y\\
	B_z - E_x\\
	- B_x - E_z\\
\end{pmatrix} \ , \quad
 \mathbf{R}_{z}\vec{f}  = \frac{1}{\sqrt{2}} \begin{pmatrix}
	\sqrt{2}E_z\\
	- B_y + E_x\\
	B_x + E_y\\ 
	\sqrt{2}B_z\\
	B_y + E_x\\
	- B_x + E_y\\
\end{pmatrix} \ .
\end{align}
Equations \eqref{eq:dtrx} and \eqref{eq:Rf} imply backward propagation in $x$-direction of the field components $ (E_{y}-B_{z}) $ and $ (E_{z}+B_{y}) $, forward propagation of the field components $ (E_{y}+B_{z}) $ and $ (E_{z}-B_{y}) $, and non-propagation of the first and fourth component.
The components for the other propagation directions can be read off analogously.

\section{From PDE to ODE}\label{sec:PDEtoODE}
For demonstration purposes only the $x$-direction is considered in the following.
In order to transform the partial differential equation into an ordinary one, finite difference approximations are introduced.
The spatial derivative $ \partial_{x} $ is replaced by finite differences
\begin{align}
\partial_{x}\left(\mathbf{R}_{x}\vec{f}\right)(x,y,z)  \approx \mathcal{D}_x \left(\mathbf{R}_{x}\vec{f}\right)(x,y,z) \ ,
\end{align}
where
\begin{align} \label{DefinitionDiscreteDerivativeWithStencil}
	{\mathcal{D}_{x}} \left(\mathbf{R}_{x} \vec{f} \right)  \left(x,y,z\right) =   \sum_{\nu} \frac{1}{\Delta_{x}} S_{\nu} \left(\mathbf{R}_{x} \vec{f}\right)\left( x+\nu\Delta_x,y,z\right) \ .
\end{align}
The $S_\nu$ are stencil matrices and $\Delta_x$ is the spatial resolution.
The stencil matrices $S_{\nu}$ are derived with the help of a Taylor expansion of a generic function $g(x + \nu \Delta_x)$ around $\nu \Delta_x = 0$.
The Taylor expansion up to accuracy order six is given by
\begin{align}
\begin{aligned}
g(x + \nu \Delta_x) = g(x) &+ \left(\nu \Delta_x\right)g'(x) + \frac{1}{2} \left(\nu \Delta_x\right)^2 g''(x) + \frac{1}{6} \left(\nu \Delta_x\right)^3 g'''(x) + \frac{1}{24} \left(\nu \Delta_x\right)^4 g^{(4)}(x) \\
&+ \frac{1}{120} \left(\nu \Delta_x\right)^5 g^{(5)}(x) + \frac{1}{720} \left(\nu \Delta_x\right)^6 g^{(6)}(x) + \mathcal{O}\left(\left(\nu \Delta_x\right)^7\right) \ .
\end{aligned}
\end{align}
For the finite differences approximations of derivatives, forward and backward steps can be made use of.
For the values $\nu \in \left\lbrace -3,...,3\right\rbrace$ the expansion in matrix form reads
\begin{align} \label{StencilMatrixSixthOrder}
	\begin{pmatrix}
		g(x - 3 \Delta_x) \\
		g(x - 2 \Delta_x) \\
		g(x -  \Delta_x) \\
		g(x ) \\
		g(x +  \Delta_x) \\
		g(x + 2 \Delta_x) \\
		g(x + 3 \Delta_x) \\
	\end{pmatrix} 
	= 
	\begin{pmatrix}
		1 & -3 & 9/2 & -9/2 & 27/8 & -81/40 & 81/80 \\
		1 & -2 & 2 & -4/3 & 2/3 & -4/15 & 4/45 \\
		1 & -1 & 1/2 & -1/6 & 1/24 & -1/120 & 1/720 \\
		1 & 0 & 0 & 0 & 0 & 0 & 0 \\
		1 & 1 & 1/2 & 1/6 & 1/24 & 1/120 & 1/720 \\
		1 & 2 & 2 & 4/3 & 2/3 & 4/15 & 4/45 \\
		1 & 3 & 9/2 & 9/2 & 27/8 & 81/40 & 81/80 \\
	\end{pmatrix}
\cdot
	\begin{pmatrix}
		g(x) \\
		\Delta_x \, g'(x) \\
		\Delta^2_x \, g''(x) \\
		\Delta^3_x \, g'''(x) \\
		\Delta^4_x \, g^{(4)}(x) \\
		\Delta^5_x \, g^{(5)}(x) \\
		\Delta^6_x \, g^{(6)}x) \\
	\end{pmatrix} \ 
\end{align}
and can be extended to larger ranges of $\nu$ and inverted to obtain approximations for the derivatives taking into account more distant points.
Recall that only the first derivative is required for the merged equation of motion \eqref{eq:NumPDE}. 

To illustrate how the finite difference approximation of the first derivative is calculated, $\nu \in \left\lbrace 0,1\right\rbrace$ is considered for simplicity.
It is obtained
\begin{align}
	\begin{pmatrix}
		g(x ) \\
		g(x +  \Delta_x) \\
	\end{pmatrix} 
	= 
	\begin{pmatrix}
		1 & 0 \\
		1 & 1 \\
	\end{pmatrix}
\cdot
	\begin{pmatrix}
		g(x) \\
		\Delta_x \, g'(x) \\
	\end{pmatrix} \ ,
\end{align}
which leads to the first derivative of $g$ given by
\begin{align}\label{ForwardFirstDerivative}
	g'(x) \approx g'_{\nu \in \left\lbrace 0,1\right\rbrace} = \frac{g(x + \Delta_x) - g(x)}{\Delta_x} \ .
\end{align}
The derivative $g'_{\nu \in \left\lbrace 0,1\right\rbrace}$ in \eqref{ForwardFirstDerivative} is called first order forward discrete derivative.
In the same way the first order backward discrete derivative $g'_{\nu \in \left\lbrace -1 ,0\right\rbrace}$ is obtained  with the help of
\begin{align}
	\begin{pmatrix}
		g(x -  \Delta_x) \\
		g(x ) \\
	\end{pmatrix} 
	= 
	\begin{pmatrix}
		1 & -1 \\
		1 & 0 \\
	\end{pmatrix}
\cdot
	\begin{pmatrix}
		g(x) \\
		\Delta_x \, g'(x) \\
	\end{pmatrix} \ ,
\end{align}
where
\begin{align} \label{BackwardFirstDerivative}
	g'_{\nu \in \left\lbrace -1,0\right\rbrace} = \frac{g(x) - g(x - \Delta_x)}{\Delta_x} \ .
\end{align}

In order to obtain specific dispersion properties and numerical stability of the discretized version of Equation \eqref{eq:dtrx} it is necessary to introduce forward and backward finite differences for $\mathcal{D}_x$, as discussed further below.
With Equation \eqref{DefinitionDiscreteDerivativeWithStencil} the full expression for the spatial derivatives for distinct propagation directions is formulated
\begin{align}\label{eq:finite_diff}
	\partial_{x} \vec{f}(x,y,z) = \mathbf{R}_{x}^{\text{T}} \partial_{x} \mathbf{R}_{x}\vec{f}(x,y,z) \approx \mathbf{R}_{x}^{\text{T}} \sum_{\nu} \frac{1}{\Delta_{x}} S_{\nu} \left(\mathbf{R}_{x} \vec{f}\right)(x + \nu \Delta_x,y,z) = \mathbf{R}_{x}^{\text{T}} \mathcal{D}_{x} \left( \mathbf{R}_{x} \vec{f} \right) (x,y,z) \ .
\end{align}
Setting aside non-propagation of modes, the first three elements of the rotated field vector $\mathbf{R}_x \vec{f}$ represent forward-propagating electromagnetic modes, while the last three components of the latter represent backward-propagating ones, as discussed at the end of Section \ref{sec:Maxwell}.
This understanding is adopted for the discussion of dispersion relations in Section \ref{sec:Dispersion}.

For the nonlinear vacuum the system of ODEs is obtained from the merged equation of motion \eqref{eq:NumPDE} and Equation \eqref{eq:finite_diff} to be
\begin{align}\label{PropEquation}
\frac{\partial_{t}}{c} \vec{f}&= \left( \mathbf{1}_{6} + \mathbf{A} \right)^{-1} \sum_{j\in\{x,y,z\}} \mathbf{Z} _{j} \mathbf{R}_j^{\text{T}} \mathcal{D}_{j} \mathbf{R}_j \vec{f} \ .
\end{align}
The inversion of $ \left( \mathbf{1}_{6} + \mathbf{A} \right) $ is an expensive operation.
An approximation in terms of a truncated geometric series is not satisfying since also larger nonlinear corrections ought to be taken into account.
The problem of inverting the matrix can be reduced from a 6$\times$6-dimensional problem to a 3$\times$3-dimensional one by making use of the block form in \eqref{eq:AZ} such that \cite{Pons2018}
\begin{align}
	\left( \mathbf{1}_6 + \mathbf{A} \right)^{-1} = \begin{pmatrix}
		\mathbf{1}_{3} + \mathbf{J}_{\vec{P}} \left(\vec{E}\right) & \mathbf{J}_{\vec{P}} \left(\vec{B}\right) \\
		\mathbf{0}_{3} & \mathbf{1}_{3} \\ 
	\end{pmatrix} \ = \ 
	\begin{pmatrix}
		\mathbf{C}^{-1} & - \mathbf{C}^{-1} \mathbf{J}_{\vec{P}} \left(\vec{B}\right) \\
		\mathbf{0}_{3} & \mathbf{1}_{3} \\ 
	\end{pmatrix} \ ,
\end{align}
with
\begin{align}
	\mathbf{C} = 
	\mathbf{1}_{3} + \mathbf{J}_{\vec{P}} \left(\vec{E}\right) \ .
\end{align}
This 3$\times$3 matrix can be inverted explicitly.
An external numerical solver introduced in Subsection \ref{sec:ode_solver} is responsible for the time integration of the ODE system.

\section{Dispersion relations}\label{sec:Dispersion}

\subsection{Simple examples}
To investigate dispersion effects in the finite differences approximation an elementary example is worked out in the following.
For the values $\nu \in \left \lbrace -1,0,1 \right\rbrace $ the expansion of $g(x)$ in matrix form is given by
\begin{align}\label{StencilMatrixThirdOrder}
\begin{pmatrix}
g(x -  \Delta_x) \\
g(x ) \\
g(x +  \Delta_x) \\
\end{pmatrix} 
= 
\begin{pmatrix}
1 & -1 & 1/2 \\
1 & 0 & 0 \\
1 & 1 & 1/2 \\
\end{pmatrix}
\cdot
\begin{pmatrix}
g(x) \\
\Delta_x \, g'(x) \\
\Delta^2_x \, g''(x) \\
\end{pmatrix} \ .
\end{align}
It has to be noted that Equation \eqref{StencilMatrixThirdOrder} is not restrictive to one specific kind of finite difference method.
To obtain an expression for the first derivative the matrix in \eqref{StencilMatrixThirdOrder} has to be inverted.
This leads to
\begin{align}\label{InvertedStencilMatrixThirdOrder}
\begin{pmatrix}
g(x) \\
\Delta_x \, g'(x) \\
\Delta^2_x \, g''(x) \\
\end{pmatrix}  = 
\begin{pmatrix}
0 & 1 & 0 \\
-1/2 & 0 & 1/2 \\
1 & -2 & 1 \\
\end{pmatrix} 
\cdot
\begin{pmatrix}
g(x -  \Delta_x) \\
g(x ) \\
g(x +  \Delta_x) \\
\end{pmatrix} \ .
\end{align}

Assuming that $g(x)$ is one of the two polarizations of an electromagnetic mode that propagates in positive ($ + $) or negative ($ - $) $x$-direction in 1D, the corresponding equation of motion reads
\begin{align} \label{PDESimple}
\left( \partial_t \pm c \, \partial_x \right) g_\pm(x;t)=0 \ .
\end{align}
Making use of \eqref{InvertedStencilMatrixThirdOrder}, Equation \eqref{PDESimple} becomes for symmetric forward and backward differentiation
\begin{align}\label{NewPDESimple}
\partial_t g_\pm(x;t) \pm  \frac{c}{2 \Delta_x} \left( g_\pm(x+\Delta_x; t ) - g_\pm(x-\Delta_x;t)\right) =0 \ ,
\end{align}
where the spatial derivative of $g$ is accurate up to second order in $\Delta_x$. Equation \eqref{NewPDESimple} can be solved analytically by a plane wave ansatz
\begin{equation}\label{eq:plane_wave_exa}
	g_\pm(x;t)= e^{-i\omega t + ik_\pm x} \ ,
\end{equation}
where $ k_+ > 0 $ and $ k_-<0 $ to obtain
\begin{align}
-&\left(  \mathfrak{Re} (\omega) + i \, \mathfrak{Im}(\omega) \right) \pm \frac{c }{ \Delta_x}  \sin \left(k_\pm \Delta_x \right) =0 \, ,  \\
\Rightarrow \ &\omega  =  \pm \frac{c }{\Delta_x} \sin(k_\pm \Delta_x)\ .
\end{align}

However, the derivative in \eqref{PDESimple} can also be approximated for a forward-propagating mode $ g_+ $ by a first order forward finite difference.
In this case, Equation \eqref{NewPDESimple} can be written as
\begin{align}\label{NewPDESimpleForward}
\partial_t g_{+}(x;t) +  \frac{c}{ \Delta_x} \left( g_{+}(x+\Delta_x; t ) - g_{+}(x;t)\right) =0 \ ,
\end{align}
where the solution of \eqref{NewPDESimpleForward} with $ \omega \in \mathbb{C} $ reads
\begin{align}
-& \omega + \frac{c }{ \Delta_x}   \left( \sin (k_+\Delta_x) +i - i \cos(k_+\Delta_x) \right)=0 \ ,  \\
\Rightarrow \ &\mathfrak{Re} (\omega) =  \frac{c}{\Delta_x} \sin(k_+\Delta_x)\, , \\
\Rightarrow \ &\mathfrak{Im}(\omega) = \frac{c }{\Delta_x} \left( 1- \cos (k_+\Delta_x) \right) \ .
\end{align}

Further, the derivative in \eqref{PDESimple} for the backward-propagating part can be approximated by a first order backward finite difference.
In this case, Equation \eqref{NewPDESimple} can be approximated for the backward-propagating case as
\begin{align}
\partial_t g_{-}(x;t) -  \frac{c}{ \Delta_x} \left( g_{-}(x;t ) - g_{-}(x-\Delta_x;t)\right) =0 
\end{align}
with the solution
\begin{align}
-&\omega - \frac{c }{ \Delta_x}   \left(\sin (k_-\Delta_x) -i + i \cos(k_-\Delta_x) \right)=0 \ ,  \\
\Rightarrow \ &\mathfrak{Re}(\omega) = - \frac{c }{\Delta_x} \sin(k_-\Delta_x)\ , \\
\Rightarrow \ &\mathfrak{Im}(\omega) = \frac{c }{\Delta_x} \left( 1- \cos (k_-\Delta_x) \right) \ .
\end{align}

\begin{SCfigure}
	\includegraphics[width=0.4\linewidth]{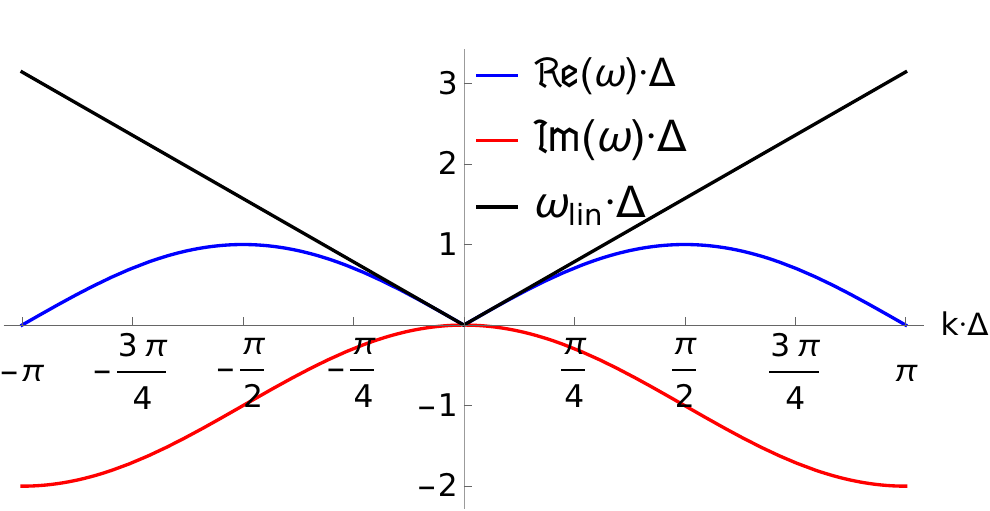}
	\caption[Dispersion relations for varying finite differences schemes at low order]{Dispersion relations obtained with low-order finite differences for a simple plane wave.
	Here $c=1$.
	\textit{Top left}: for a second order symmetric forward and backward finite difference scheme.
	\textit{Top right}: for a first order forward finite difference for the forward-propagating mode and a first order backward finite difference for the backward-propagating mode.
	\textit{Bottom}: for a first order forward finite difference for the backward-propagating mode and a first order backward finite difference for the forward-propagating mode.
	The linear vacuum is given by the black line as reference with the vacuum speed of light $ c  $ set to unity.
    Note that all stencils result in a symmetric dispersion relation.
    The phase velocity is the same for both directions.}
	\label{fig:dispersion_exa}
\end{SCfigure}

The dispersion relations connect a simulated wave with a specified wavelength to its phase velocity on the lattice.
While in the case of symmetric differentiation there is no imaginary part of $ \omega $, it does show up for biased differentiation.
The derived dispersion relations are shown in the top row of Figure \ref{fig:dispersion_exa}.
The Nyquist frequency $ f_\text{Ny} = \Delta_x^{-1}/2 $, corresponding to $ k \cdot \Delta_x = \pi $, marks the point where $ \mathfrak{Re}(\omega) =0$ for $k\neq 0$ in all cases. 

\begin{figure*}
	\centering
		\includegraphics[width=0.38\linewidth]{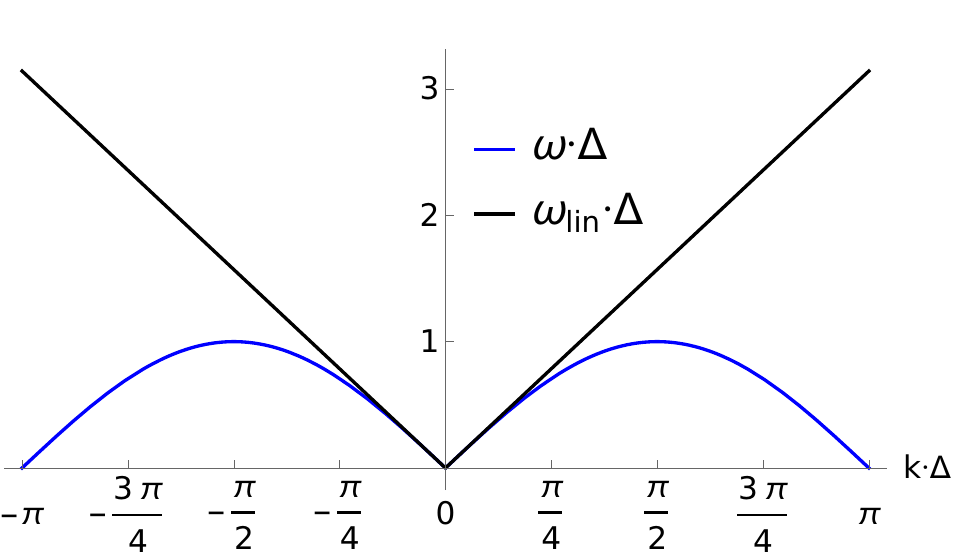}
	\qquad\qquad
		\includegraphics[width=0.38\linewidth]{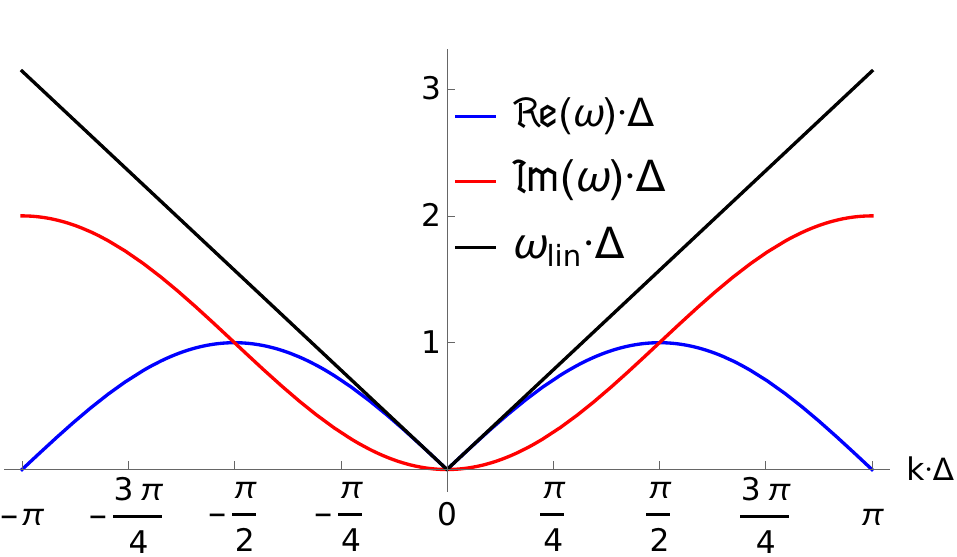}
\end{figure*}

\subsection{The scheme at fourth order}
The imaginary part amplifies the wave and thus makes this scheme unstable.
By differentiating biased against the propagation direction, however, the imaginary part switches signs and has a damping effect.
The calculations are straightforward and the result is shown at the bottom of Figure \ref{fig:dispersion_exa}.
This behavior can be tuned with higher order schemes in order to get a more vacuum-like real part and to defer the imaginary part such that it becomes relevant only for short wavelengths and damps only those.
As the real part of $ \omega $ eventually decreases in any scheme as it reaches the Nyquist frequency, the damping of those high-frequency waves serves as anti-aliasing effect.
This is shown below.

Now, the unbalanced coefficients for the first order discrete derivative in fourth order accuracy in $\Delta_x$, $g'_{\nu \in \left\lbrace -3,1\right\rbrace}$ and $g'_{\nu \in \left\lbrace -1,3\right\rbrace}$ in the above notation, are given with the help of Equation \eqref{StencilMatrixSixthOrder}.
This results in 
\begin{align} \label{CoefficientsForBackFirstDerivative}
\begin{aligned}
\Delta_x \, g'_{\nu \in \left\lbrace -3,1\right\rbrace} &= 
\begin{pmatrix}
-1/12 & 1/2 & -3/2 & 5/6 & 1/4\\
\end{pmatrix}
\begin{pmatrix}
g(x - 3 \Delta_x) \\
g(x - 2 \Delta_x) \\
g(x -  \Delta_x) \\
g(x ) \\
g(x +  \Delta_x) \\
\end{pmatrix}  \, ,\\
\ \\
\Delta_x \, g'_{\nu \in \left\lbrace -1,3\right\rbrace} &= 
\begin{pmatrix}
-1/4 & -5/6 & 3/2 & -1/2 & 1/12\\
\end{pmatrix}
\begin{pmatrix}
g(x - \Delta_x) \\
g(x ) \\
g(x + \Delta_x) \\
g(x + 2\Delta_x) \\
g(x +  3 \Delta_x) \\
\end{pmatrix}  \, .
\end{aligned}
\end{align}
To make the connection to Equation \eqref{eq:dtrx}, $ g $ is replaced by the components of $ \mathbf{R}_{x} \vec{f} $.
As explained above, $ \mathbf{R}_{x} \vec{f} $ has components propagating in different directions that are differentiated with biases in the respective opposite direction.
This means that the first three components of $ \mathbf{R}_{x} \vec{f} $ are differentiated in the same way as $ g $ in the first line of \eqref{CoefficientsForBackFirstDerivative} and the last three components of $ \mathbf{R}_{x} \vec{f} $ are differentiated as $ g $ in the second line of \eqref{CoefficientsForBackFirstDerivative}.
Following these instructions for Equation \eqref{DefinitionDiscreteDerivativeWithStencil} yields
\begin{align}
	{\mathcal{D}_{x}} \left(\mathbf{R}_{x} \vec{f}\right) \left(x,y,z\right) =   \sum_{\nu = -3}^{3} \frac{1}{\Delta_{x}} S_{\nu} \left(\mathbf{R}_{x} \vec{f}\right)\left( x+\nu\Delta_x,y,z\right) \ , 
\end{align}
with the fourth order stencils given by
\begin{align}\label{eq:stencilsnonlinear}
	\begin{aligned}
		S^4_{+3} &= \text{diag}(1/12,1/12,1/12,0,0, 0) \ ,            &  S^4_{-3} &= \text{diag}(0,0,0,-1/12,-1/12,-1/12) \ ,\\
		S^4_{+2} &= \text{diag}(-1/2,-1/2,-1/2,0,0, 0) \ ,       & S^4_{-2} &= \text{diag}(0,0,0,1/2,1/2, 1/2) \ ,\\
		S^4_{+1} &= \text{diag}(3/2,3/2,3/2,1/4,1/4,1/4) \ ,   & S^4_{-1} &= \text{diag}(-1/4,-1/4,-1/4,-3/2,-3/2,-3/2) \ ,\\
		S^4_{0} &= \text{diag}(-5/6,-5/6,-5/6,5/6,5/6,5/6)\ .
	\end{aligned}
\end{align}

In consequence of the symmetry and redundancy, it is instructive to rewrite the stencil matrices in terms of its components applied to forward- and backward-propagating modes.
The fourth order stencil matrices can be expressed as
\begin{equation}
S^4_\nu = \text{diag}(s^4_b[\nu],s^4_b[\nu],s^4_b[\nu],s^4_f[\nu],s^4_f[\nu], s^4_f[\nu]) \ ,  
\end{equation}
where the components applied to forward- and backward-propagating modes $s^4_{f}$ and $s^4_{b}$ are given by
\begin{equation}\label{eq:stencil_components}
\begin{aligned}
	s^4_f \big\rvert_{\nu=-3,...,1}= &  \left\{-\frac{1}{12},\frac{1}{2},-\frac{3}{2},\frac{5}{6},\frac{1}{4}\right\} \ , \\
	s^4_b \big\rvert_{\nu=-1,...,3} = & \left\{-\frac{1}{4},-\frac{5}{6},\frac{3}{2},-\frac{1}{2},\frac{1}{12}\right\} \ ,
\end{aligned}
\end{equation}
and $s^4_{f/b}=0$ for out-of-range values of $\nu$.
Due to the obvious symmetry, the stencil matrix can also be written as
\begin{equation}
	S^4_\nu = \text{diag}(-s^4_f[-\nu],-s^4_f[-\nu],-s^4_f[-\nu],s^4_f[\nu],s^4_f[\nu], s^4_f[\nu]) \ ,  
\end{equation}
for the whole range $\nu=-3,...,3$.

\subsubsection{Dispersion relations at order four and thirteen}
The solver has implementations of the scheme up to order thirteen.
In the present work the currently maximal available accuracy is used.
By going seven steps of $ \Delta_x $ forward and six steps backward in the forward-biased differentiation and vice versa in the backward-biased case, the unbalance is kept very small.
The components of the low-bias stencils up to order thirteen are listed in \ref{app:stencils}.

For a single plane wave the Heisenberg--Euler Lagrangian reduces to the Maxwell Lagrangian, as in this case $ \mathcal{F} = \mathcal{G} = 0 $ holds.
The vacuum is thus the linear one.
Inserting a plane wave, without loss of generality propagating on the x-axis,
\begin{equation}
\vec{E}(x;t)=\vec{A} \, e^{-i(\omega t-kx)} \ ,
\end{equation}
into the propagation equation \eqref{PropEquation} yields
\begin{equation}
	-i \omega \mathbf{1}_6= \mathbf{Z}_x^{\text{lin}} \, \mathbf{R}_x^{\text{T}} \sum_\nu \frac{1}{\Delta_x} S_\nu e^{i \nu k \Delta_x} \mathbf{R}_x \ .
\end{equation}
Since the stencil matrices are diagonal, this can be written as
\begin{equation}
	-i \omega \mathbf{1}_6=  \mathbf{Z}_x^{\text{lin}} \, \mathbf{R}_x^{\text{T}} \mathbf{R}_x \frac{1}{\Delta_x}  \sum_\nu S_\nu e^{i \nu k \Delta_x}  \ ,
\end{equation}
which, when multiplied with $\mathbf{R}_x$ from the left, gives
\begin{equation}
	-i \, \mathbf{R}_x  \, \omega =  \text{diag}(0,1,1,0,-1, -1)\, \mathbf{R}_x   \frac{1}{\Delta_x}  \sum_\nu S_\nu e^{i \nu k \Delta_x}  \ .
\end{equation}
Inserting the stencil matrices expressed in terms of their components applied to forward- and backward-propagating modes $s_f$ and $s_b$ results in
\begin{equation}
	-i \, \mathbf{R}_x \, \omega = \text{diag}(0,1,1,0,-1, -1)\, \mathbf{R}_x   \frac{1}{\Delta_x}  \sum_\nu \text{diag}\left(s_b[\nu],s_b[\nu],s_b[\nu],s_f[\nu],s_f[\nu], s_f[\nu]\right) e^{i \nu k \Delta_x}  \ .
\end{equation}
It can be seen that the equation above holds irrespective of the axis of propagation since $\mathbf{R}_x$ can be canceled.
Hence, the equation becomes
\begin{equation}\label{eq:Disp}
	\omega \Delta = i \, \text{diag}(0,1,1,0,-1, -1)  \sum_\nu \text{diag}\left(s_b[\nu],s_b[\nu],s_b[\nu],s_f[\nu],s_f[\nu], s_f[\nu]\right) e^{i \nu k \Delta}  \ .
\end{equation}
There are the distinct cases of forward- and backward-propagating modes, and non-propagating ones.
It is obtained for

\begin{figure}
	\centering
		\begin{subfigure}{.5\textwidth}
        \includegraphics[width=0.95\linewidth]{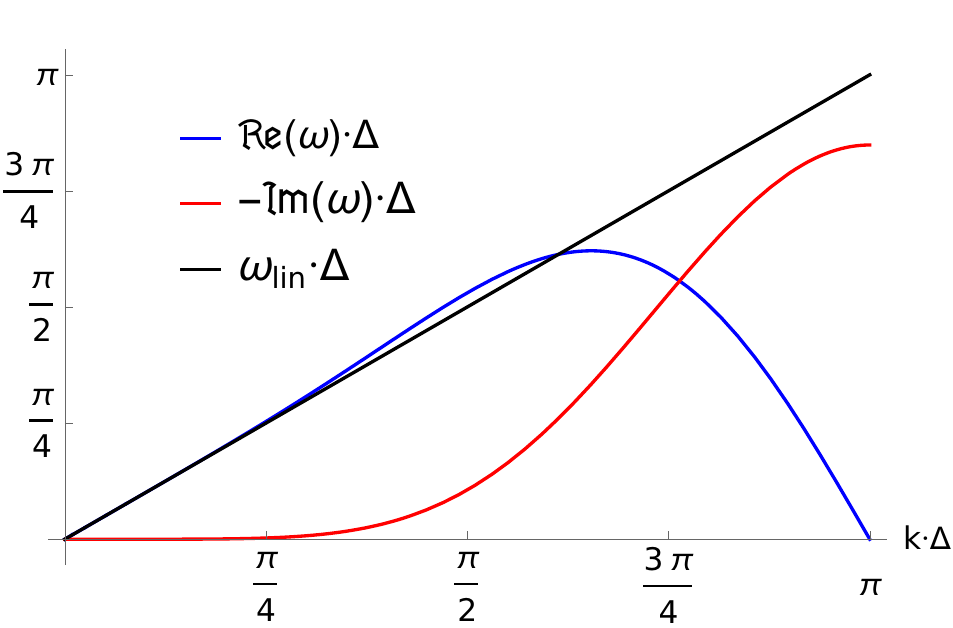}
	\end{subfigure}%
	\begin{subfigure}{.5\textwidth}
		\includegraphics[width=0.95\linewidth]{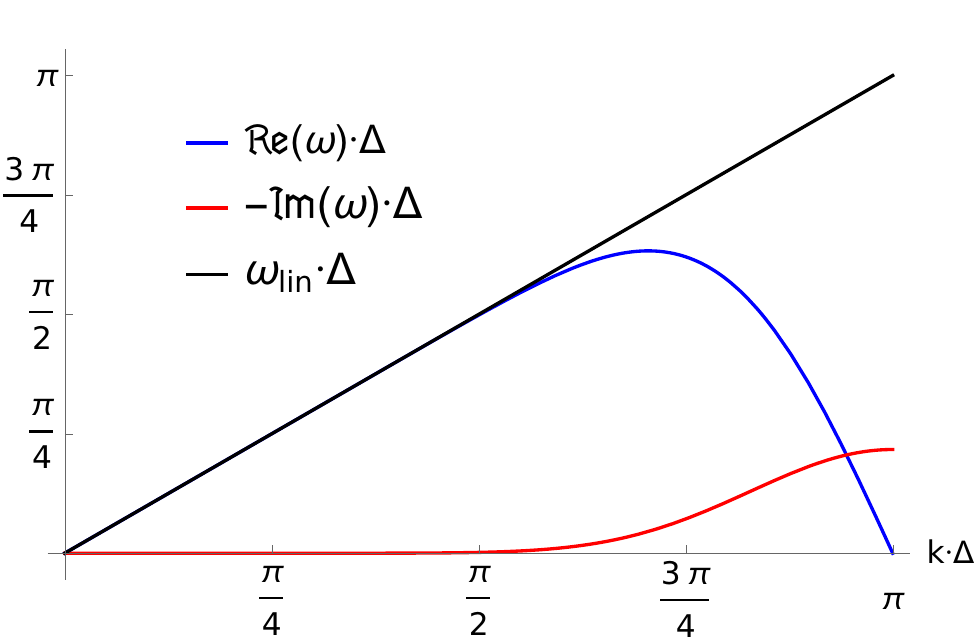}
	\end{subfigure}
	\caption[Dispersion relation at fourth and thirteenth order]{Dispersion relations of the numerical scheme for minimally biased finite differences at order four (\textit{left}) and order thirteen (\textit{right}).
		The black lines represent the real vacuum dispersion relation with $c=1$.
		$\Delta$ is the grid spacing, the physical distance between lattice points.
		For better visibility of the symmetric plots only the values for $ k \geq 0 $ are shown.}
	\label{fig:dispersion}
\end{figure}

\begin{itemize}
	\item a forward-propagating mode (implying $k>0$):
	\begin{equation}
		\omega \Delta = \sum_\nu s_f[\nu] \left( -i \cos(\nu k \Delta) + \sin(\nu k \Delta )  \right) \, ;
	\end{equation}
	\item a backward-propagating mode (implying $k<0$):
\begin{equation}
	\omega \Delta = \sum_\nu s_b[\nu] \left( i \cos(\nu k\Delta ) - \sin(\nu k \Delta)  \right) = \sum_\nu s_f[-\nu] \left( -i \cos(\nu k\Delta ) + \sin(\nu k \Delta)  \right) \, ;
\end{equation}
    and
   \item a non-propagating mode (implying $k=0$) :
   \begin{equation}
   	\omega = 0 \ .
   \end{equation}
\end{itemize}
The stencil components to all orders up to order thirteen are listed in \ref{app:stencils} and plots of the dispersion relations are given.
The results for order four and thirteen are visualized in Figure \ref{fig:dispersion}, the latter being the ones used for simulations of this work.
The minimal resolvable distance $\Delta$ is the spacing between the lattice points, the resolution of the lattice.
It is the decisive parameter to tune the dispersion effects for a given wavelength and is given by the ratio of physical length and lattice points.

The real parts of $\omega$ start in the vicinity of the black line and deviate from it for shorter wavelengths or smaller grid resolutions.
There is less deviation at higher orders.
The imaginary part of $\omega$ starts with values close to zero for large wavelengths/high grid resolutions and decreases until the Nyquist frequency is reached.
The negative imaginary part in this scheme has the positive effect to promptly annihilate those modes that deviate strongly from the vacuum dispersion relation.

At higher orders, the damping effect of the imaginary part is deferred to shorter wavelengths and is overall smaller.
It can further be seen that for higher orders the real part stays closer to the linear vacuum for smaller wavelengths.
Higher orders also defer and decrease the rise of the imaginary part.
A relatively strong bias as in the fourth order scheme \eqref{CoefficientsForBackFirstDerivative} causes a superluminal phase-velocity in a given $ k \cdot \Delta $ range, a critical regime where the dispersion relation deviates remarkably from the linear vacuum.
The well balanced order thirteen scheme has a smaller critical regime and is well-behaved for a larger range of wavelengths.

For small wavelengths the curve showing the real part of $ \omega $ falls off and the imaginary part of $\omega$ causes a damping.
As a result, there are two $ k $'s for each $\mathfrak{Re}(\omega)$.
At the wavelength corresponding to the Nyquist frequency, $k\cdot \Delta = \pi $, $\mathfrak{Re}(\omega)$ becomes zero as in the simple scenarios above.
The high-$k$ values cause nonphysical standing waves, as $\mathfrak{Re}(\omega)/k \rightarrow 0$, which would cause a self-heating of the system.
The damping takes care of their annihilation.
The superluminosity which can be seen in the fourth order finite difference differentiation is thus an acceptable add-on to the favor of a damping of nonphysical modes.
With increasing accuracy and balancing as in the order thirteen scheme there is no superluminal range left.

\subsubsection{Damping of modes}
\begin{figure}
	\centering
	\begin{subfigure}{.5\textwidth}
		\includegraphics[width=0.85\linewidth]{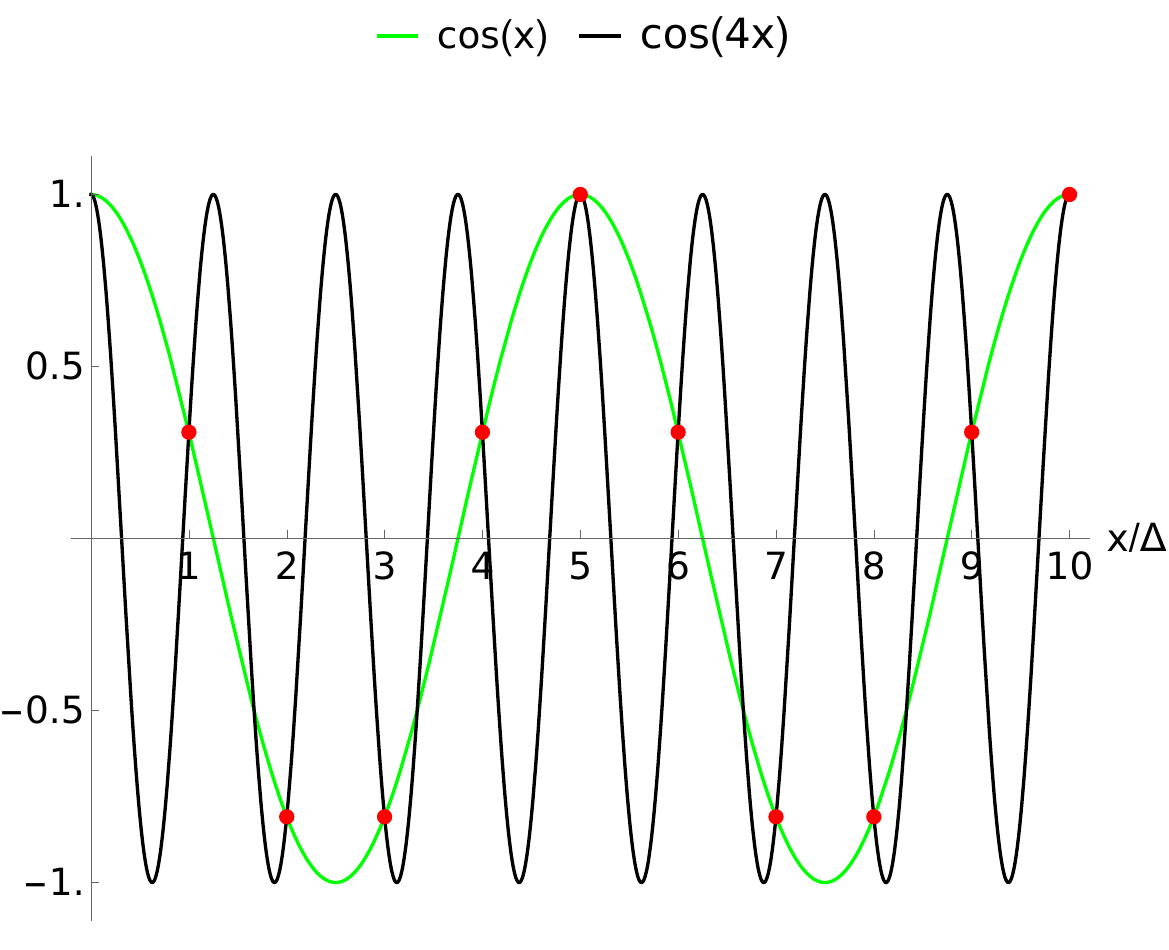}
		\vspace*{0.3cm}
		\caption*{Effect of overshooting the Nyquist frequency.
			\textit{Left}: two cosine functions that, when periodically evaluated at multiples of the distance $ \Delta = 0.4\pi $, have the same values.
			\textit{Right}: simulations to verify the behavior on the grid.
			On a 1D line with point spacing $ \Delta = 0.1 \si{\micro \metre} $
			two waves with wavelengths $ 5 \Delta $ and $ 5/4 \Delta $, respectively, are simulated, corresponding to the analytical scenario on the left.
			It can be seen that the discretization makes no difference between the two.
			The simulated waves are shown in the initial state (\textit{top right}) and after they have propagated the distance of 90 periods (\textit{bottom right}).
			The damping effect initiated by the imaginary part of $ \omega $ in the dispersion relation is the same for both waves.}
					\vspace*{1cm}
	\end{subfigure}%
	\begin{subfigure}{.5\textwidth}
		\includegraphics[width=0.87\linewidth]{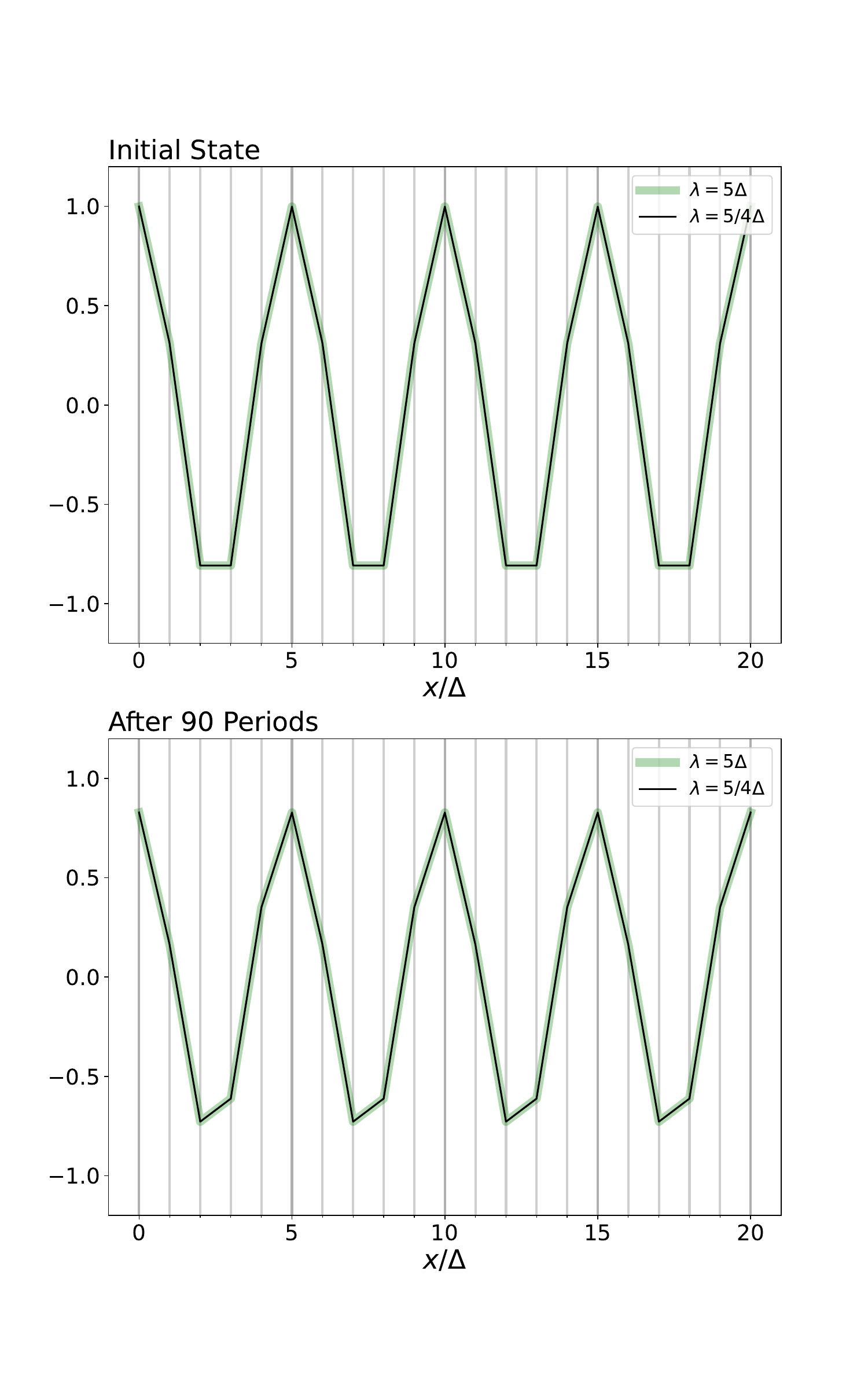}
	\end{subfigure}
	\vspace*{-1.2cm}
\caption[On the Nyquist frequency]{On the Nyquist frequency.}
\label{fig:nyquist}
\end{figure}

In the numerical investigation it is demonstrated that the damping has noticeable effects already at an earlier stage than a first look at the above plot would suggest.
Note that the Nyquist frequency, on the other hand, $f_\text{Ny} = \Delta^{-1}/2$ (corresponding to $k\cdot \Delta=\pi$), is not the limiting factor for wave modeling in this scheme.
It has to be stayed well below the Nyquist limit for accurately time-evolved waves.
Note also that the simulation of a high-frequency wave does not overshoot the Nyquist limit, as it would be sampled as a wave with lower frequency, see Figure \ref{fig:nyquist}.
The relevant frequency scale of the dispersion relation in Figure \ref{fig:dispersion} thus ranges only to the Nyquist limit and the scheme is generally stable for any frequency.

It has to be noted that energy is not conserved on the grid as a consequence of the amplitude damping during the propagation.
The imaginary part of $\omega$ and therefore the damping effect might be small if the grid resolution is high, but there is a trade-off since high grid resolutions are computationally more expensive.

To visualize the effects of the dispersion relation at varying wavelengths the propagation of a plane wave in one direction of a two-dimensional square grid with side length \SI{80}{\micro\metre} divided into $ 1024 \times 1024 $ points is investigated.
The resolution is therefore given by $ \Delta=\SI{80}{\micro \metre}/1024 $, $ \Delta^{-1} = \SI{128e5}{\per \metre}  $.

\begin{figure}
	\centering
	\begin{subfigure}{.5\textwidth}
		\includegraphics[width=\linewidth]{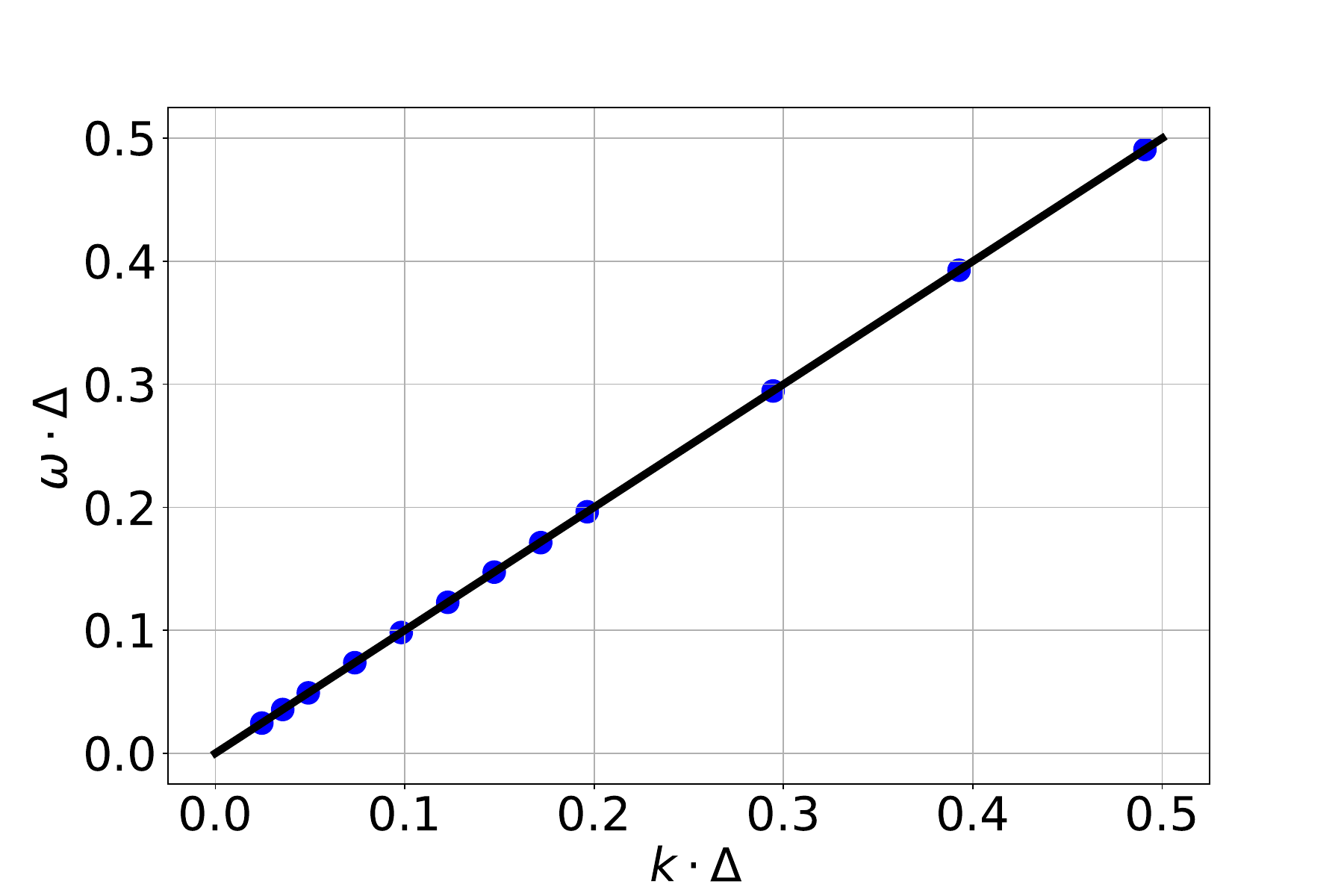}
		\label{fig:dispersion_numerical}
	\end{subfigure}%
	\begin{subfigure}{.5\textwidth}
		\includegraphics[width=\linewidth]{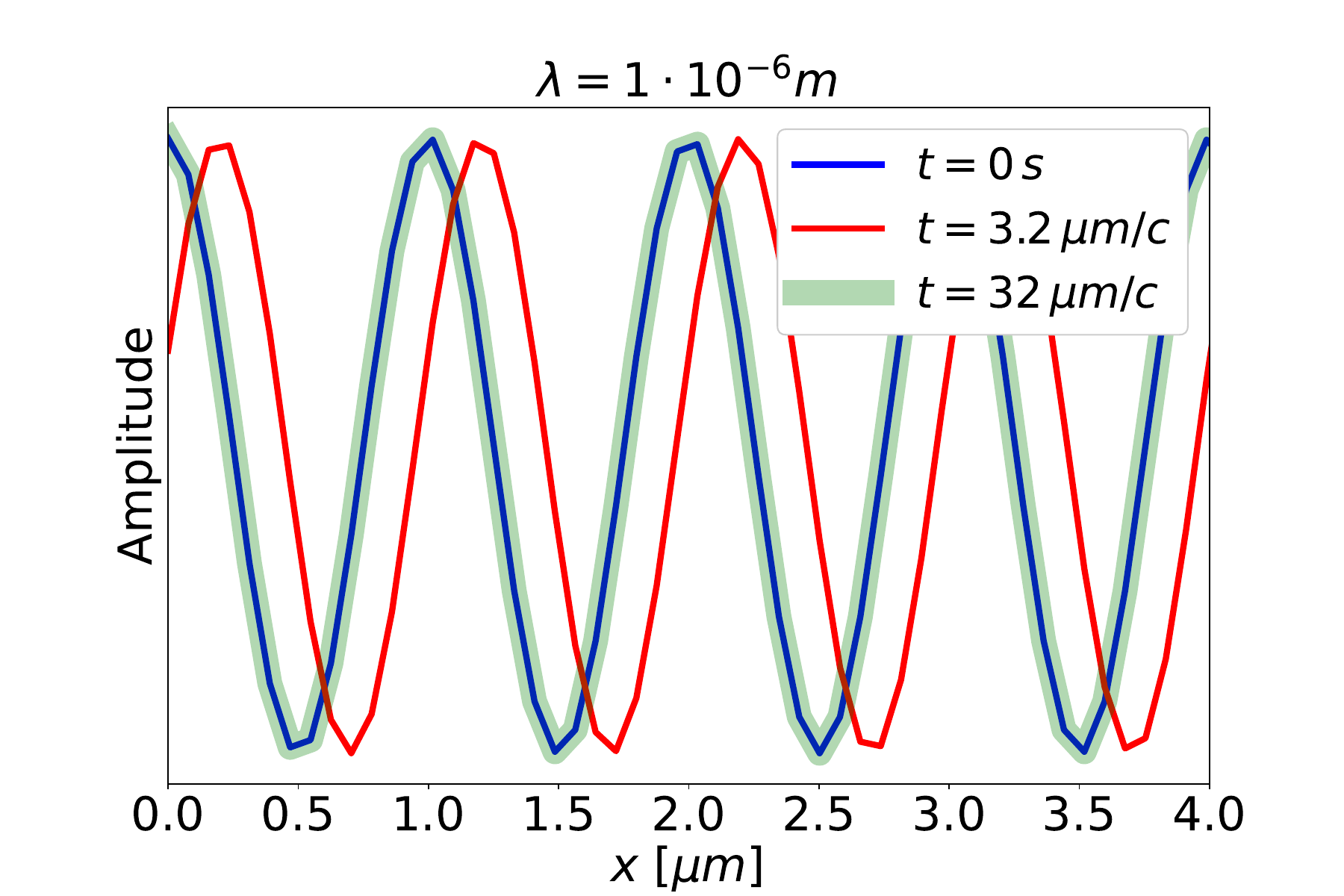}
		\label{fig:dispersion_k1e6}
	\end{subfigure}
	\caption[Numerical tests of the dispersion for low frequencies.]{Numerical tests of the dispersion for low frequencies.
		\textit{Left}: numerical results of dispersion relation tests.
		Simulation results (\textit{blue dots}) are in agreement with the free vacuum (\textit{black line}) for wavelengths from \SI{20}{\micro \metre} to \SI{1}{\micro \metre}.
		\textit{Right}: the plane wave corresponding to the rightmost point of the left figure with $\lambda=\SI{1}{\micro \metre}$ at different time steps.
		The wave is not that accurately modeled anymore but still does not lose energy in the given amount of time -- the amplitude stays the same.
		Even after 32 periods the wave is perfectly overlapping with its initial state.}
	\label{fig:Disp_and_k1e6}
\end{figure}

\begin{figure}
	\centering
	\begin{subfigure}{.5\textwidth}
		\includegraphics[width=\linewidth]{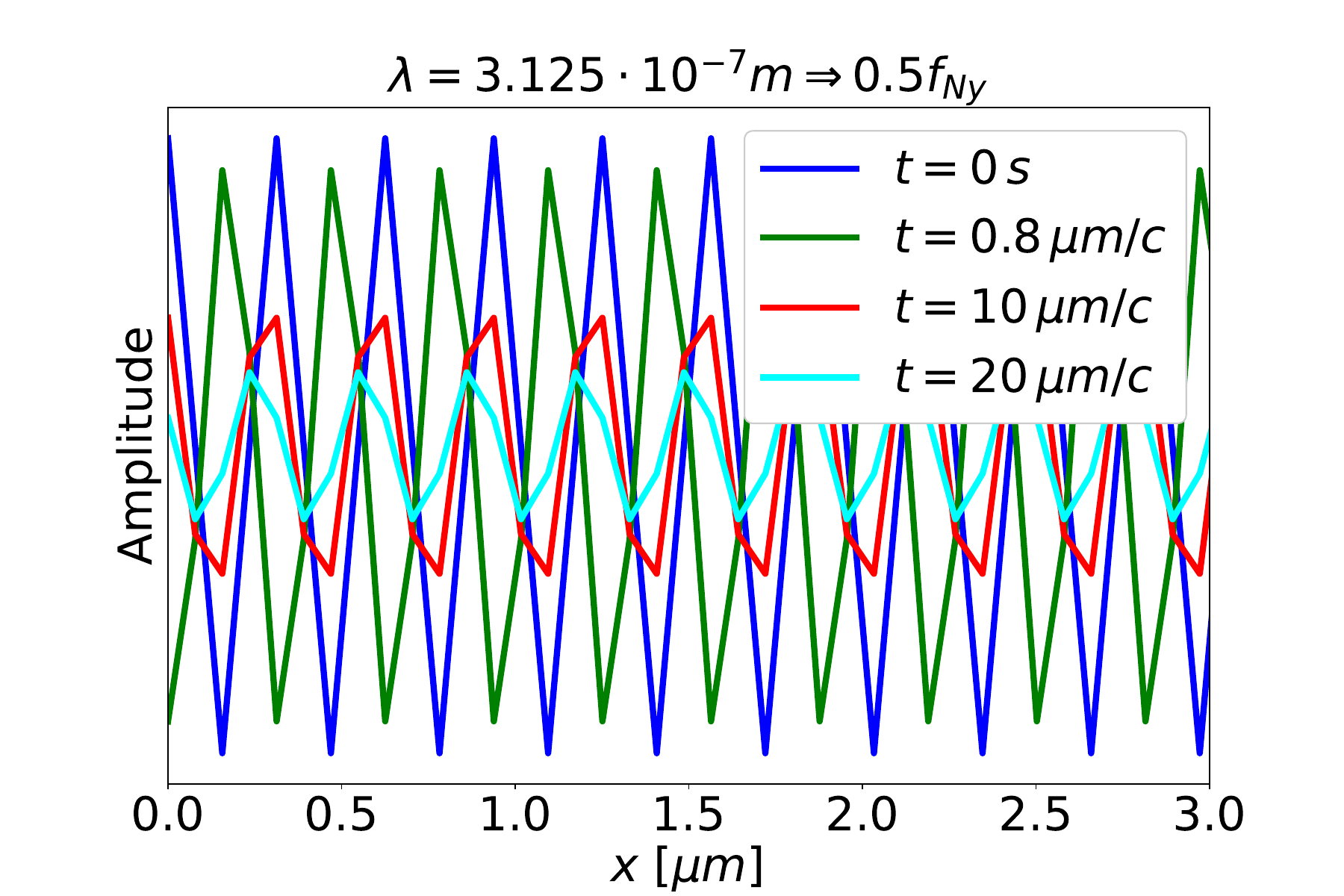}
	\end{subfigure}%
	\begin{subfigure}{.5\textwidth}
		\includegraphics[width=\linewidth]{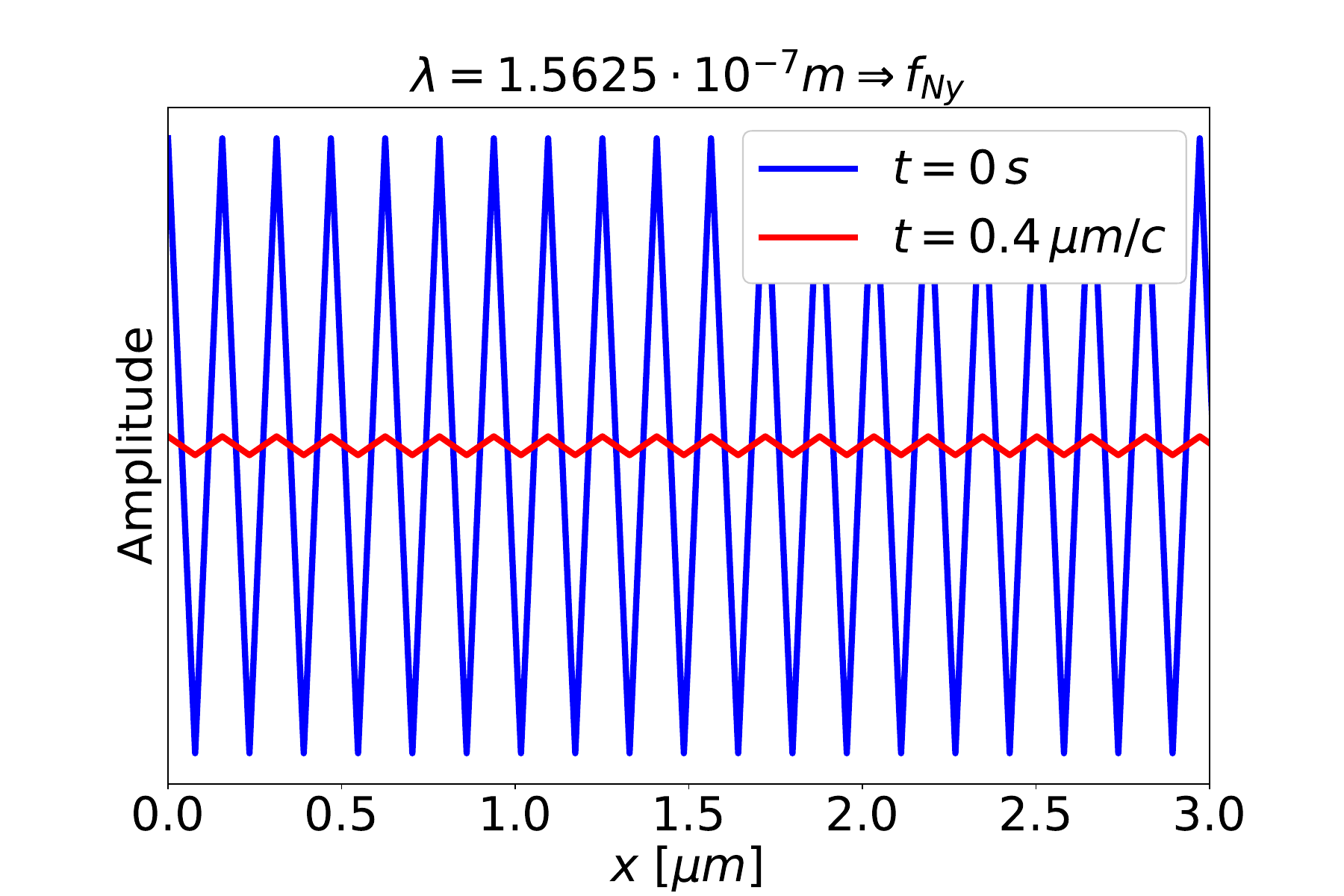}
	\end{subfigure}
	\caption[Numerical tests of the dispersion for high frequencies.]{Numerical tests of the dispersion for high frequencies.
		\textit{Left}: a plane wave with $\lambda=\SI{312.5}{\nano \metre}$ at different time steps.
		The grid resolution is given by $ \Delta = \SI{80}{\micro \metre}/1024 $ in the propagation direction.
		In the free vacuum the corresponding frequency is to be associated to half the Nyquist frequency.
		32 periods have passed after \SI{10}{\micro \metre}/c and 64 periods after \SI{20}{\micro \metre}/c.
		The damping and also the barely subluminal phase velocity are, after a long enough propagation time, noticeable even at this wavelength.
		The initial peak position has clearly shifted after \SI{20}{\micro \metre}/c.
		\textit{Right}: a plane wave with $\lambda=\SI{156.25}{\nano \metre}$ at different time steps with the same grid settings.
		In the free vacuum the corresponding frequency is to be associated to the Nyquist frequency.
		The wave is standing and the damping is strong, resulting in a rapid annihilation.}
	\label{fig:Nyquist}
\end{figure}

From Figure \ref{fig:Disp_and_k1e6} it can be deduced that the waves are well-behaved and quite well-modeled for $k\cdot \Delta \lesssim 0.5$.
From Figure \ref{fig:Nyquist}, it can be inferred that already at half the Nyquist frequency the damping is non-negligible for relevant time scales.
The Nyquist frequency in this scenario is given by $f_\text{Ny}=\SI{64e5}{\per \metre} $ ($\lambda=\SI{1.5625e-7}{\m}$).
A video demonstrating the evolution of a plane wave on the lattice with a wavelength corresponding to half the Nyquist frequency can be found in the \textit{Mendeley Data} repository \cite{Lindner2022}.
The conclusion is that for a proper modeling and to avert damping effects, one is obliged to adapt the grid resolutions to the lowest wavelength such that $\Delta \lesssim 1/12 \, \lambda$.
While this relation is not a hard limit and can be relaxed in many cases without detriments, it poses a safe rule of thumb for long-time simulations.
A sufficiently fine grid resolution is pertinent for a clean analysis of the polarization flipping and harmonic generation effects in the present work.

\section{Implementation and scaling}\label{sec:scaling}

\subsection{Solving the ODE and processing the data}\label{sec:ode_solver}
For the numerical solution of the nonlinear system of ODEs for $\vec{f}$ in \eqref{PropEquation}
the \textit{CVODE} solver is used, which is part of the \textit{SUNDIALS} family of solvers \cite{Haireretal1993,Hindmarshetal2021,Hindmarshetal2005,Gardneretal2022}.
The implicit Adams method (Adams-Moulton formula) in conjunction with a fixed-point iteration is utilized.
 
The numerical time integration error is controllable with \textit{CVODE} by setting relative and absolute integrator tolerances per user-defined step.
The solver adapts its internal time step sizes according to the system's dynamics, guided by the tolerances.
Larger time steps are performed in quiet regions and shorter steps in highly dynamic regions.
Errors per step accumulate to a global error.

Integrator error tolerances are set to below $ 10^{-12} $ for the present work.
Since \textit{CVODE} integrates with high accuracy, the size of a time step can be defined as is convenient and in this work varies in the range of \SI{1}{\femto \second} to \SI{6}{\femto \second}, or \SI{0.3}{\micro\metre}/c to \SI{2}{\micro\metre}/c .
The computation time to achieve a given numerical integration error threshold is approximately independent of the stencil order, but the correct physical solution is rather approached with a higher stencil order.

Output data are written at each user-defined time step optionally into convenient comma-separated values (\textit{CSV}) files or a compact and fast binary file.
A \textit{CSV} output file contains six columns for the field components $E_x \, , \ E_y \, , \ E_z \, , \ B_x \, , \ B_y$, and $B_z$;
each column holding the grid values of the corresponding field component.
A binary output file aligns all six field components for every lattice point.
The post-processing of the data for this work is done with the help of \textit{Python} scripts, \textit{Jupyter notebooks} employing the \textit{SciPy} library, and with the help of {\it Mathematica} and \textit{Paraview} \cite{Kluyveretal2016,Virtanenetal2020,Mathematica,Ahrensetal2005}.

\subsection{Parallelization of the algorithm}\label{sec:Parallelization}

\begin{figure}
	\centering
	\includegraphics[width=.5\textwidth]{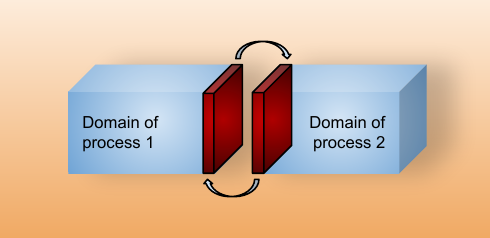}
	\caption[Ghost-cell exchange]{Ghost-cell exchange.
	The blue boxes represent sub-lattices (patches), batches of the interdependent grid data, each being processed on single cores.
	The processes (compute cores) exchange their boundary cells' values with the neighboring processes.
	The boundary regions, whose required depths depend on the order of the finite differences scheme, are indicated in red.
	This exchange is performed sequentially for each dimension.
    Here it is shown for one boundary region between two processes.}
	\label{fig:halo}
\end{figure}

For faster computation and scalability the code is parallelized.
To this end a Cartesian domain decomposition of the lattice is performed, allowing individual compute cores to process patches of the lattice.
The number of these sub-lattices is determined by the user who may subdivide each dimension of the lattice into a number of equally sized parts.
The finite differences scheme used to discretize derivatives requires values from neighboring sub-lattices when it is applied at the boundaries of a patch.
For this purpose, ghost cells are placed at the boundaries and updated via message passing making use of \textit{MPI} \cite{Clarkeetal1994,mpi}, see Figure \ref{fig:halo}.

Since the sub-lattices form a Cartesian grid, the communication scheme is conveniently implemented in an \textit{MPI} virtual Cartesian topology.
Output to \textit{CSV} format is written to one file per process (patch), whereas output in binary form is written to one single file per step, making use of \textit{MPI IO}.

\begin{figure}
	\centering
   	\includegraphics[width=0.65\linewidth]{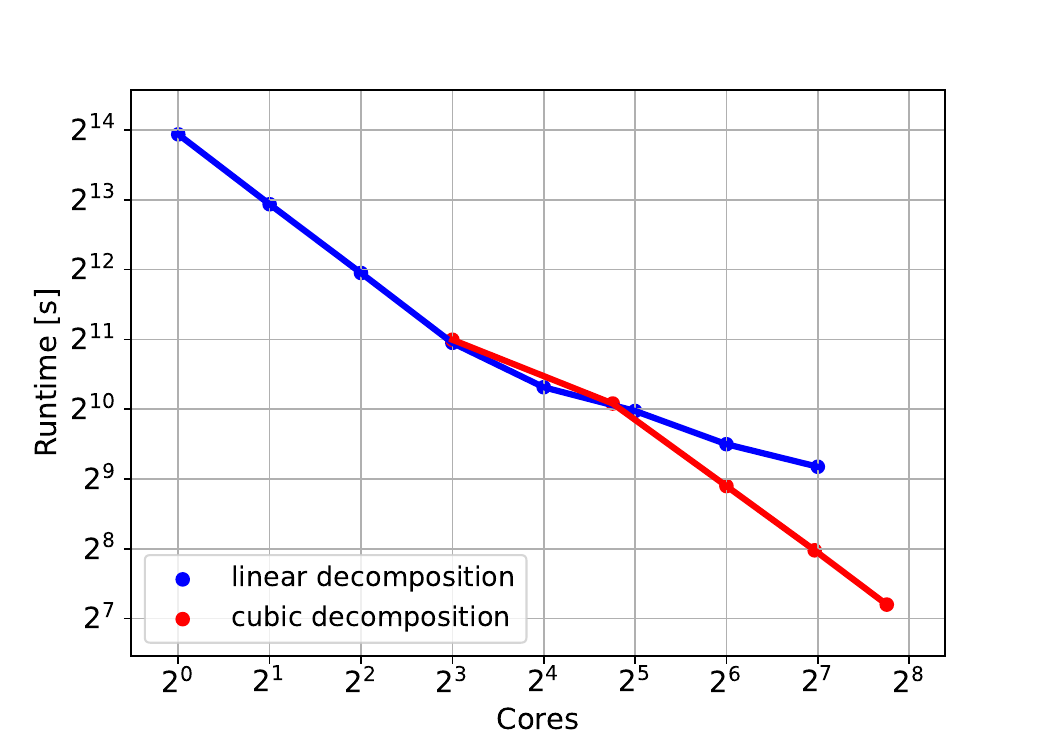}
   	\caption[Strong scaling test]{Strong scaling test for a relatively cheap simulation in 3D.
   		The overall lattice keeps its size and is split into smaller and smaller patches to be processed by single cores.
   		The runtimes nearly halve for a doubling of cores for up to eight cores.
   		The lattice is sliced into sub-lattices in one dimension (\textit{blue line and dots}).
   		The runtime speedup decreases where one memory domain (socket) is fully occupied with sixteen cores.
   		The code for this benchmark configuration is memory-bound as this typical behavior at the socket saturation shows.
   		The intra-socket scaling is not linear since the memory as a shared resource on the socket does not scale along with additionally used cores.
   		Slicing only in one dimension leads to a communication overhead when the sub-lattices become too narrow. 
   		This is remedied in this scenario by an equal slicing in every dimension such that the patches are cubic (\textit{red line and dots}).
   		The scaling across nodes is then again optimal.
   		Each setting ran twice and is averaged in order to take into account minor runtime variations.
   		More tests prove that cubic patches form the most efficient decomposition.}
   	 \label{fig:scaling_strong}
\end{figure}

\begin{figure}
	\centering
 \includegraphics[width=0.65\linewidth]{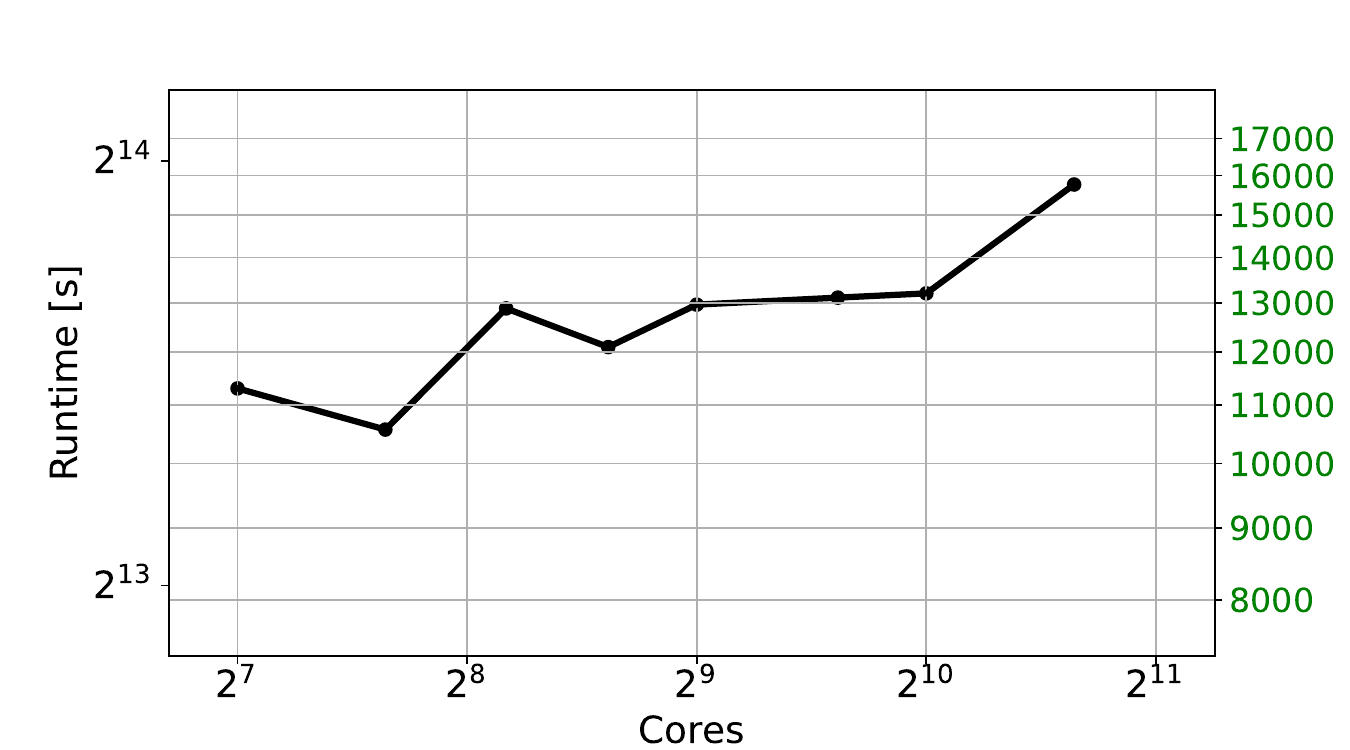}
 \caption[Weak scaling test]{Weak scaling test for a simulation such as to obtain 3D results shown in Section \ref{sec:2dsims}.
 	While the strong scaling test uses a fixed overall lattice, a weak scaling test is obtained by keeping the patch size constant and thereby increase the size of the overall lattice with the number of patches and compute cores.
 	Hence, the problem size is increased along with the number of parallel workers.
 	For such 3D simulations the code becomes more and more communication-bound.
 	Runtime variations and scaling issues are mainly attributable to data transfer via the node interconnect and resource contention thereon.
 	The weak scaling can be considered satisfying with the only large increase in runtime observed at the last run.
 	This one occupies 50 of a total of 153 nodes on the cluster. 
 	These simulations ran only once each, in order to save resources.}
 \label{fig:scaling_weak}
\end{figure}

Since efficient parallelization is paramount for expensive 3D simulations, strong and weak scaling tests proving the basic parallelization capability are conducted.
For the tests no output data are generated, yet the \textit{MPI IO} variant is highly efficient.
The employed high-performance computing system contains sixteen cores per memory domain (here a socket) and two sockets per node.
The results of the scaling tests are visualized in Figures \ref{fig:scaling_strong} and \ref{fig:scaling_weak}.

Runtimes and bottlenecks vary strongly with the particular setting for the problem under consideration.
Therefore, it is hard to give universal benchmarks and provide general scaling properties.
Bottlenecks have been investigated using the \textit{Intel\textsuperscript{\textregistered} oneAPI} HPC Toolkit \cite{oneAPIhpc}.
For simulation configurations as used in the present work the code is compute-bound in 1D, memory-bound in 2D, and \textit{MPI}-bound in 3D.

Remarkable speedups in the latter case of full three-dimensional simulations can thus be achieved through a reduction of the communication overhead by lowering the order of the numerical scheme.
This is because the required depth of the ghost layers decreases, e.g., from seven to three from order thirteen to order 4, as can be seen from the stencils in Section \ref{sec:Dispersion}.
To further ease the messaging pressure and memory requirements at large scales, additional parallelism by multi-threading with the help of \textit{OpenMP} \cite{Mattson2019,openmp} is used on top of the multi-processing.
As mentioned above, the stencil order is on the other hand not decisive for computation speed alone.


\if false
\subsection{Computational complexity scaling of the algorithm}

3D simulations run in a cube $ \mathcal{C} $ with $N=N_x\cdot N_y\cdot N_z$ equally spaced points,

\begin{equation*}
\mathcal{C}=\{x_{x,y,z}=(n_x\Delta_x,n_y\Delta_x,n_z\Delta_x) | n_i=1,...,N_i \}\subset\mathbb{R}^3 \ .
\end{equation*}
The application of the algorithm is subdivided into three computations.
The calculation of the spatial derivatives of $\vec{f}$ as well as the right hand side of the ODE \eqref{PropEquation} followed by its integration.

In the first step the derivatives are approximated by finite differences.
These have to be performed explicitly for each point and for each dimension.
For the calculation of the derivatives the computational load $\mathbf{C}_\text{D}$ is obtained to be

\begin{equation*}
\mathbf{C}_\text{D}\propto N_x \cdot N_y \cdot N_z  \cdot D \, ,
\end{equation*}
where $D$ is the number of spatial dimensions.

Second, the computation of the right hand side of Equation \eqref{PropEquation} has to be performed for each lattice point.
The dimension-dependent summation causes an extra computational load, increasing $D$ by 1.
The computational load for the calculation of the matrix equation \eqref{PropEquation} $\mathbf{C}_\text{M}$ is given by

\begin{equation*}
\mathbf{C}_\text{M}\propto N_x \cdot N_y \cdot N_z \cdot(D+1)\, .
\end{equation*}

In the last step for the solution of \eqref{PropEquation} the computational load of the used ODE solver has to be taken into account.
Hereby, the integration of the ODE highly depends on a wide range of considerations, such as involved frequencies or the magnitude of the nonlinearities. 
The computational load of the {\it CVODE} solver $\mathbf{C}_\text{S}$ is

\begin{equation*}
\mathbf{C}_\text{S}\propto N_x \cdot N_y \cdot N_z \cdot \Delta^{-1}.
\end{equation*}

Thus, for the upper limit of the problem with regard to the scaling behavior it is obtained

\begin{equation*}
	\mathbf{C}\propto (N_x \cdot N_y \cdot N_z) \cdot(D+1) \cdot \Delta^{-1}\, .
\end{equation*}
The linear dependence on the grid size coincides with what is found in the middle part of the weak scaling test.
\fi

\section{Phase velocity in a strong background}\label{sec:phasevel}

A probe plane wave is propagated along the $x$-axis,
\begin{equation}
    \vec{E}(x;t) = \vec{A}_p \cos (kx-kvt) \ ,
\end{equation}
through a linearly polarized strong electromagnetic background with field strength $A_b$.
The polarization of latter breaks the isotropy of space, giving rise to different refractive indices \cite{Toll1952,BaierBreitenlohner1967,Bialynicka-BirulaBialynicki-Birula1970,TsaiErber1975}
\begin{equation}\label{eq:RefIndizes}
n_{\pm} = 1 + \frac{\alpha}{45\pi} (11\pm3) \frac{A_b^2}{E_{\textrm{cr}}^2} = 1 + \delta n_{\pm}
\end{equation}
for a probe polarization orthogonal (+) and parallel (-) to the background polarization.
Since Equation \eqref{eq:RefIndizes} only takes into account the four-photon interaction contribution, i.e., it neglects all but the first nonlinear term in the weak-field expansion \eqref{approxLHE}, the results are verified turning off six-photon processes in the simulations.

The resulting phase velocity change $ v_{\textrm{nli}} $ from the vacuum speed of light, given by
\begin{equation}\label{eq:phase_vel}
	\frac{v_{\textrm{nli}}}{c} = \frac{v}{c} -1 = \frac{1}{n_\pm} -1 = - \frac{\delta n_{\pm}}{1+\delta n_{\pm}} \ ,
\end{equation}
can be extracted with the help of a Fourier analysis of a time-propagated wave.
When the passed time of propagation is chosen to be an integer multiple of $\lambda$, it is obtained with $ l \in \mathbb{N} $ \cite{Pons2018}
\begin{equation}
    \vec{E}(x;l\cdot \lambda/c) = \vec{A}_p \cos (2\pi x/\lambda-2\pi l \, v_{\textrm{nli}}) =  \vec{A}_p \cos (2\pi (x/\lambda-l \, v_{\textrm{nli}}) ) \ .
\end{equation}
The nonlinear phase velocity contribution can then be extracted from the phase after a spatial Fourier transformation evaluated at $\lambda^{-1}$.
This results in
\begin{equation}\label{eq:vnli_num}
\frac{v_{\textrm{nli}}}{c} = - \frac{1}{2\pi l} \ \textrm{arg}\left(\textrm{FT}[E(x;t_l)](\lambda^{-1}) \right) \quad\text{with}\quad c\cdot t_l=l \cdot \lambda \ . 
\end{equation}
The spatial Fourier transformation in Equation \eqref{eq:vnli_num} can be replaced by a Fast Fourier Transformation in the analysis.

To analyze the phase velocity variation numerically, the background field strength is varied. In a second step the relative polarization of the waves is changed from parallel to orthogonal.
The configurations are given in Table \ref{tab:phase_vel}.
The total simulation time is chosen to be \SI{200}{\micro\metre}/c, conveniently divided into 100 steps of \SI{2}{\micro\metre} -- the chosen wavelength of the probe wave.

\begin{SCtable}[]
\caption[Parameters for phase velocity change simulations in a strong background]{Settings for phase velocity variation tests.
	\\
	The background amplitudes and the relative polarizations are varied.
	The large wavelength of the background manifests itself as an ever-persistent static background.
	\label{tab:phase_vel}}
\centering
\renewcommand{\arraystretch}{1.3}  
\begin{tabular}{!{\vrule width 3\arrayrulewidth}c|c|c!{\vrule width 3\arrayrulewidth}}
	\noalign{\hrule height 3\arrayrulewidth}
	\textbf{Grid} &	Length & \SI{100}{\micro\metre} 
	\\ \hline
	& Lattice Points & 1000 
	\\
	\noalign{\hrule height 3\arrayrulewidth}
	\textbf{Background} & $\vec{A}$ & $(0,3,0) \times \SI{e-6}{} \, E_{\textrm{cr}} $ up to (0,0.9,0)$\, E_{\textrm{cr}} $
	\\\hline
	& $\lambda$ &  \SI{1}{\peta\metre}
	\\
	\noalign{\hrule height 3\arrayrulewidth}
	\textbf{Probe} & $\vec{A}$ &$(0,1,0) \times \SI{e-6}{} \, E_{\textrm{cr}} $ and $(0,0,1) \times \SI{e-6}{} \, E_{\textrm{cr}} $
	\\\hline
	& $\lambda$ & \SI{2}{\micro\metre}
	\\
	\noalign{\hrule height 3\arrayrulewidth}
\end{tabular}
\end{SCtable}

\begin{figure}
	\centering
	\includegraphics[width=0.65\linewidth]{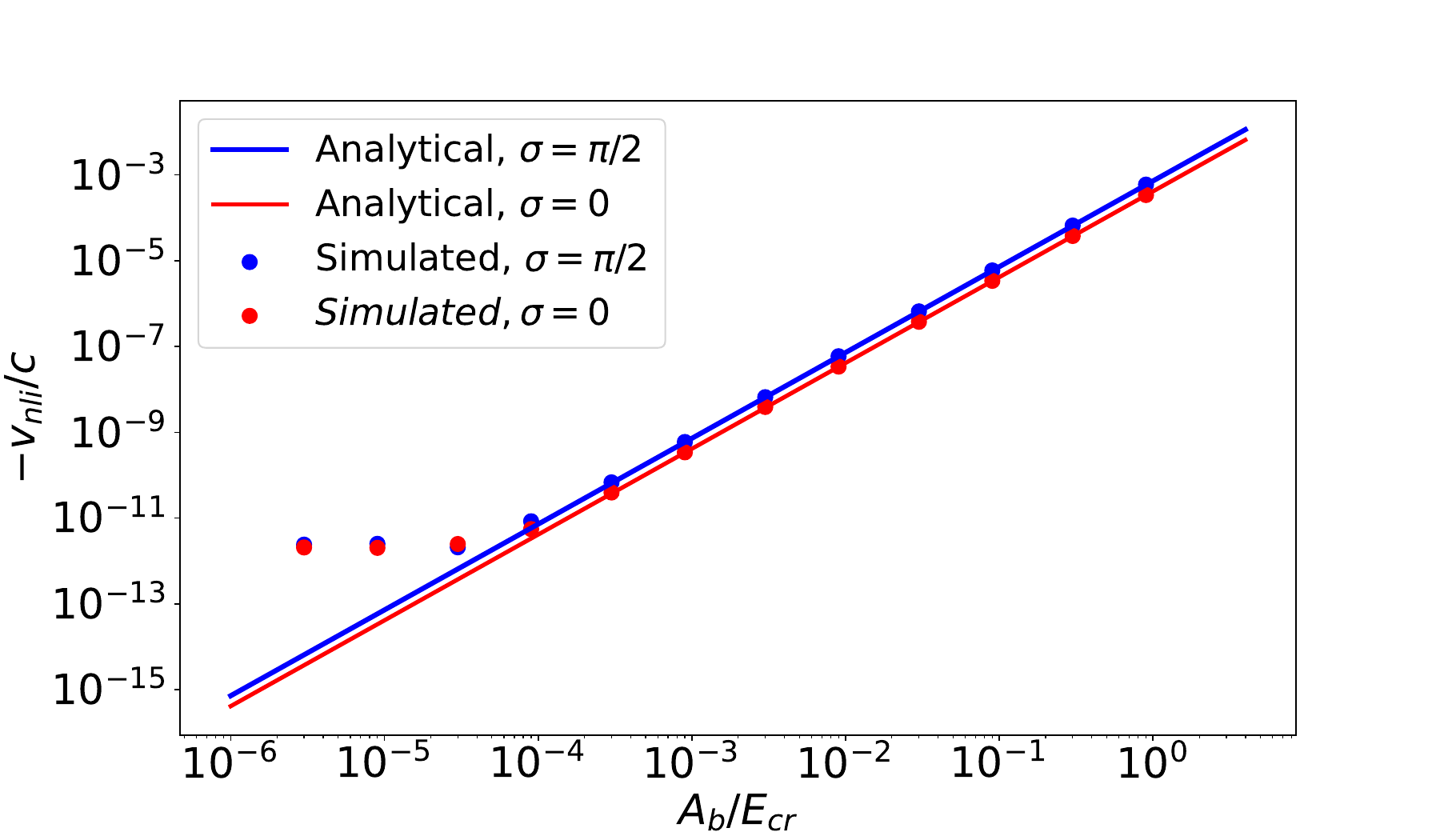}
	\caption[Phase velocity change in a strong background]{
	Nonlinear contribution to the phase velocity slow-down in a background with varying field strength.
	The simulation results converge to a value of $v_{\textrm{nli}} \approx \SI{2e-12}{} $ for small background field strengths.
	This is the phase velocity reduction caused by numerical errors, which are getting larger than the physical effect for background field strengths $A_b< 10^{-4} E_{\textrm{cr}} $.
    Deviations from the analytical expectation are higher for lower background field strengths.
    The error at $A_b=3\times 10^{-4} E_{\textrm{cr}} $ is still 6.1\% for parallel and 4.5\% for orthogonal relative polarization of the probe.
    This difference originates in the probe wave not being a perfect ``probe'' as in the idealized theoretical scenario and thus is contributing with its own polarization of $A_p = 3 \times 10^{-6} E_{\textrm{cr}} $.
    The values for background field strengths larger than $ 10^{-4} E_{\textrm{cr}} $ have a mean absolute percentage error of 1.8\% for parallel relative polarization and 1.2\% for orthogonal relative polarization of the probe.}
	\label{fig:phase_vel}
\end{figure}

It can be seen that the nonlinear interactions give note to themselves in a reduction of the phase velocity of the simulated waves in Figure \ref{fig:phase_vel}.
For sufficiently large background field strengths the numerical values are in very good agreement with the analytical predictions.

\section{Polarization flipping -- vacuum birefringence}
\label{sec:flipping}

Polarization, or helicity, flipping of a fraction of photons in a probe pulse propagating through a strong background pump pulse is a result of vacuum birefringence.
The origin of the effect is again the breaking of the isotropy of space by the polarization of the strong background.
The refractive indices from above,
\begin{equation}\label{RefIndizes}
n_-=1+\frac{8\alpha}{45\pi}\frac{E^2}{E_{\textrm{cr}}^2}\quad\text{and}\quad n_+=1+\frac{14\alpha}{45\pi}\frac{E^2}{E_{\textrm{cr}}^2} \ ,
\end{equation}
generate a difference in optical path length for the probe pulse polarization components parallel and orthogonal to the pump polarization, which results in \textit{bi}refringence.
On a microscopic level, a portion of the probe pulse's quanta flip their polarization.
Macroscopically, the overall polarization experiences a tiny rotation.

\begin{figure}
	\centering
	\includegraphics[width=0.7\linewidth]{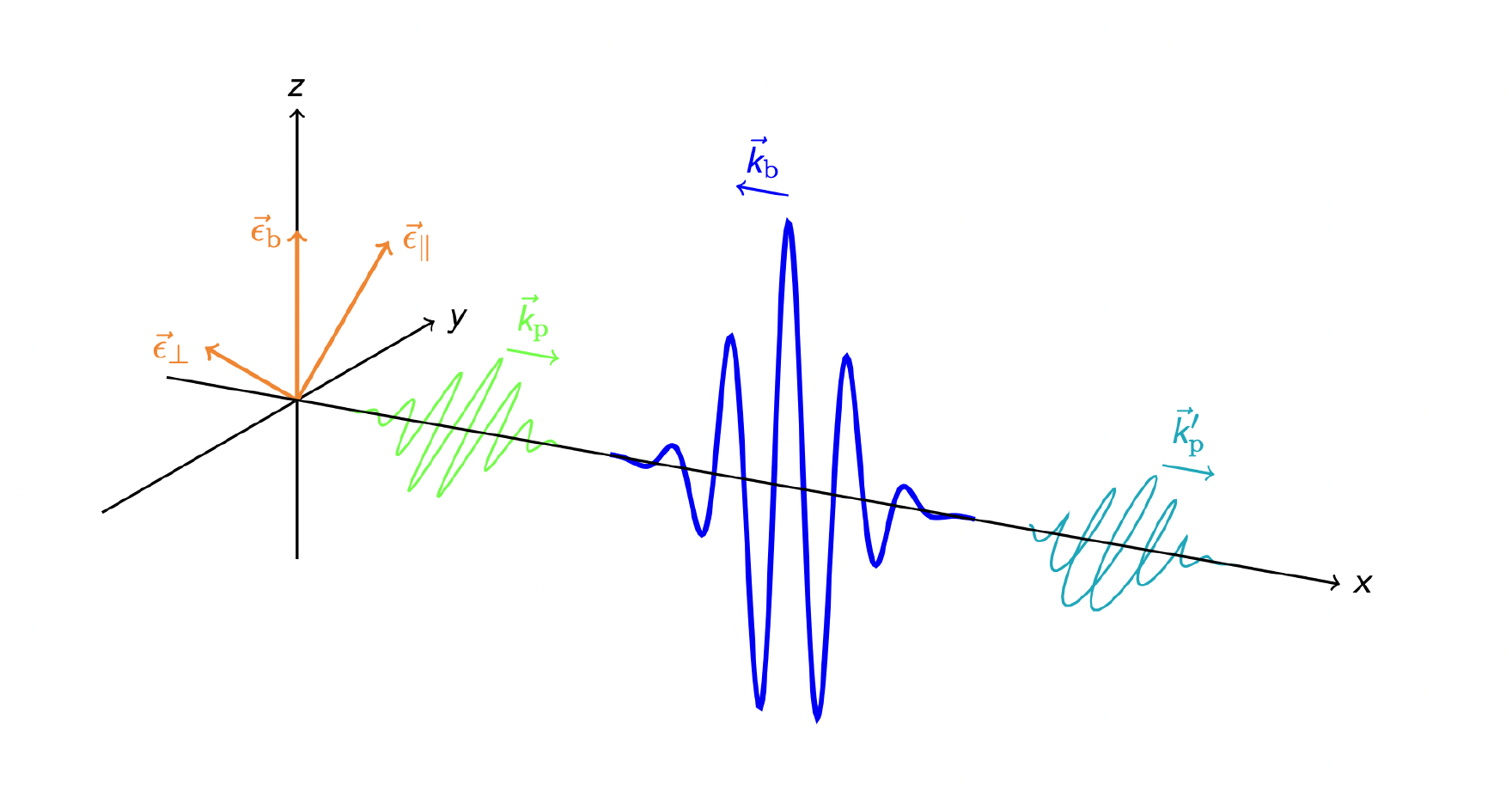}
	\caption[Depiction of polarization flipping]{Qualitative depiction of the electric fields in a coaxial probe--pump experiment for the measurement of vacuum birefringence.
	The probe (\textit{green}) traverses the counter-propagating pump (\textit{blue}), experiencing a polarization rotation as a result of different refractive indices.
	The originally linearly polarized probe is afterwards marginally elliptical (\textit{turquoise}).
	The effect is depicted greatly exaggerated for visibility.
	This effect is attributable the fact that the isotropy of space for the charged particle--antiparticle fluctuations in the vacuum is broken by the polarization of the strong background pulse.
	The coupling of these fluctuating particles in turn to the probe pulse results in different refractive indices for the polarization modes of the probe, as given in Equation \eqref{RefIndizes}.}
	\label{fig:BireSetup}
\end{figure}

\begin{SCfigure}[]
	\centering
	\includegraphics[width=0.7\textwidth]{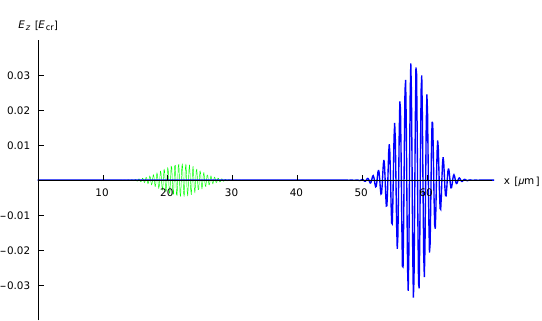}
	\caption[Initial setup for birefringence in 1D]{Sketch of the pulse configuration for the simulation of polarization flipping as shown in Figure \ref{fig:BireSetup}.
		On the left is the weaker probe pulse which propagates to the right; on the right is the strong pump pulse propagating leftwards.
		Note that in actual simulations (parameters in Table \ref{tab:pol_flip_scaling}) the probe field strength and wavelength are significantly smaller.
		Adjustments are made here for better visibility.
		\label{fig:pol_flip_initial}}
\end{SCfigure}

A typical probe--pump scenario devised for the observation of helicity flips is sketched in Figure \ref{fig:BireSetup}.
A probe pulse propagates through a strong low-frequency pump field in which spatial isotropy is broken.
While propagating through a pump field a fraction of probe pulse photons flips their polarization by 90\textdegree \ which results in a tiny ellipticity of an initially linearly polarized probe pulse.
A corresponding simulation configuration showing one polarization direction is shown in Figure \ref{fig:pol_flip_initial}.

In order to backtest the numerical solver, firstly, in Subsection \ref{subsec:PolFlip-Exp}, parametric checks of the flipping probability as derived in \cite{Dinuetal2014} are performed.
The settings given in Table \ref{tab:pol_flip_scaling} are chosen to this end with Gaussian pulses given by Equation \eqref{eq:1dgaussian} below.
With that result being verified, secondly, in Subsection \ref{subsec:extrapol}, the parametric scaling properties are made use of to extrapolate results to wavelenghts in the x-ray regime.
This is compared to a calculation for a realistic polarization flipping scenario calculated in \cite{Karbsteinetal2015}.
The parameters for that setup are listed in Table \ref{tab:GiesPulsParameter}.

In both cases the normalized vectors parallel and orthogonal to the initial probe polarization are given by
\begin{equation}
 \vec{\varepsilon}_\parallel = (0,1/\sqrt{2},1/\sqrt{2}) \quad \text{and}  \quad \vec{\varepsilon}_\perp = (0,-1/\sqrt{2},1/\sqrt{2}) \ .
\end{equation}
For the simulations 1D Gaussian pulses are used in the form
\begin{equation}\label{eq:1dgaussian}
	\vec{E}=\vec{A}\, e^{- (\Vec{x}-\Vec{x}_0)^2/\tau^2} \cos \left( \Vec{k} \cdot \Vec{x} \right) \ ,  \quad\text{with}
	\quad \Vec{k} = \frac{2\pi}{\lambda} \, \Hat{k} \ , \quad \text{and}
	\quad \Vec{B} = \Hat{k} \times \frac{\vec{E}}{c} \ ,
\end{equation}
where the vector $ \vec{A} $ comprises amplitude and polarization, $ \Vec{x}_0 $ is the center of the pulse, $ \tau $ its width, $ \lambda $ the wavelength, and $\Hat{k}$ the unit propagation direction vector.

Pulses are implemented in space without explicit time dependence.
With spatial derivatives via the finite difference scheme the ODE in time \eqref{PropEquation} is formulated, which is solved by \textit{CVODE} for the time evolution.
For convenience, the normalized vector $ \hat{k} $ indicating the propagation direction is stated in the parameter tables.

Since analytical estimates for polarization flipping such as in \cite{Karbsteinetal2015} make use of a photon picture, while the numerical simulations presented in the present paper propagate coherent modes,
the mapping
\begin{equation}\label{eq:flip_number}
\frac{N_\perp}{N}=\frac{\hbar\omega N_\perp}{\hbar\omega N}=\frac{\text{E}_\perp}{\text{E}_{\textrm{tot}}} \, 
\end{equation}
is used.
The energies in the respective polarization directions are proportional to the electric field strength projections squared,
\begin{equation}
\text{E}_\perp \sim \sum_{x_i\in \mathcal{C}}\left(\vec{E}(x_i)\cdot \vec{\varepsilon}_\perp\right)^2 \ ,\quad\text{E}_\parallel \sim \sum_{x_i\in \mathcal{C}}\left(\vec{E}(x_i)\cdot \vec{\varepsilon}_\parallel\right)^2 \ ,\quad \text{E}_{\textrm{tot}} = \text{E}_\perp+\text{E}_\parallel \ .
\label{EnergyComputation}
\end{equation}
All other factors in \eqref{eq:flip_number} cancel out.
It has to be mentioned that Equation \eqref{eq:flip_number} implies that the frequencies of the signal photons equal that of the probe pulse.
However, as to be shown in Section \ref{sec:harmonics}, the nonlinear interaction results in a small fraction of photons with altered frequency.

\subsection{Vacuum birefringence -- parametrical dependencies}\label{subsec:PolFlip-Exp}

\begin{figure}
	\centering
	\begin{subfigure}{.45\textwidth}
		\includegraphics[width=\linewidth]{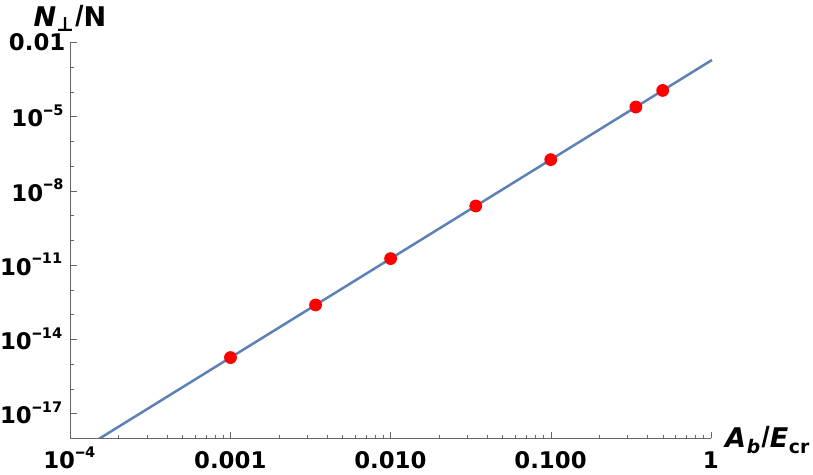}
	\end{subfigure}\quad\quad
	\begin{subfigure}{.45\textwidth}
		\includegraphics[width=\linewidth]{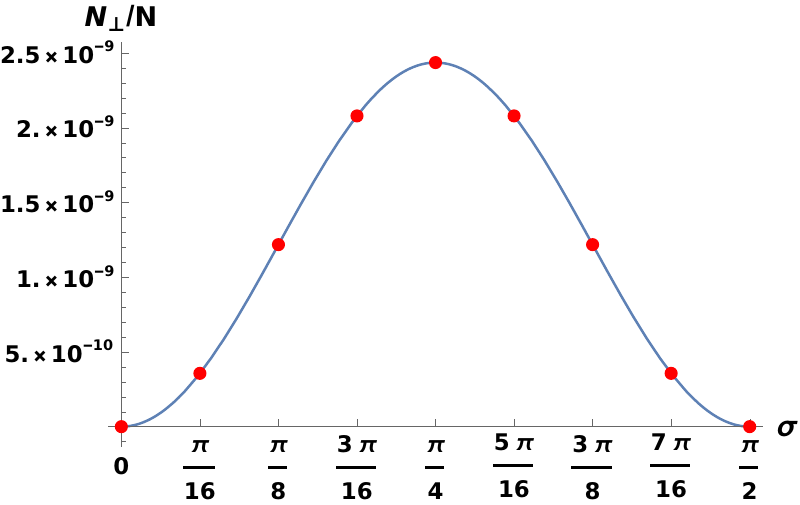}
	\end{subfigure}
\\[0.5cm]
	\begin{subfigure}{.45\textwidth}
		\includegraphics[width=\linewidth]{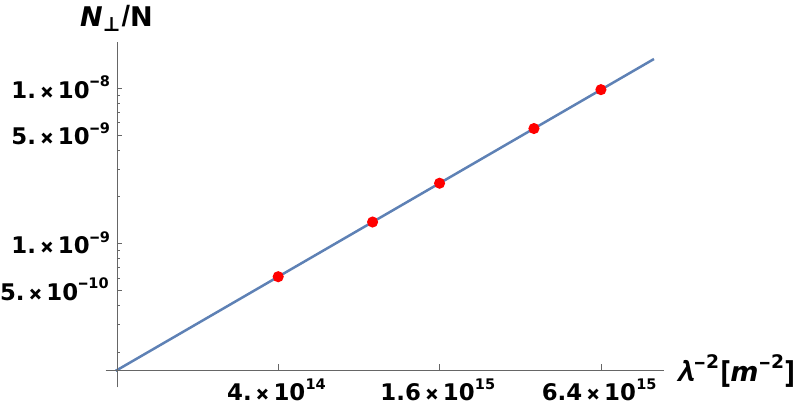}
    \end{subfigure}\quad\quad
    \begin{subfigure}{.45\textwidth}
		\includegraphics[width=\linewidth]{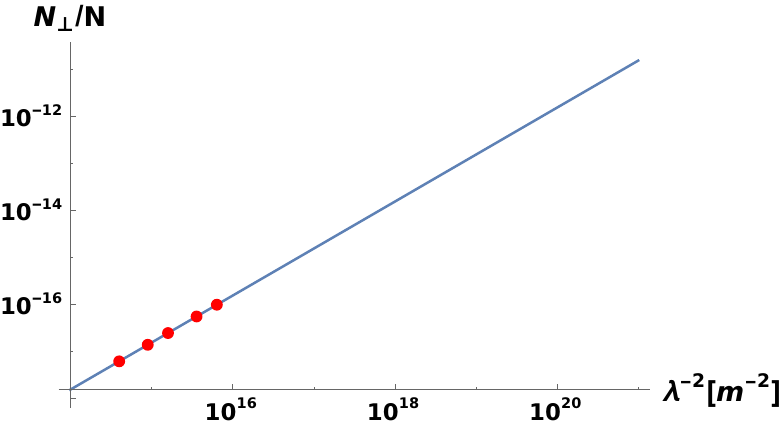}
    \end{subfigure}
\\[0.3cm]
\caption[Parametric scaling of the polarization flipping probability]{Scaling of the polarization flipping probability with variations of the parameters given in Table \ref{tab:pol_flip_scaling}.
The solid lines are the analytical curves obtained with the help of \eqref{eq:Pflip}.
The red dots are simulation results.
\textit{Top left}: varying the background field strength.
\textit{Top right}: varying the relative polarization angle of probe and pump.
\textit{Bottom left}: varying the probe wavelength. 
\textit{Bottom right}: combining the scaling of the pump field strength and probe wavelength;
$ A_b $ is scaled down by a factor of 100 compared to the parameter in Table \ref{tab:pol_flip_scaling} and
simultaneously an extrapolation to frequencies in the x-ray regime is performed.
This last way of combining scaling properties is useful in order to extrapolate to relevant probe frequency regimes while keeping the numerical accuracy high and the computational load low.
}
\label{fig:pol_flip_scaling}
\end{figure}

For a probe pulse coaxially counter-propagating to a plane wave background field, the polarization flipping probability, taking into account again only up to four-photon interactions, in the low-energy approximation, is given by \cite{Dinuetal2014}
\begin{equation}\label{eq:Pflip}
	P_{\textrm{flip}} = \frac{N_\perp}{N} = \frac{\alpha^2}{255 \, \lambda_p^2} \sin^2(2\sigma) \left( \int dx \ \frac{A_b(x)^2}{E_{\textrm{cr}}^2} \right)^2 \ ,
\end{equation}
where $\sigma$ is the initial angle between the probe and pump polarizations, $\lambda_p$ the wavelength of the probe pulse, and $A_b$ the amplitude of the background pulse.
The propagation direction of the probe is assumed to be perpendicular to the pump polarization and the probe field strength to be negligible compared to the pump.
The probability directly translates to the flip ratio,
\begin{equation}
    N_\perp = P_{\textrm{flip}} \cdot N \ .
\end{equation}
Formula \eqref{eq:Pflip} yields all the parametric dependencies for the probability of polarization flips and indirectly excludes other parameters.
There is a strong dependence on the optical path of the pump pulse and on the probe wavelength.
Notably, the ratio is independent of the shapes of the pulses.
Limitations of the above formula for focused background pulses are discussed in the following subsection.
In 1D simulations the background can be modeled as a Gaussian pulse.
To investigate the scaling properties of the numerical solver the settings in Table \ref{tab:pol_flip_scaling} are used, where only those parameters affecting \eqref{eq:Pflip} are actually relevant.
A time-resolved flipping process for those parameters is depicted in Figure \ref{fig:FlipTimeEvol}.
The results of the parametric scaling tests are visualized in Figure \ref{fig:pol_flip_scaling}.
There is perfect agreement between the 1D simulation results and formula \eqref{eq:Pflip}.
With these scaling properties being verified in the algorithm, in the following subsection the analytical result in case (a) of \cite{Karbsteinetal2015}, where a small probe pulse traverses a strong pump field, is compared to an extrapolation of simulation results.

\begin{table}
\begin{varwidth}[c]{0.45\linewidth}
		\caption[Parameters for parametric scaling tests of birefringence]{Parameters for probe and pump beams chosen to test the parametric dependencies in Equation \eqref{eq:Pflip}.
		The probe wavelength, the pump field strength, and their relative polarizations are varied to obtain the parametric scaling results of Figure \ref{fig:pol_flip_scaling}.}
	\label{tab:pol_flip_scaling}
	\centering
	\renewcommand{\arraystretch}{1.3}  
  \begin{tabular}{!{\vrule width 3\arrayrulewidth}c|c|c!{\vrule width 2\arrayrulewidth}}
	\noalign{\hrule height 3\arrayrulewidth}
	\textbf{Grid} &	Length & \SI{80}{\micro\metre}
	\\ \hline
	& Lattice Points & $60 \times \SI{e3}{} $ 
	\\
	\noalign{\hrule height 3\arrayrulewidth}
	\textbf{Pump} & $ \vec{A} $ & $(0,0,34) \times \SI{e-3}{} E_{\textrm{cr}}  $
	\\\hline
	& $ \hat{k} $ &  (-1,0,0) 
	\\\hline
	& $ \lambda $  & \SI{800}{\nano\metre}
	\\\hline
	& $ \vec{x}_0 $  & \SI{58}{\micro\metre}
	\\\hline
	& $ \tau $  & \SI{3.5}{\micro\metre}
	\\
	\noalign{\hrule height 3\arrayrulewidth}
	\textbf{Probe} & $ \vec{A}$  & $(0,50,50) \times \SI{e-6}{} E_{\textrm{cr}}  $
	\\\hline
	& $ \hat{k} $ &  (1,0,0) 
	\\\hline
	& $ \lambda $  & \SI{25}{\nano\metre}
	\\\hline
	& $ \vec{x}_0 $  &  \SI{22}{\micro\metre}
	\\\hline
	& $ \tau $  & \SI{4.0}{\micro\metre}
	\\
	\noalign{\hrule height 3\arrayrulewidth}
\end{tabular}
\end{varwidth}
\hfill
\begin{minipage}[c]{0.5\linewidth}
	\centering
	\includegraphics[width=1.0\textwidth]{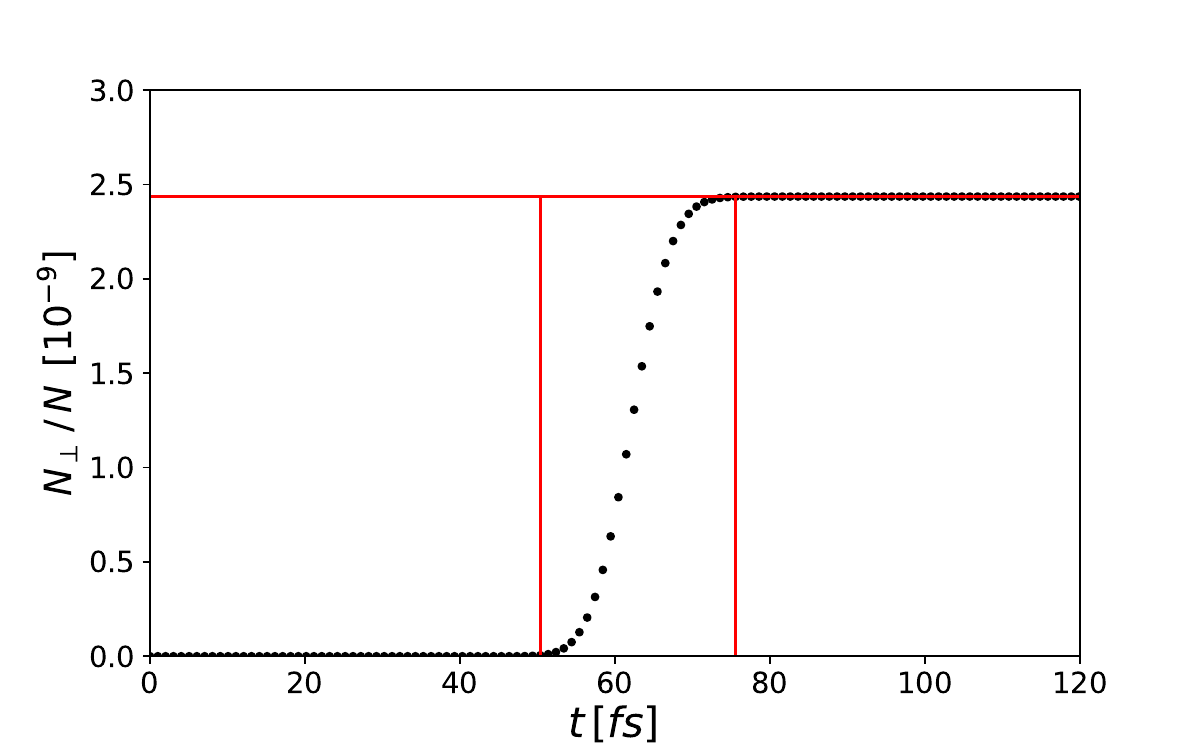}
	\captionof{figure}[Polarization flipping time evolution 1]{Time evolution of the polarization flipping ratio for the parameters presented in Table \ref{tab:pol_flip_scaling}.
	A simulation time of \SI{30}{\micro\metre} divided into 100 steps, so \SI{1}{\femto\second} per step is used.
	The settings used here provide an interaction time of \SI{25}{\femto\second}, indicated by the red vertical lines.
	The red horizontal line corresponds to the asymptotic relative flip ratio.
	\label{fig:FlipTimeEvol}}
\end{minipage}
\end{table}

Neglecting the signals for $ \sigma=0,\, \pi/2 $, where $ P_\textrm{flip}=0 $ analytically, that cannot be respected in a relative error calculation with the true values as baseline, the mean absolute percentage errors for each scaling test are below 0.1\%.

\subsection{Vacuum birefringence -- extrapolation to an analytical value in the x-ray regime}\label{subsec:extrapol}

\begin{table}
\begin{varwidth}[c]{0.45\linewidth}
		\caption[Parameters for a birefringence benchmark]{Parameters for probe and pump beam adapted to \cite{Karbsteinetal2015}.
		The pump field strength is obtained as the square root of the ratio of intensity to critical intensity. 
		The pump pulse duration is \SI{30}{\femto\second} ($2\tau$ in the $1/e^2$ criterion).
		}
	\label{tab:GiesPulsParameter}
	\centering
	\renewcommand{\arraystretch}{1.3}  
  \begin{tabular}{!{\vrule width 3\arrayrulewidth}c|c|c!{\vrule width 3\arrayrulewidth}}
	\noalign{\hrule height 3\arrayrulewidth}
	\textbf{Grid} &	Length & \SI{80}{\micro\metre}
	\\ \hline
	& Lattice Points & $80 \times \SI{e3}{} $
	\\
	\noalign{\hrule height 3\arrayrulewidth}
	\textbf{Pump} & $ \vec{A} $ & $(0,0,0.34) \times \SI{e-3}{} E_{\textrm{cr}}  $
	\\\hline
	& $ \hat{k} $ & (-1,0,0) 
	\\\hline
	& $ \lambda $  & \SI{800}{\nano \meter}
	\\\hline
	& $ \vec{x}_0 $  & \SI{58}{\micro \meter}
	\\\hline
	& $ \tau $  & \SI{4.5}{\micro \meter} 
	\\
	\noalign{\hrule height 3\arrayrulewidth}
	\textbf{Probe} & $ \vec{A}$  & $(0,50,50) \times \SI{e-6}{} E_{\textrm{cr}}  $
	\\\hline
	& $ \hat{k} $ &  (1,0,0) 
	\\\hline
	& $ \lambda $  & \SI{96}{\pico \meter}
	\\\hline
	& $ \vec{x}_0 $  & \SI{22}{\micro \meter}
	\\\hline
	& $ \tau $  & \SI{3.4}{\micro \meter}
	\\
	\noalign{\hrule height 3\arrayrulewidth}
\end{tabular}
\end{varwidth}
\hfill
\begin{minipage}[c]{0.5\linewidth}
	\centering
	\includegraphics[width=1.0\textwidth]{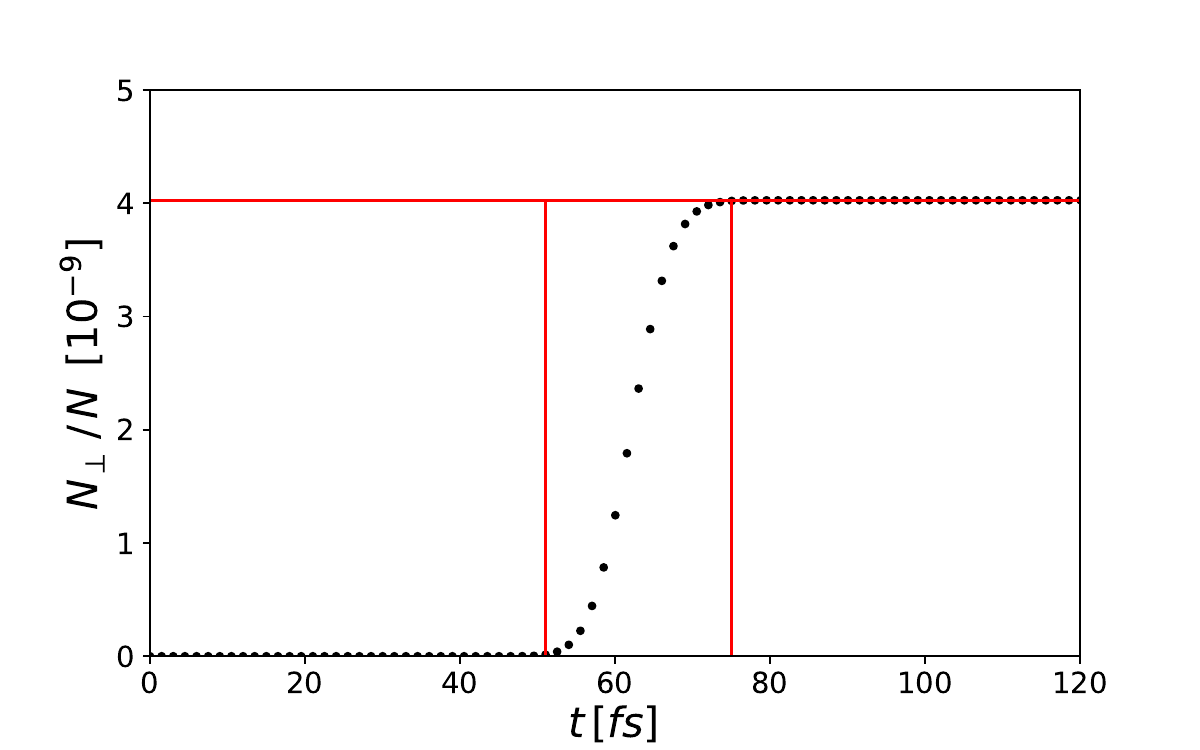}
	\captionof{figure}[Polarization flipping time evolution 1]{Time evolution of the polarization flipping probability for an adaptation of the parameters presented in Table \ref{tab:GiesPulsParameter}.
	One time step corresponds to \SI{1.5}{\femto \second}.
	The adaptations are:
	The pump field strength is magnified by a factor of 100 to 34$\, \times \SI{e-3}{} E_{\textrm{cr}}  $ in order to reduce numerical noise.
	The probe wavelength is enlarged to a computationally acceptable value of \SI{25}{\nano \meter}.
	The distance between the vertical red lines is the total interaction time $t_I=\SI{24}{\femto \second}$.
	The horizontal red line denotes the asymptotic flip ratio.
	\label{fig:FlipTimeEvolExt}}
\end{minipage}
\end{table}

Simulations of birefringence effects are computationally expensive in higher dimensions for the small wavelengths and field strengths targeted in experiments.
Making use of the scaling properties in Equation (\ref{eq:Pflip}), the phenomenon of birefringence can still be predicted for the parameters accessible in planned near future experiments by simulating numerically feasible, quasi-1D setups with consecutive extrapolation.

To this end this approach is used to reproduce an analytical result for a coaxial probe--pump setup with Gaussian laser pulses in a realistic scenario by considering case (a) of \cite{Karbsteinetal2015}.
In this case the radius of the probe pulse is taken to be much smaller than the waist of the pump beam, such that the probe does not sense the transverse structure of the pump.
This scenario thus amounts to a 1D case.
Settings adapted to the scenario described in case (a) in \cite{Karbsteinetal2015} are given in Table \ref{tab:GiesPulsParameter}.
The calculation in \cite{Karbsteinetal2015} is performed in the vacuum emission picture.

Some parameters devised for experimental verification impair the numerical approach.
First, the pump field strength is too low to extract the flipping process from the numerical noise.
To combat this, the field strength scaling properties can be used to simulate with a larger background amplitude.
Second, the probe wavelength of $\lambda_p=  \SI{96}{\pico \meter} $ is in the x-ray regime to amplify the effect, c.f. Equation \eqref{eq:Pflip}, and is therefore problematically small for modeling on a discrete grid.
An extremely fine grid would be necessary to model that pulse.
To evade computations that expensive, the wavelength scaling properties can be made use of.
The resulting extrapolation is thus a combination of two scaling methods in the way shown in the bottom right of Figure \ref{fig:pol_flip_scaling} and described in that caption.

\begin{figure}
    \centering
    \includegraphics[width=0.65\textwidth]{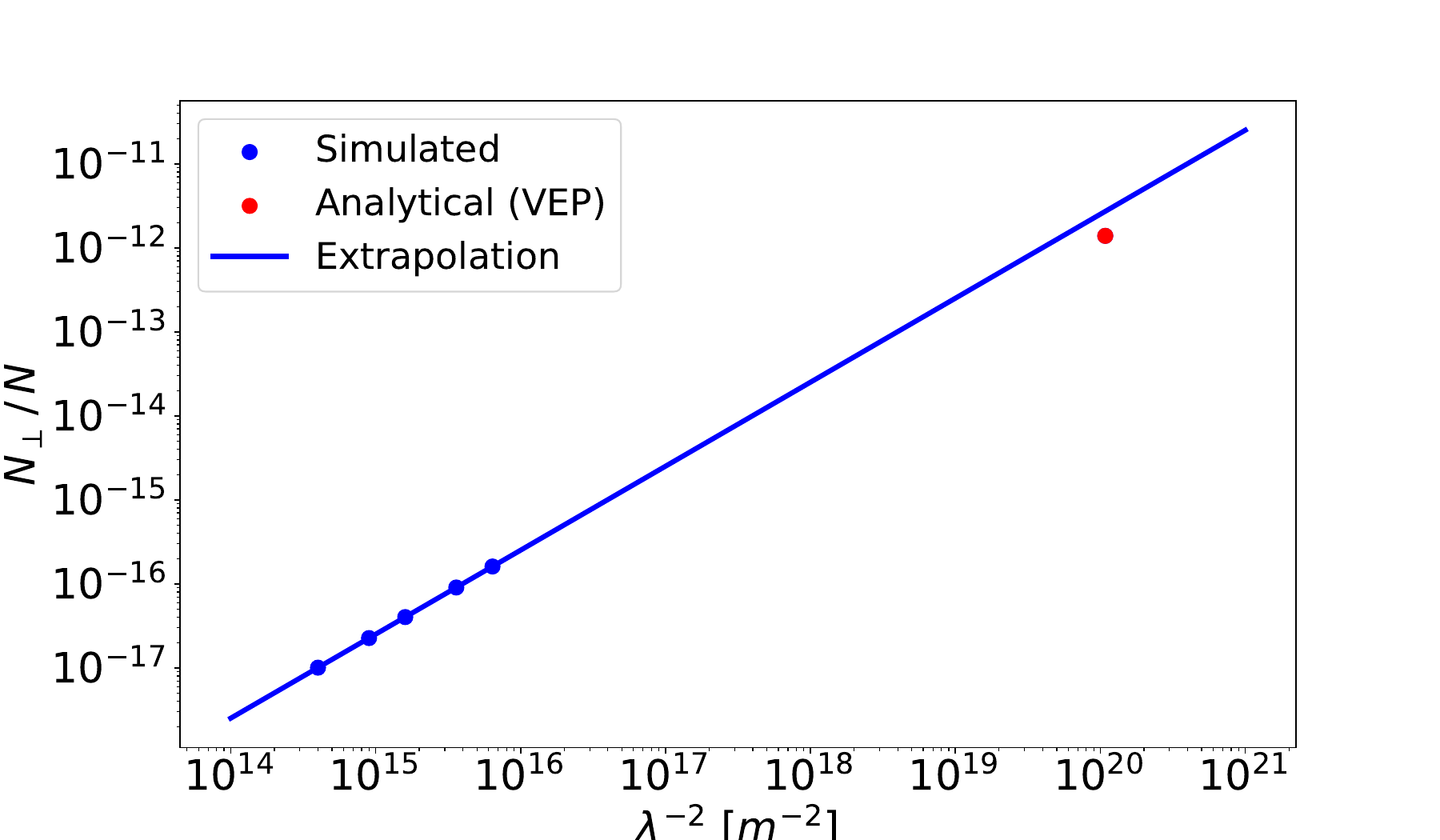}
    \caption[Extrapolation of the polarization flipping probability to the x-ray regime]{Extrapolation of polarization flipping ratios to the x-ray regime and comparison to a result
    	obtained via the vacuum emission picture in \cite{Karbsteinetal2015} (\textit{red dot}).
    The blue line is an extrapolation of simulation results (\textit{blue dots}) with various probe wavelengths.
    \\[1.5cm]}
    \label{fig:Gies_extrapol}
\end{figure}

Figure \ref{fig:Gies_extrapol} shows that the extrapolated flipping ratio of \SI{2.72e-12}{} is almost twice as high as the flipping ratio of \SI{1.39e-12}{} obtained in \cite{Karbsteinetal2015}.
This is attributable to the neglect of a longitudinally localizing term with the Rayleigh length in the pulse form in Equation \eqref{eq:1dgaussian}, c.f. the higher-dimensional Gaussian pulse in Equation \eqref{eq:2dgaussian}.
This yields a further suppression of about a factor of two.
The value obtained at the corresponding probe frequency
via formula \eqref{eq:Pflip} is \SI{2.73e-12}{} and agrees with the simulations.
Accordingly, in future higher-dimensional simulations a reconciliation of the results should be achieved.


\section{Harmonic generation}
\label{sec:harmonics}

To further crosscheck the solver, the prominent probe--pump scenario for the detection of nonlinear vacuum signatures shown in Figure \ref{hh_gen_initial_config} is considered again with two head-on colliding pulses, a strong background pulse and a weaker probe pulse.
For this analysis, the former pulse is assumed to have zero frequency.
The initial settings are listed in Table \ref{tab:hh_gen}.

\begin{table}
\begin{varwidth}[c]{0.45\linewidth}
	\caption[Parameters to simulate harmonic generation]{Initial settings to observe harmonic generation in 1D simulations, see Figure \ref{hh_gen_initial_config}.}
	\centering
	\renewcommand{\arraystretch}{1.3}  
\begin{tabular}{!{\vrule width 3\arrayrulewidth}c|c|c!{\vrule width 3\arrayrulewidth}}
	\noalign{\hrule height 3\arrayrulewidth}
	\textbf{Grid} &	Length & \SI{300}{\micro\metre}
	\\ \hline
	& Lattice Points & 4000 
	\\
	\noalign{\hrule height 3\arrayrulewidth}
	\textbf{Pump} &  $\vec{A} $ &  $(0,20,0) \times \SI{e-3}{} E_{\textrm{cr}}  $ 
	\\\hline
	&  $\hat{k}$ & (-1,0,0) 
	\\\hline
	& $\lambda$  & \SI{1}{\metre}
	\\\hline
	& $x_0$  &  \SI{200}{\micro\metre}
	\\\hline
	& $\tau$  & \SI{12.8}{\micro\metre}
	\\
	\noalign{\hrule height 3\arrayrulewidth}
	\textbf{Probe} & $\vec{A} $ &$ (0,5,0) \times \SI{e-3}{} E_{\textrm{cr}}  $
	\\\hline
	& $\hat{k}$ & (1,0,0) 
	\\\hline
	& $\lambda$  & \SI{2}{\micro\metre}
	\\\hline
	& $x_0$  & \SI{100}{\micro\metre}
	\\\hline
	& $\tau$  & \SI{10}{\micro\metre}
	\\
	\noalign{\hrule height 3\arrayrulewidth}
\end{tabular}
\label{tab:hh_gen}
\end{varwidth}%
  \hfill
\begin{minipage}[c]{0.5\linewidth}
	\centering
	\includegraphics[width=1.0\textwidth]{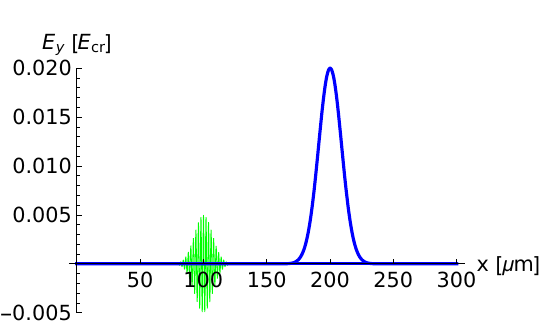}
	\captionof{figure}[Initial configuration to simulate harmonic generation in 1D]{Visualization of the pulse configuration to detect higher harmonics with a zero-frequency background pulse, see Table \ref{tab:hh_gen}.
	\label{hh_gen_initial_config} }
\end{minipage}

\end{table}

\begin{figure}
	\centering
	\begin{subfigure}{.45\textwidth}
		\includegraphics[width=\linewidth]{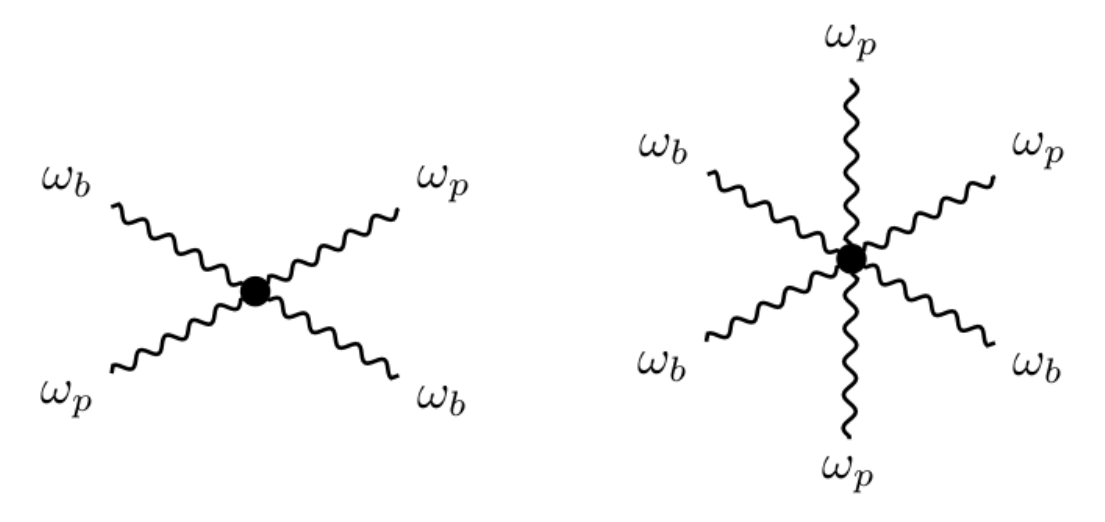}
		\caption{Effective vertices for four- and six-photon scattering.}
	\end{subfigure}
$ \qquad \qquad $
	\begin{subfigure}{.27\textwidth}
		\includegraphics[width=\linewidth]{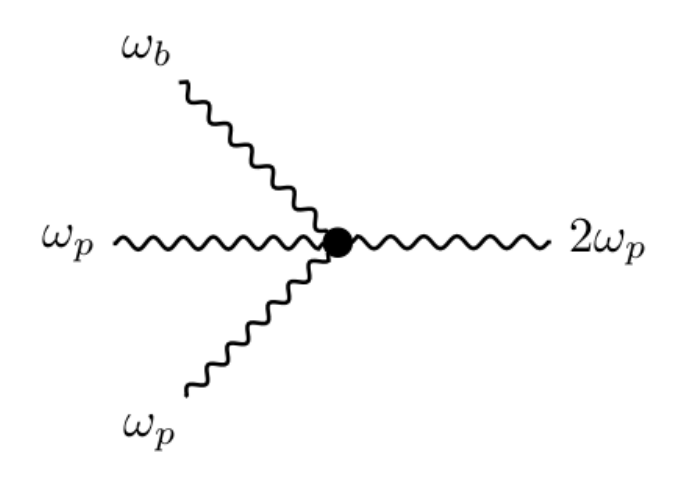}
		\caption{Example of high-harmonic generation with a zero-frequency background.}
	\end{subfigure}
	\caption[Effective vertices for harmonic generation]{Feynman diagrams for harmonic generation with a probe (\textit{subscript p}) and background (\textit{subscript b}) wave.
		In (b) there is an implicit time axis from left to right.}
	\label{fig:harmonic_generation}
\end{figure}

\begin{figure}
	\centering
	\begin{subfigure}{.5\textwidth}
		\includegraphics[width=\linewidth]{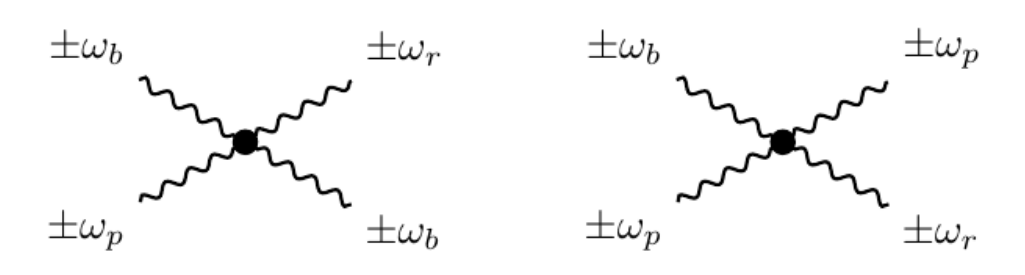}
		\caption{$\omega_r \in \omega_b,  2\omega_p\pm w_b, \omega_p, \omega_p \pm 2 \omega_b$}
	\end{subfigure}%
	\begin{subfigure}{.5\textwidth}
		\includegraphics[width=\linewidth]{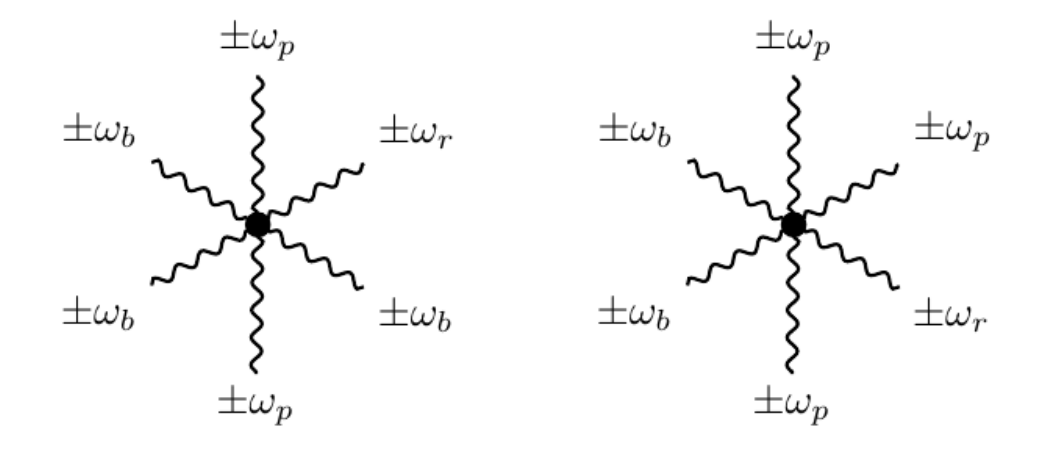}
		\caption{$w_r \in \omega_b, 3 \omega_b, \omega_p +3 \omega_b, 2 \omega_p \mp 2\omega_b, \omega_p , \omega_p \mp 2\omega_b, 3\omega_p \mp 2\omega_b $}
	\end{subfigure}
	\caption[Allowed resulting harmonics through four- and six-photon processes]{Allowed vertices of Figure \ref{fig:harmonic_generation} with resulting frequencies $ \omega_r $ for (a) four-photon and  (b) six-photon processes in a probe--pump setup \cite{Boehl2016}.
		$\pm$ indicates in-/outgoing photons. 
		Further restrictions are posed upon the asymptotic states by energy conservation.}
	\label{fig:harmonic_generation_restriction}
\end{figure}

Approximate analytical results for this scenario are derived in \cite{Boehl2016,Kingetal2014}.
The effective vertices for four- and six-photon scattering in Figure \ref{fig:harmonic_generation} (a)
can produce outgoing photons with higher frequency by photon merging.
For example, in Figure \ref{fig:harmonic_generation} (b) two probe photons and a zero-frequency background photon merge into an outgoing photon with frequency $2\omega_p$.
The possible contributions of two-wave scattering that result from the first orders of the weak-field expansion are listed in Figure \ref{fig:harmonic_generation_restriction}.

In the present case of $ \omega_b = 0 $, there are for
four-photon processes
\begin{itemize}
	\item the scattering of a background and a probe photon contributing with one photon to the zeroth harmonic ($\omega_r=0$), also called dc component, and with one to the first harmonic ($\omega_r=\omega_p$), also called fundamental harmonic;
	
	\item two background photons and one probe photon merging to produce a photon of the fundamental harmonic ($\omega_r=\omega_p$); and
	
	\item two probe photons and one background photon merging to produce a photon of the second harmonic ($\omega_r=2\omega_p$).
\end{itemize}

\begin{figure}
	\centering
	\begin{subfigure}{.32\textwidth}
		\includegraphics[width=\linewidth]{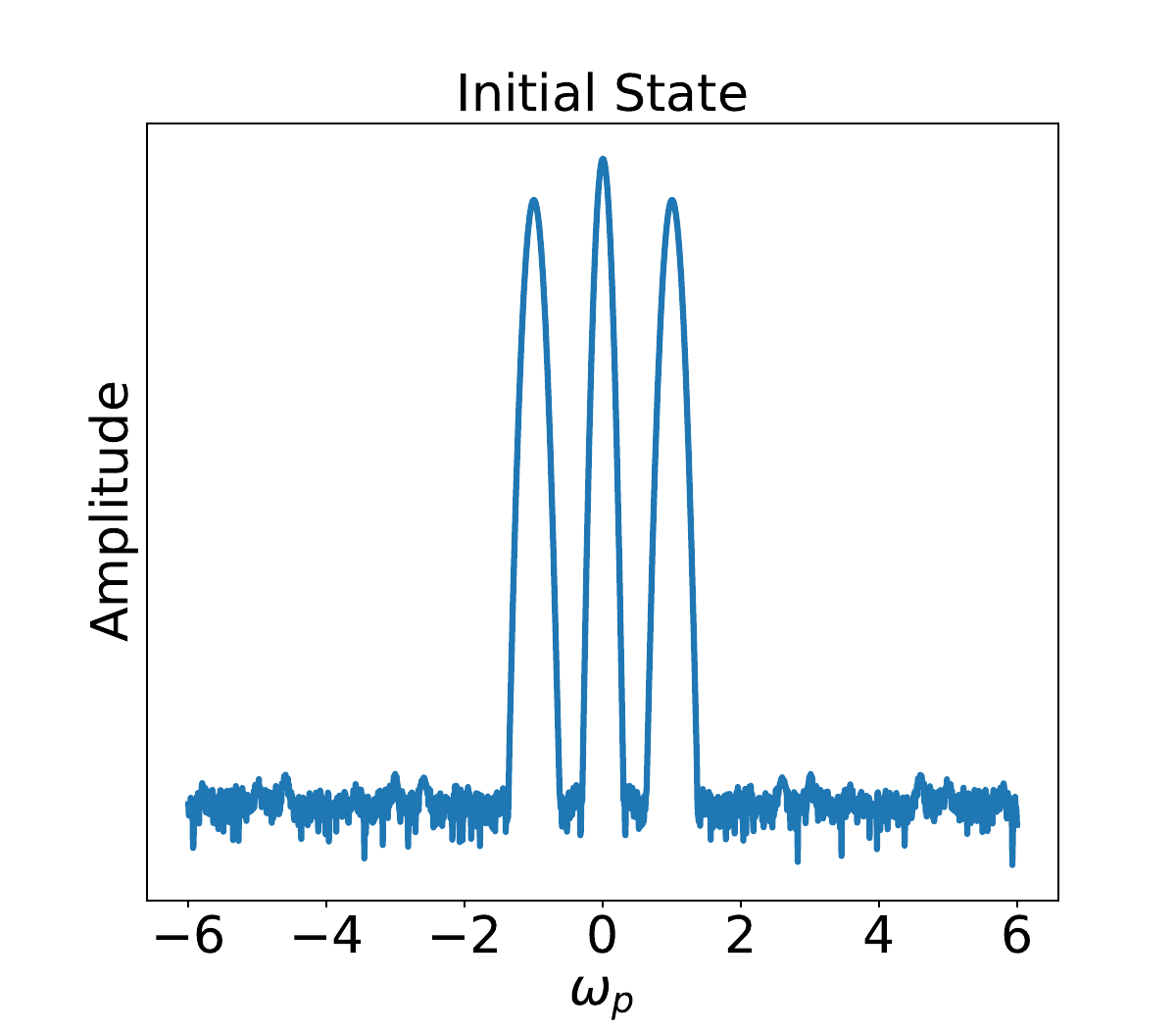}
	\end{subfigure}
	\begin{subfigure}{.32\textwidth}
		\includegraphics[width=\linewidth]{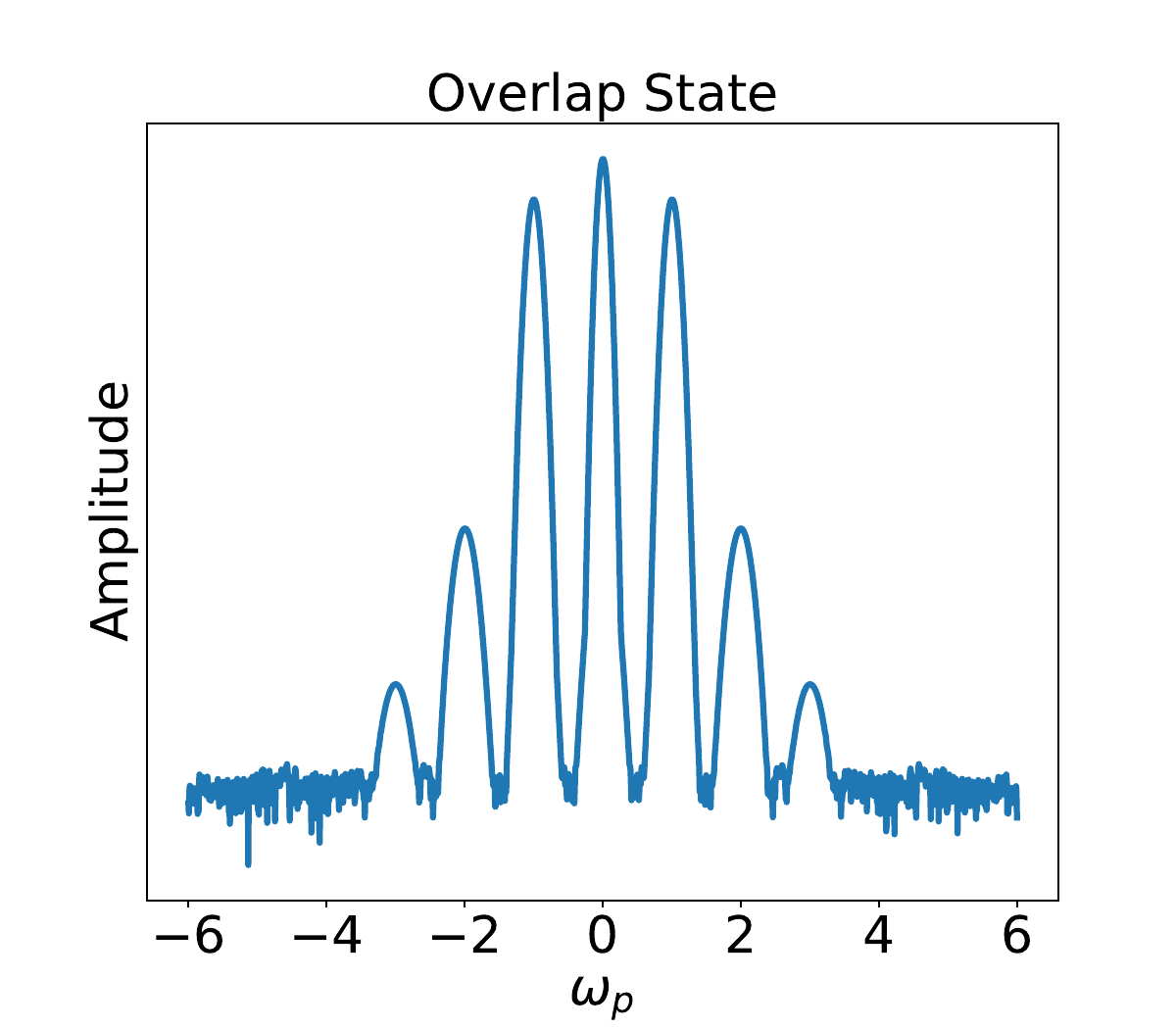}
	\end{subfigure}
	\begin{subfigure}{.32\textwidth}
		\includegraphics[width=\linewidth]{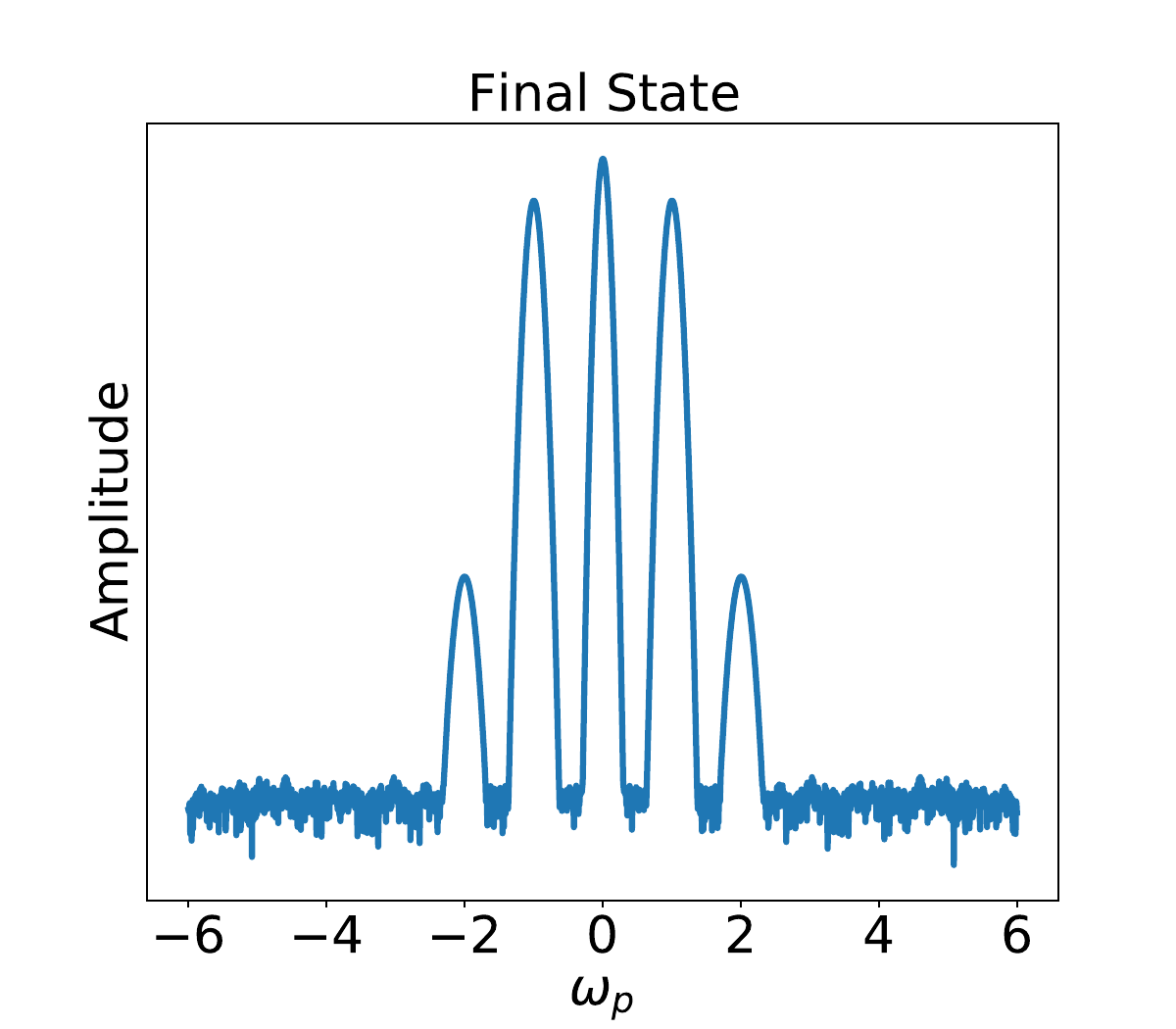}
	\end{subfigure}
	
	\begin{subfigure}{.32\textwidth}
		\includegraphics[width=\linewidth]{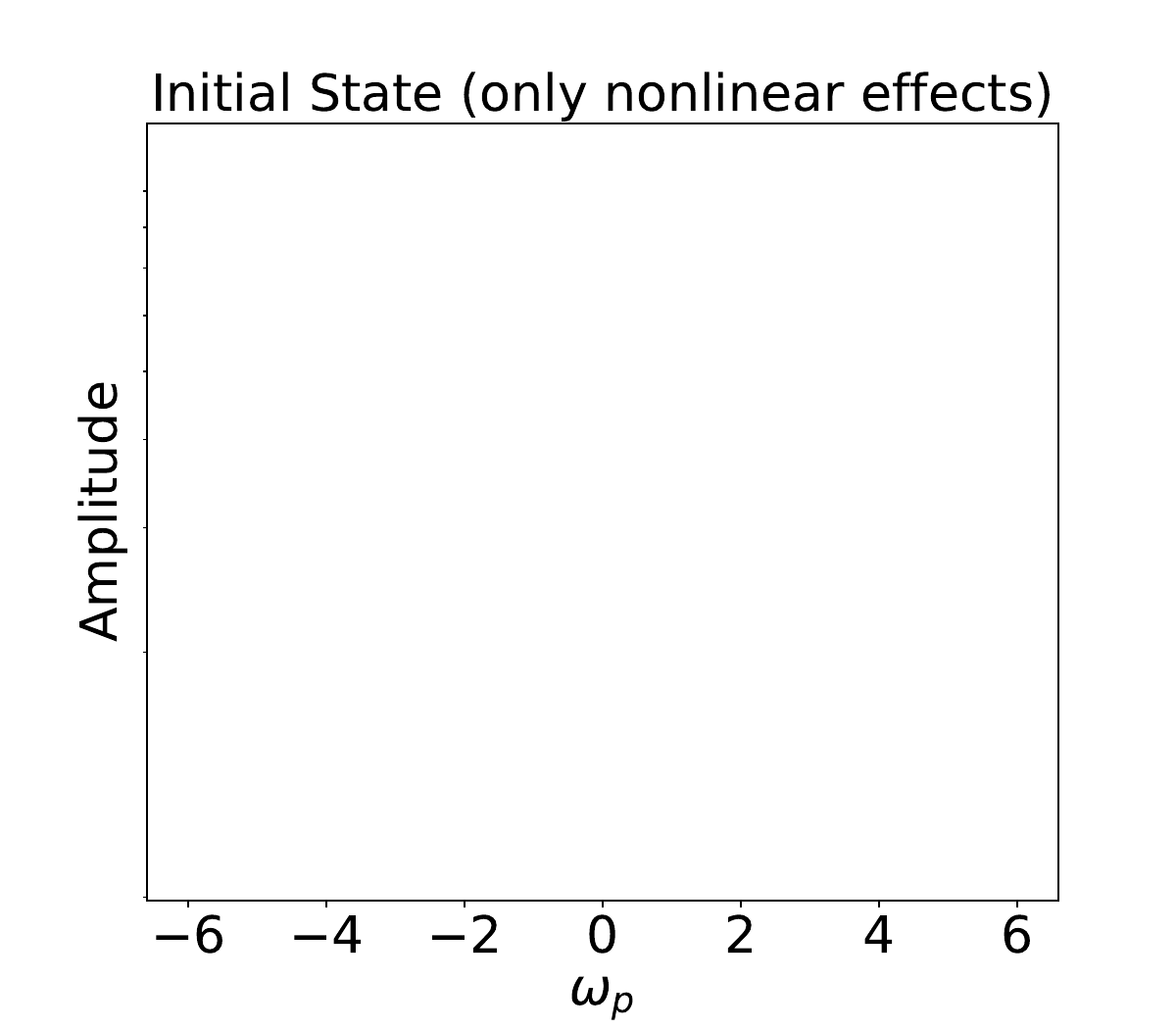}
	\end{subfigure}
	\begin{subfigure}{.32\textwidth}
		\includegraphics[width=\linewidth]{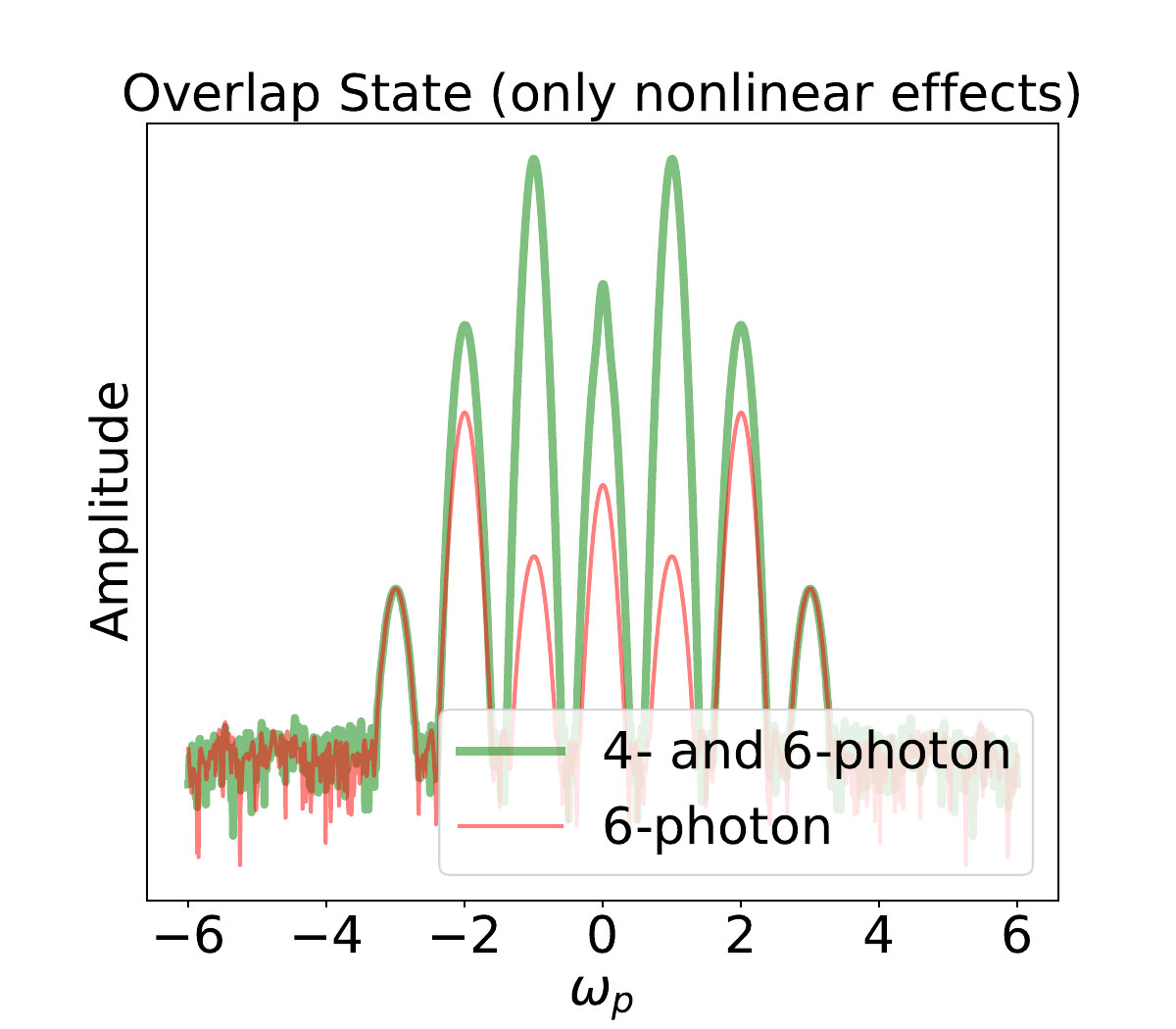}
	\end{subfigure}
	\begin{subfigure}{.32\textwidth}
		\includegraphics[width=\linewidth]{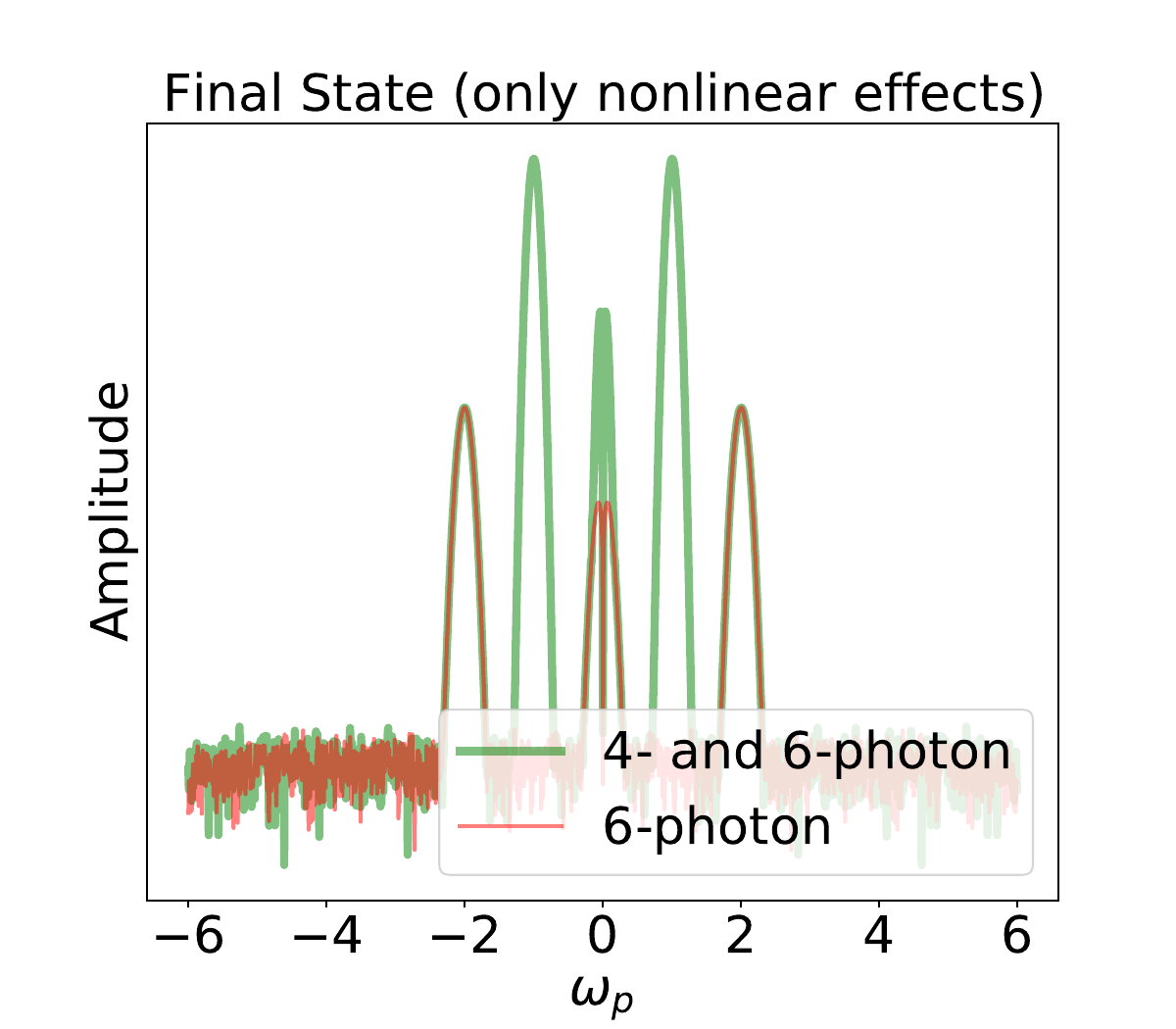}
	\end{subfigure}
	\caption[Log-scale plot of harmonics at different states in time]{Log-scale making the higher harmonics visible in frequency space.
		\textit{Top}: a full simulation in the linear vacuum supplemented by four- and six-photon nonlinear interactions.
		\textit{Bottom}: after subtraction of the linear vacuum.
		The initial state contains only the main signals of the pulses with no nonlinear interaction yet present.
		The overlap state denotes the time step where the pulses are directly overlapping, at the final state they have separated again -- the asymptotic field is left.
		It can be deduced that the third harmonic and the asymptotic part of the second harmonic are solely ascribable to six-photon processes.
	}
	\label{fig:harmonics_log}
\end{figure}

For six-photon processes it is obtained
\begin{itemize}
	\item the sheer scattering of background and probe photons contributing to the dc component and the fundamental harmonic;
	
	\item two background and two probe photons merging and producing one photon contributing to the second harmonic and one contributing to the dc component;
	
	\item two background and two probe photons merging and producing two photons contributing to the fundamental harmonic;
	
	\item two background photons and three probe photons merging and producing a photon contributing to the third harmonic ($\omega_r=3\omega_p$); and
	
	\item the merging of three background and two probe photons producing a photon contributing to the second harmonic.
\end{itemize}
A visualization of the contributions at selected points in time is provided with Figure \ref{fig:harmonics_log}.

These simulations are strictly 1D and the use of plane waves leads to strong constraints.
It can be seen that the asymptotic contribution to the second harmonic is only attributable to six-photon processes.
That is because wave-mixing -- resulting pulses that are combined of photons of both pulses -- is not allowed asymptotically.
The reason behind this is energy conservation, since for coaxial pulses and a photon resulting of wave-mixing it is found
\begin{equation}
    k^\mu_r = n_p \, \omega_p \, (1,\hat{k}) + n_b \, \omega_b \, (1,-\hat{k}) \quad\text{and}\quad (k^\mu_r)^2=0 \ \Rightarrow n_p \, n_b \overset{!}{=} 0 \ ,
\end{equation}
where $n_p$ and $n_b$ are the numbers of the contributing photons and $\hat{k}$ is the unit propagation direction vector of the probe pulse.
These states are thus only visible in the overlap position.
The six-photon process, on the other hand, can produce second harmonics without wave-mixing, see the second point for six-photon processes above.

As discussed in the context of birefringence in Section \ref{sec:flipping}, the 1D case corresponds to a simplified handling of the experimentally relevant scenario of counter-propagating pulses. 

For the highest generated harmonic, the short-lived third harmonic, the rule-of-thumb resolution limit for the grid defined at the end of Section \ref{sec:Dispersion} is slightly exceeded.
This comes without noticeable accuracy problems as is shown in the next subsections.

\subsection{Harmonic generation -- analytical results}

Analytical methods in \cite{Kingetal2014,Boehl2016} contain a derivation of iterative solutions to the nonlinear equations of motion for zero-frequency backgrounds.
With the probe (p) and background (b) pulses as time-dependent 1D Gaussian pulses with the parameters of Table \ref{tab:hh_gen} the pulses are obtained to be, compared to \eqref{eq:1dgaussian},
\begin{equation}
\vec{E}_p(x;t)= \vec{\epsilon}_p \, A_p\, e^{- (k_p^\mu x_{\mu p})^2/(\omega_p\tau_p)^2} \cos ( k_p^\mu x_{\mu p}) \quad\text{and}\quad \vec{E}_b(x;t)= \vec{\epsilon}_b \, A_b\, e^{- (k_b^\mu x_{\mu b})^2/(\omega_b\tau_b)^2} \ ,
\end{equation}
with $k_j^\mu x_{j\mu} = \omega_j t - \vec{k}_j\vec{x}_j$.
The shifted coordinates of probe and background field read
\begin{equation}
    \vec{x}_p = (x-\SI{100}{\micro\metre},0,0) \quad\text{and}\quad \vec{x}_b = (x-\SI{200}{\micro\metre},0,0) \ .
\end{equation}
Writing a combined electric field as
\begin{equation}
    \vec{E} =  \vec{E}_p + \vec{E}_b \ ,
\end{equation}
the inhomogeneous wave equation in 1D can be written as
\begin{equation}
    (\partial_t^2/c^2-\partial_x^2) \, \vec{E} = T[\vec{E}] \ .
\end{equation}
The source term $T[\vec{E}]$ comprises the nonlinear Heisenberg--Euler interactions.
Decomposing the electric field into
\begin{equation}
    \vec{E} =  \vec{E}^{(0)} +  \vec{E}^{(1)} + ... \ ,
\end{equation}
where $\vec{E}^{(0)}$ solves the free vacuum wave equation $(\partial_t^2/c^2-\partial_x^2) \, \vec{E}^{(0)} = 0$,
an iterative procedure is obtained \cite{Kingetal2014,Boehl2016} in which
\begin{equation}\label{eq:nonlinwave1}
    (\partial_t^2/c^2 - \partial_x^2) \, \vec{E}^{(1)} = T[\vec{E}^{(0)}] \ .
\end{equation}
With the polarization for both fields given by $ \vec{\epsilon} = (0,1,0)  $
and defining the shorthands
\begin{equation}
\begin{aligned}
     \kappa_p &= \frac{k_p^\mu x_{\mu p}}{\omega_p \tau_p} \quad \text{and} \quad \kappa_b = \frac{k_b^\mu x_{\mu b}}{\omega_b \tau_b} \ ,
\end{aligned}
\end{equation}
it is then obtained for the solution to the nonlinear wave equation \eqref{eq:nonlinwave1} at the first iterative order for \cite{Kingetal2014,Boehl2016}
\begin{itemize}
	\item the dc component:
	\subitem the overlap field
	\begin{equation}\label{eq:dc_overlap}
		\vec{E}^{(1)}_{0,o} = - \frac{8\alpha}{180\pi} \, A_p^2 \, \vec{E}_b(x;t) \, e^{-2\kappa_p^2} \, \vec{\epsilon}
	\end{equation}
	\subitem and the asymptotic field
	\begin{equation}
		\vec{E}^{(1)}_{0,a} =\frac{8\alpha}{180\pi} \, A_p^2 \, \sqrt{\frac{\pi}{2}} \, \frac{\tau_p \kappa_b}{\tau_b} (1+\textrm{erf}(\sqrt{2}\kappa_b)) \, A_b \, e^{-\kappa_b^2} \, \vec{\epsilon} \ ;
	\end{equation}
	
	\item the fundamental harmonic:
	\subitem the overlap field
	\begin{equation}
		\vec{E}^{(1)}_{1,o} = - \frac{8\alpha}{90\pi}  \, A_p e^{-\kappa_p^2} \, \vec{E}_b(x;t)^2 \, \cos(k_p^\mu x_{p\mu}) \, \vec{\epsilon} 
	\end{equation}
	\subitem and the asymptotic field
	\begin{equation}
		\vec{E}^{(1)}_{1,a} =  \frac{8\alpha}{90\pi} \, A_p e^{-\kappa_p^2} \, A_b^2 \, \sqrt{\frac{\pi}{2}} \omega_p \tau_b \, \frac{1+\textrm{erf}(\sqrt{2}\kappa_b)}{2} \, \sin(k_p^\mu x_{p\mu}) \, \vec{\epsilon} \ ;
	\end{equation}
	
	\item the second harmonic:
	\subitem the overlap field
	\begin{equation}
		\vec{E}^{(1)}_{2,o} = -A_p^2 e^{-2\kappa_p^2} \,  \left[ \frac{8\alpha}{180\pi} \, \vec{E}_b(x;t)^2 \, + \frac{96\alpha}{630\pi} \, \vec{E}_b(x;t)^3 \right]  \cos(2k_p^\mu x_{p\mu}) \, \vec{\epsilon} 
	\end{equation}
	\subitem and the asymptotic field
	\begin{equation}
		\vec{E}^{(1)}_{2,a} = \frac{96\alpha}{315\pi} \, A_p^2 e^{-2\kappa_p^2}  \,  A_b^3 \, \sqrt{\frac{\pi}{3}} \omega_p \tau_b \, \frac{1+\textrm{erf}(\sqrt{3}\kappa_b)}{2} \,   \sin(2k_p^\mu x_{p\mu}) \, \vec{\epsilon} \ ;
	\end{equation}
	and
	
	\item the third harmonic:
	\subitem the overlap field
	\begin{equation}
		\vec{E}^{(1)}_{3,o} = - \frac{96\alpha}{1260\pi} \, A_p^3 e^{-3\kappa_p^2} \, \vec{E}_b(x;t)^2 \, \cos(3k_p^\mu x_{p\mu}) \, \vec{\epsilon}
	\end{equation}
	\subitem and the asymptotic field
	\begin{equation}\label{eq:3rd_asymptotic}
		\vec{E}^{(1)}_{3,a} = 0 \ .
	\end{equation}
	
\end{itemize}
Using the values from Table \ref{tab:hh_gen} the solid lines in Figure \ref{fig:harmonics_numerical} are obtained.
Those are employed to scrutinize the correctness of the simulation results.

\begin{figure}
	\centering
	\includegraphics[width=\linewidth]{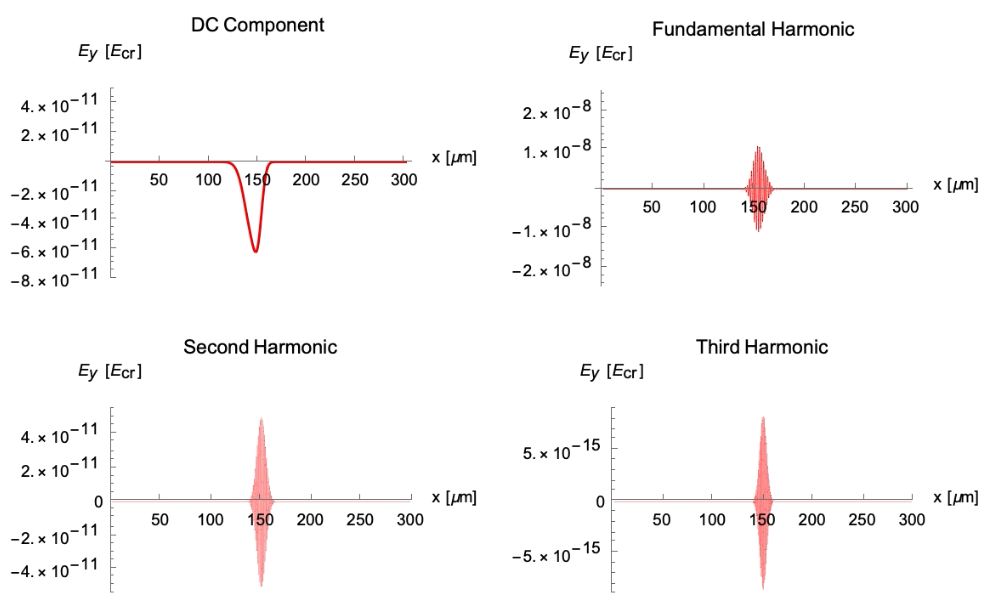}
	\caption[Thumbnail of an animation of nonlinearly generated harmonics]{
		Thumbnail of animations available in the \textit{Mendeley Data} repository \cite{Lindner2022} showing the
		time evolution of the various harmonics in 1D as a consequence of a pulse collision with the specifications of Table \ref{tab:hh_gen} with nonlinear four-photon and six-photon interactions.}
	\label{vid:1d_harmonics}
\end{figure}

A \textit{Mathematica} \cite{Mathematica} analysis of the analytical results for harmonic generation can be found in \cite{Lindner2022b}.
Animations of the arising and evolution of the various harmonics are provided in the \textit{Mendeley Data} repository \cite{Lindner2022}
(thumbnails and description in Figure \ref{vid:1d_harmonics}).

\subsection{Harmonic generation -- simulation results}

To put the focus on the nonlinear contributions to the generation of harmonics, each setting is simulated three times with varying combinations of interactions included:
once including only the nonlinear effects of four- and six-photon contributions;
once including only six-photon processes;
once excluding all nonlinear interactions, i.e., keeping only the linear vacuum.
By subtracting the dynamics in the linear vacuum from the full dynamics, the higher order processes of the weak-field expansion are extracted.
Furthermore, he simulations of six-photon processes permit to isolate their sole contribution, as is shown in Figure \ref{fig:harmonics_log}.
It is a fruitful feature of the simulation code that the contributions of four- and six-photon diagrams can be turned on and off to make them separately visible.

\begin{figure}
	\centering
	\begin{subfigure}{.5\textwidth}
		\includegraphics[width=\linewidth]{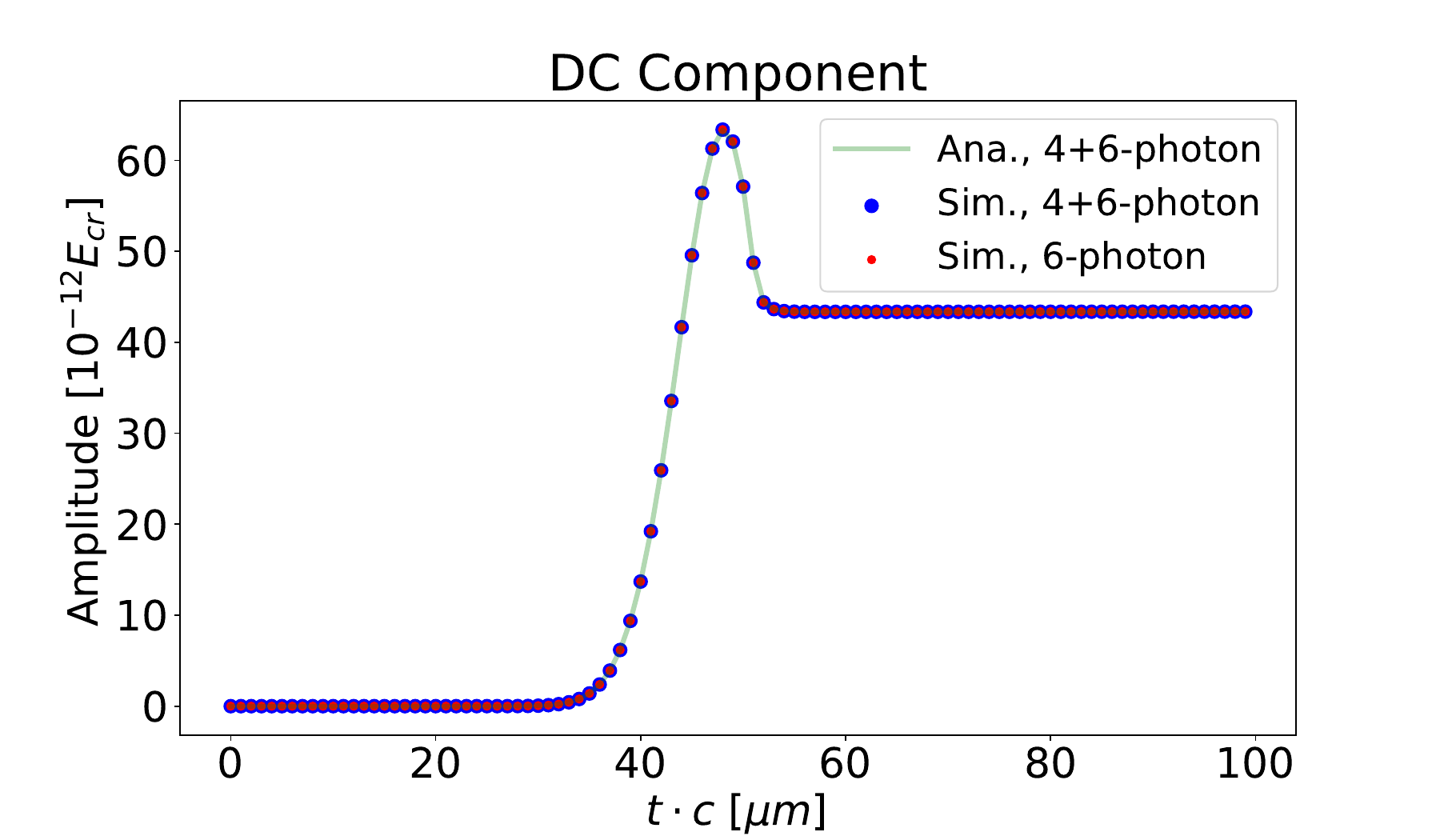}
	\end{subfigure}%
	\begin{subfigure}{.5\textwidth}
		\includegraphics[width=\linewidth]{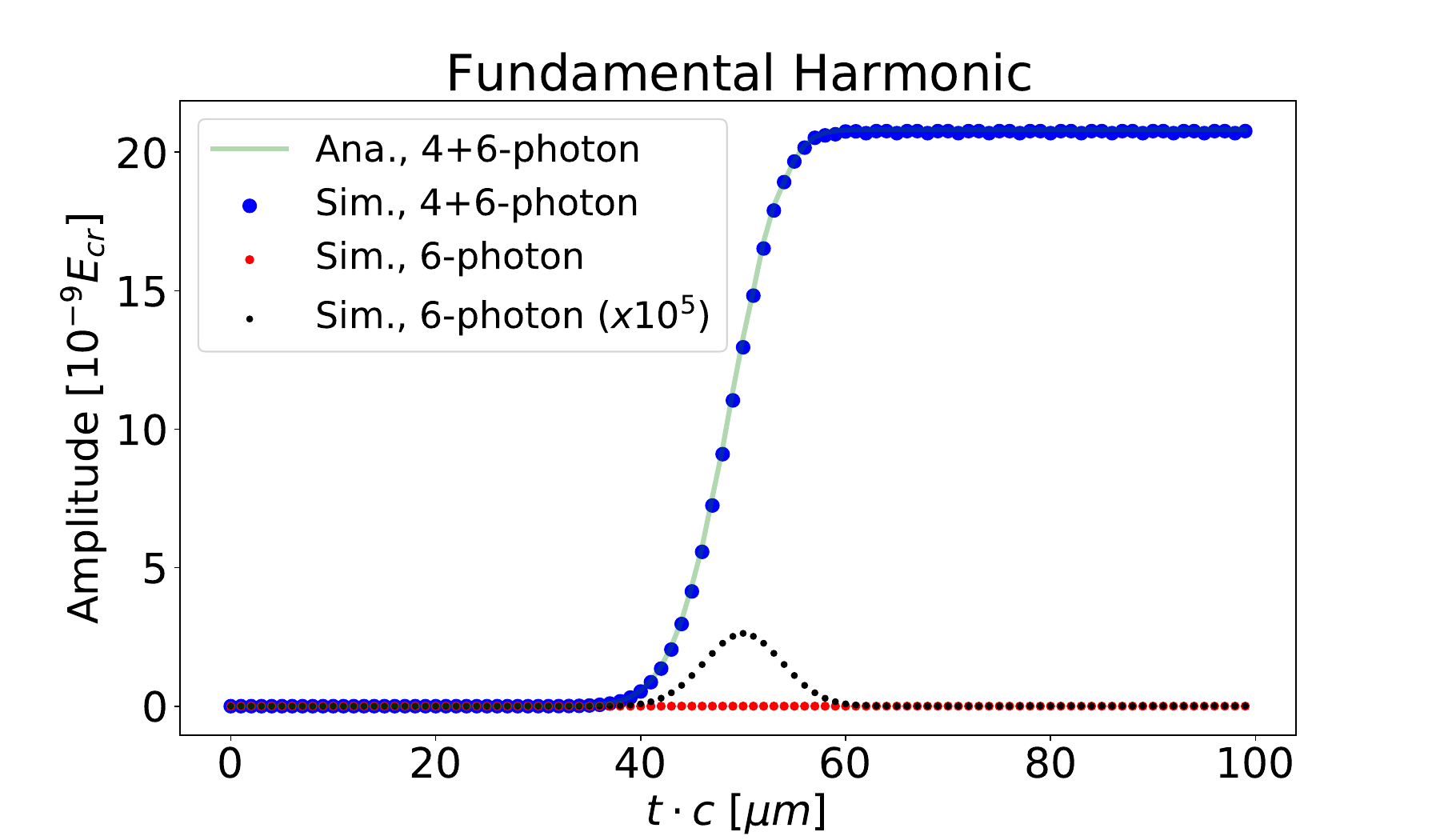}
	\end{subfigure}
	\begin{subfigure}{.5\textwidth}
		\includegraphics[width=\linewidth]{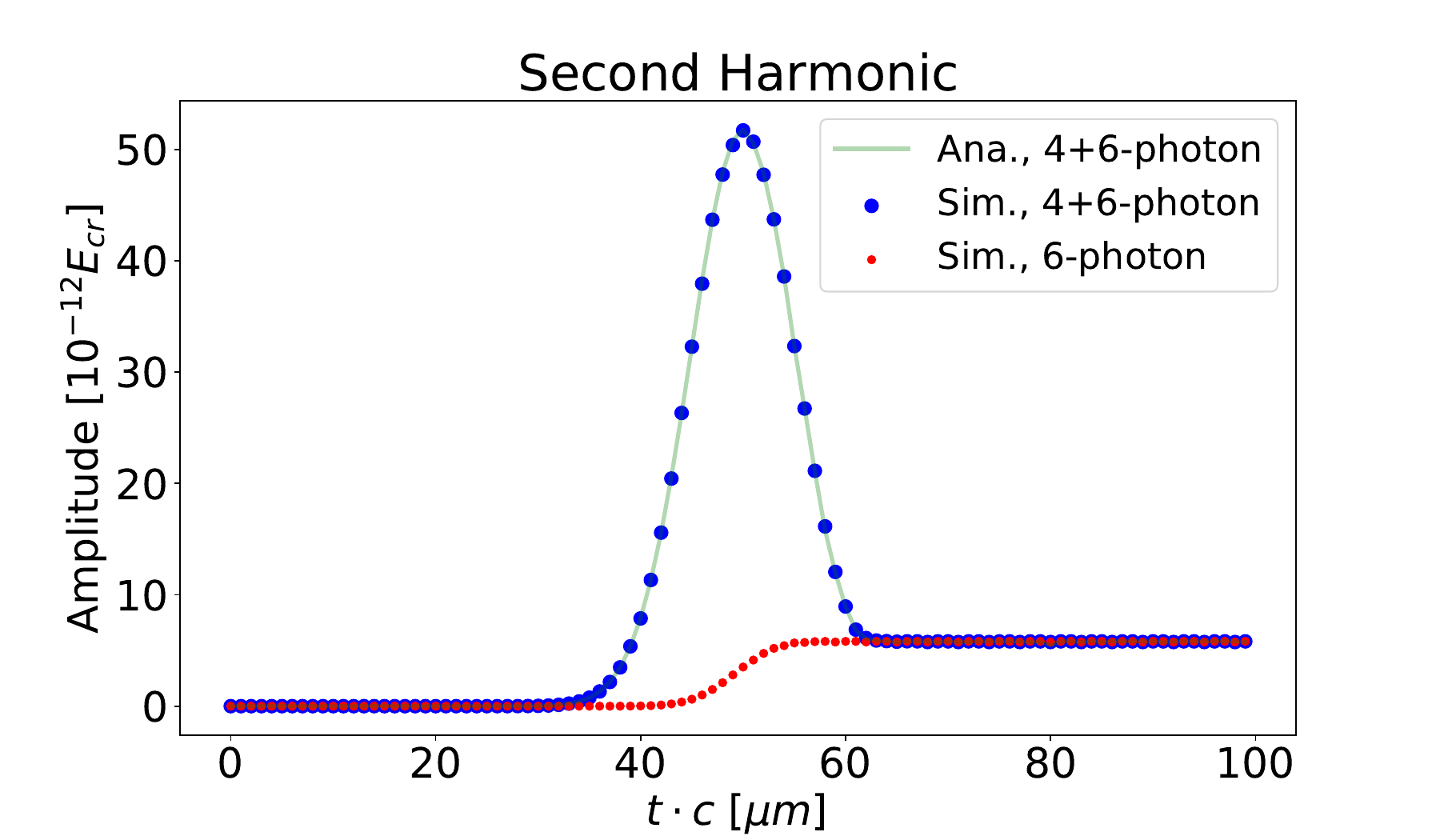}
	\end{subfigure}%
	\begin{subfigure}{.5\textwidth}
		\includegraphics[width=\linewidth]{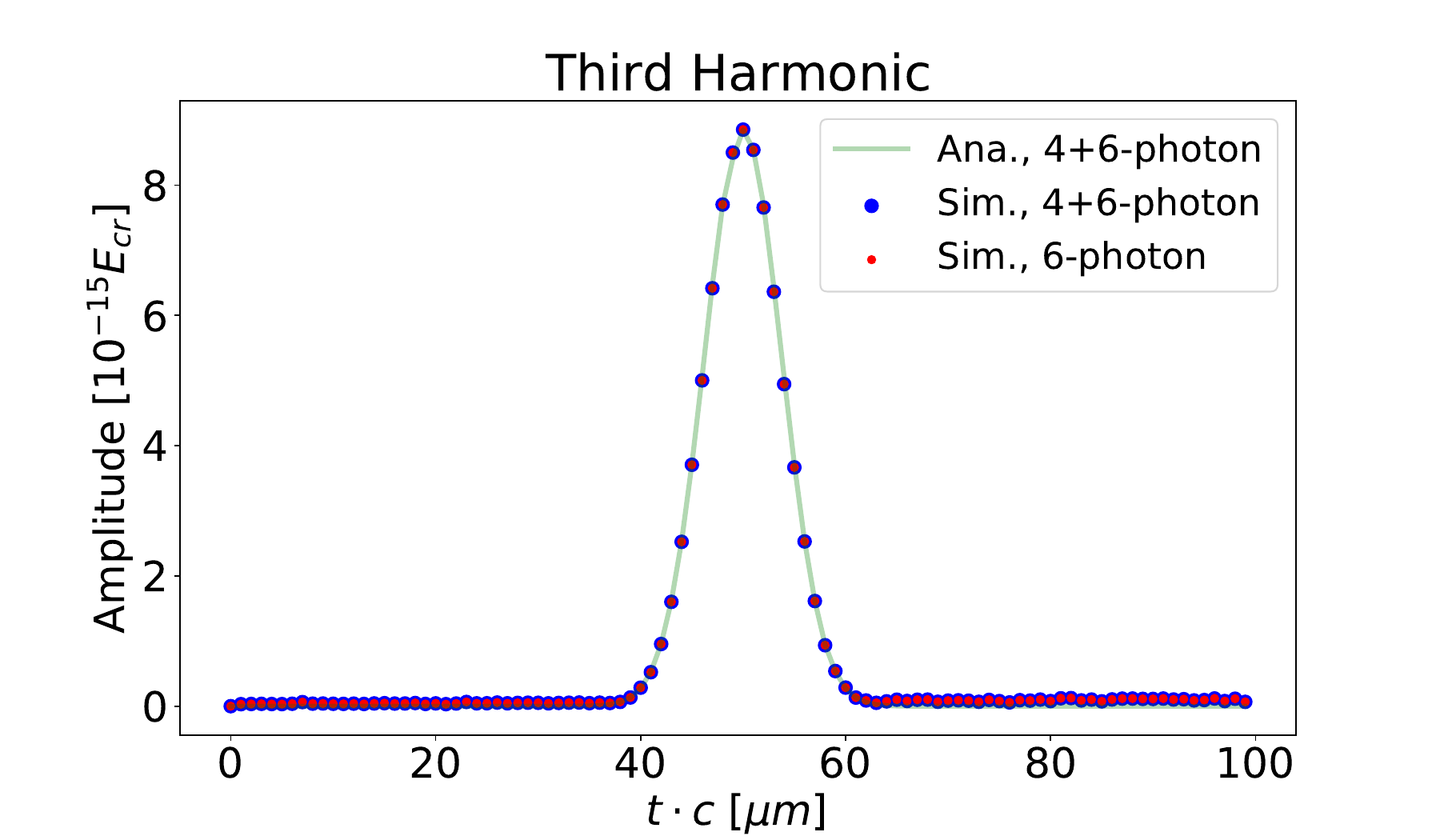}
	\end{subfigure}
	\caption[Amplitude evolution of nonlinearly generated harmonics]{Amplitude evolution of the nonlinearly generated harmonics.
		Shown are the simulation results for harmonics caused by four- and six-photon processes combined (\textit{blue dots}) and by only six-photon processes (\textit{red dots}), all without the linear vacuum contributions. 
		The analytical approximations summarized in \eqref{eq:dc_overlap}-\eqref{eq:3rd_asymptotic} are underlying (\textit{light green curve}).}
	\label{fig:harmonics_numerical}
\end{figure}

In order to extract the amplitudes of the various arising harmonics and their time evolution, their respective frequencies have to be filtered in Fourier space and then transformed back to position space \cite{Pons2018}.
The amplitude is then obtained as the maximum norm.
Its time evolution can be observed, in this case with a chosen resolution of a time step of \SI{1}{\micro\metre}/c. 
The results for the zeroth harmonic, first harmonic, second, and third harmonic are displayed in Figure \ref{fig:harmonics_numerical}.

As can be seen in Figure \ref{fig:harmonics_numerical}, there is good agreement between the analytical approximation and the simulation results.
Small systematic errors are unavoidable as a consequence of the back and forth Fourier transformation and slicing of frequency ranges.
The mean absolute percentage errors of the simulation results are calculated to be less than 1\% in the regions where the amplitudes are non-vanishing.

Asymptotic states are constrained by energy-momentum conservation, while the overlap state has a richer spectrum.
The overlap spectrum becomes even more pronounced and versatile when the pulses collide at a non-zero angle in higher dimensions.
First such results are demonstrated in Section \ref{sec:2dsims}, where also the zero-frequency background restriction is relaxed.
Signals, which are degenerate in the case of a non-zero-frequency pulse, split up in that case.

Simulations in higher dimensions provide a powerful means to analyze varying collision configurations.
Situations that pose no further difficulty to the numerical code are considerably hard to cope with analytically.
Employing simulations of the solver it is possible to track harmonic frequencies in time and space for scenarios of arbitrary pulse parameters -- with the only restrictions posed by the applicability of the Heisenberg--Euler weak-field expansion and computational feasibility.

\section{Higher-dimensional simulations}\label{sec:2dsims}

For simulations in 2D a special adaptation of 3D Gaussian pulses is used to model the diffusion behavior.
The pulse is assumed to propagate along the $z$-axis. 
The widening of the beam with respect to the longitudinal coordinate $z$ is given by the waist
\begin{equation}
    w(z) = w_0 \sqrt{1+ \left(\frac{z-z_0}{z_R}\right)^2} \ 
\end{equation}
with $w_0$ the waist of the beam at position $z_0$, where the amplitude is $1/e$ of the initial value.
The cross-sectional area in 1D is a point, in 2D is a line, and in 3D is an area.
The field intensity scales with $w_0/w(z)$ and with the surface area $\sim z^2$ in 3D.
Lateral dispersion has to be taken into account for the 2D Gaussian pulses, where the surface scales as $\sim z$.
Hence, the factor $w_0/w(z)$ appearing as prefactor in the Gaussian pulses gets a square root in the lower-dimensional case.
The pulse can thus be written as
\begin{equation}\label{eq:2dgaussian}
    \vec{E}(r,z(t)) = A \, \vec{\epsilon} \, \sqrt{\frac{w_0}{w(z)}} \, e^{-(r/w(z))^2} \, e^{-((z-z_\tau)/\tau_z)^2} \, \cos \left( \frac{k\, r^2}{2R(z)} + \zeta(z) -kz \right) \ .
\end{equation}
It is further defined
\begin{itemize}
	\item the parameter $A$ determining the peak pulse amplitude and the polarization $\vec{\epsilon}$ ;
	
	\item the distance to the propagation axis (here taken to be $ z $) $r = \sqrt{x^2+y^2}$ ;
	
	\item the wavenumber $k=2\pi/\lambda $ ;
	
	\item the pulse width in $z$-direction $\tau_z$ (pulse duration) and the envelope center $z_\tau \,$; and 
	\item the Rayleigh length $z_R = \pi w_0^2 / \lambda$ as the longitudinal distance from $z_0$ at which the waist has increased by a factor of $\sqrt{2}$ ,
	which is contained in
	\subitem $ \bullet $ the Gouy phase $\zeta(z) = \arctan (z/z_R)\,$, and
	\subitem $ \bullet $ the radius of curvature $R(z) = z (1+(z_R/z)^2) $ .
	
\end{itemize}
The settings used for 2D simulations, which are  confined to the x-y-plane for propagation, are listed in Table \ref{tab:2d_sims}.

\begin{SCtable}[]
	\centering
	\caption[Parameters to simulate harmonic generation in 2D]{Settings for 2D simulations with two conceptually equal Gaussian pulses.\\
	The wavelength is obtained via $\lambda=\pi w_0^2/z_R$.	
	The Rayleigh length and waist are chosen such that the wavelength equals one micrometer. \label{tab:2d_sims}}
	\renewcommand{\arraystretch}{1.3}  
	\begin{tabular}{!{\vrule width 3\arrayrulewidth}c|c|c!{\vrule width 3\arrayrulewidth}}
		\noalign{\hrule height 3\arrayrulewidth}
		\textbf{Grid} &	Square Size & \SI{80}{\micro\metre} $\times$  \SI{80}{\micro\metre}
		\\ \hline
		& Lattice Points & 1024$\times$1024 
		\\ \noalign{\hrule height 3\arrayrulewidth}
		\textbf{Pulse 1}
		& $\vec{\epsilon}$ &  (0,0,1) 
		\\\hline
		& $A$ & 50$\times \SI{e-3}{} E_{\textrm{cr}}  $
		\\\hline
		& $\hat{k}$ & (-1,0,0)
		\\\hline
		& $\lambda$ & \SI{1}{\micro\metre}
		\\\hline
		& $w_0$  &  \SI{2.3}{\micro\metre}
		\\\hline
		& $z_R$  &  \SI{16.619}{\micro\metre}
		\\\hline
		& $z_\tau$  & \SI{20}{\micro\metre}
		\\\hline
		& $\tau_z$  &  \SI{4.5}{\micro\metre}
		\\
		\noalign{\hrule height 3\arrayrulewidth}
		\textbf{Pulse 2} & same parameters as for pulse 1, &
		\\
		& but varying propagation direction & see Figures \ref{fig:kspace_0}-\ref{fig:kspace_135_orthogonal_pol}
		\\
		& and polarization &
		\\
		\noalign{\hrule height 3\arrayrulewidth}
	\end{tabular}
\end{SCtable}

\begin{figure}[H]
	\centering
	\includegraphics[width=\linewidth]{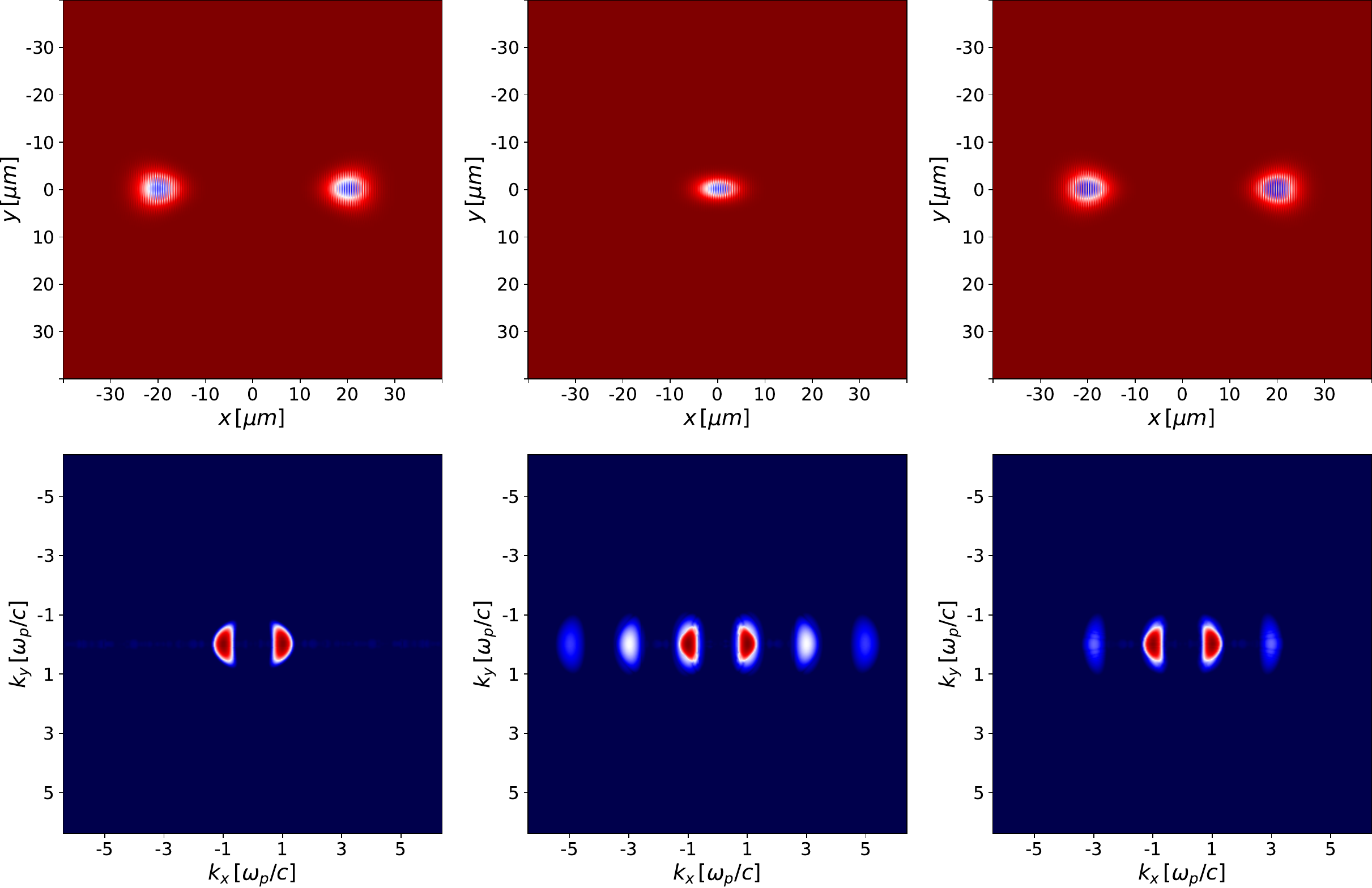}
	\caption[Harmonics in 2D for collinear pulses]{Coaxially colliding pulses with the same polarization.
		The left plots show the initial state, those in the middle the overlap state, and the right ones the final state.
		\textit{Top}: position space. \textit{Bottom}: frequency space.}
	\label{fig:kspace_0}
\end{figure}

\begin{figure}[H]
	\centering
	\includegraphics[width=\linewidth]{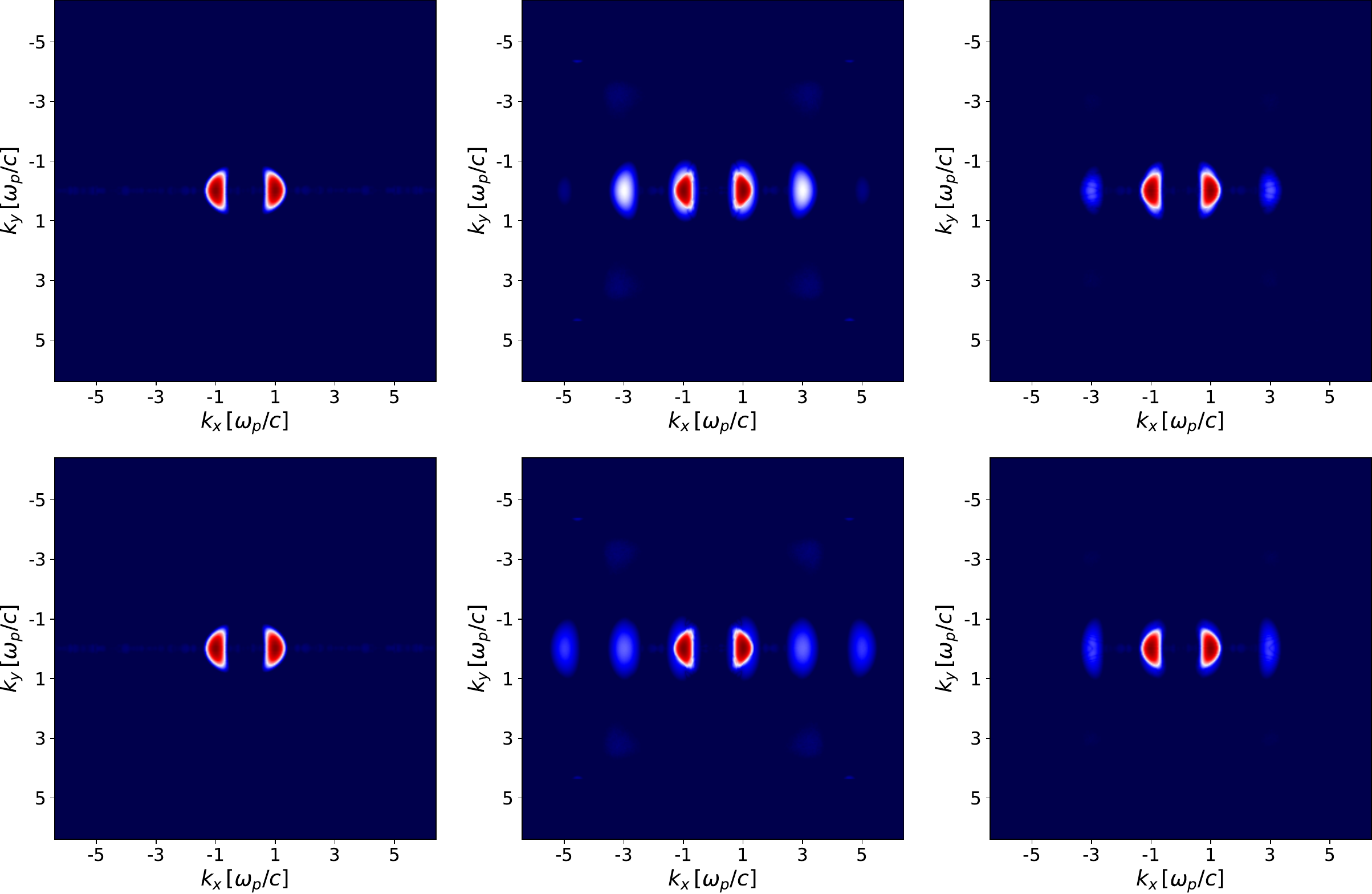}
	\caption[Harmonics in 2D for collinear pulses (four- and six-photon processes distinguished)]{Frequency space of coaxially colliding pulses with the same polarization.
	The left plots show the initial state, those in the middle the overlap state, and the right ones the final state.
	\textit{Top}: only four-photon diagrams included. \textit{Bottom}: only six-photon diagrams included.}
	\label{fig:kspace_0_46}
\end{figure}

For coaxial pulse collisions there is a correspondence to the 1D scenario with respect to the generated harmonics.
Comparing a head-on collision in 1D (Figure \ref{fig:harmonics_log}) and 2D (Figure \ref{fig:kspace_0}), a similar frequency spectrum is found at the different states in time.
The contributions of the different orders in the weak-field expansion are demonstrated again (Figure \ref{fig:kspace_0_46}).
The reasoning is the same as in Section \ref{sec:harmonics}.
The main difference stems from the fact that the two conceptually equal pulses have the same non-zero frequency $\omega_p$.
There is still a degeneracy of signals present by virtue of the equal frequencies.

A feature which is not present in 1D simulations is a lateral broadening of the pulses in frequency space, which arises during the interaction and remains.
The presence of a lateral beam profile of the pulses is a prerequisite to invoke this outcome.
This results in outgoing signal photons with transverse momentum components, effectively yielding a diffraction effect \cite{DiPiazzaetal2006}.
Diffraction spreading opens up the opportunity to detect signal photons off the beam axis with a background free measurement.
The scattering of polarization-flipped signal photons outside the forward cone of a probe beam may thus constitute an essential key ingredient for the
detection of vacuum birefringence \cite{Karbsteinetal2015}.

Correspondingly, from the position space point of view, the transversal momenta might imply a slight focusing of the pulses.
This can be explained with the lensing effect of a power pulse which creates a refractive index influencing the propagation speed and direction, as detailed in Section \ref{sec:phasevel}.
With a lower refractive index at the outer waist regions of at least one of the pulses, light passing through the strong-field zone experiences a phase velocity change comparable to the one in a convex lens.
Focusing of light by light is an interesting topic to be further investigated with the help of adequate simulations with tailored pulse parameters.

\begin{figure}
	\centering
	\includegraphics[width=\linewidth]{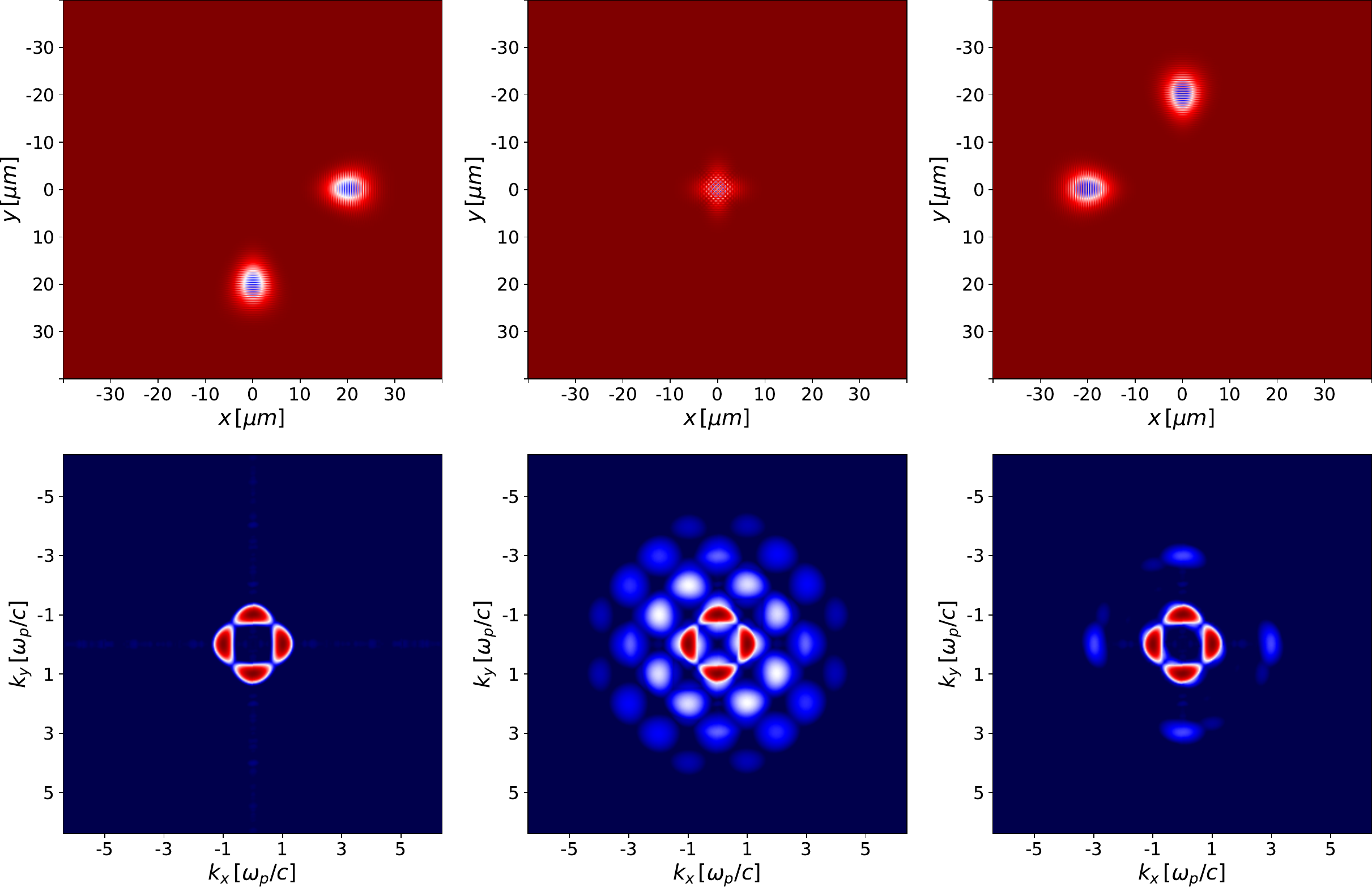}
	\caption[Harmonics in 2D for perpendicular pulses]{Perpendicularly colliding pulses with equal polarization.
	The left plots show the initial state, those in the middle the overlap state, and the right ones the final state.
	\textit{Top}: position space. \textit{Bottom}: frequency space.}
	\label{fig:kspace_90}
\end{figure}

\begin{figure}
	\centering
	\includegraphics[width=\linewidth]{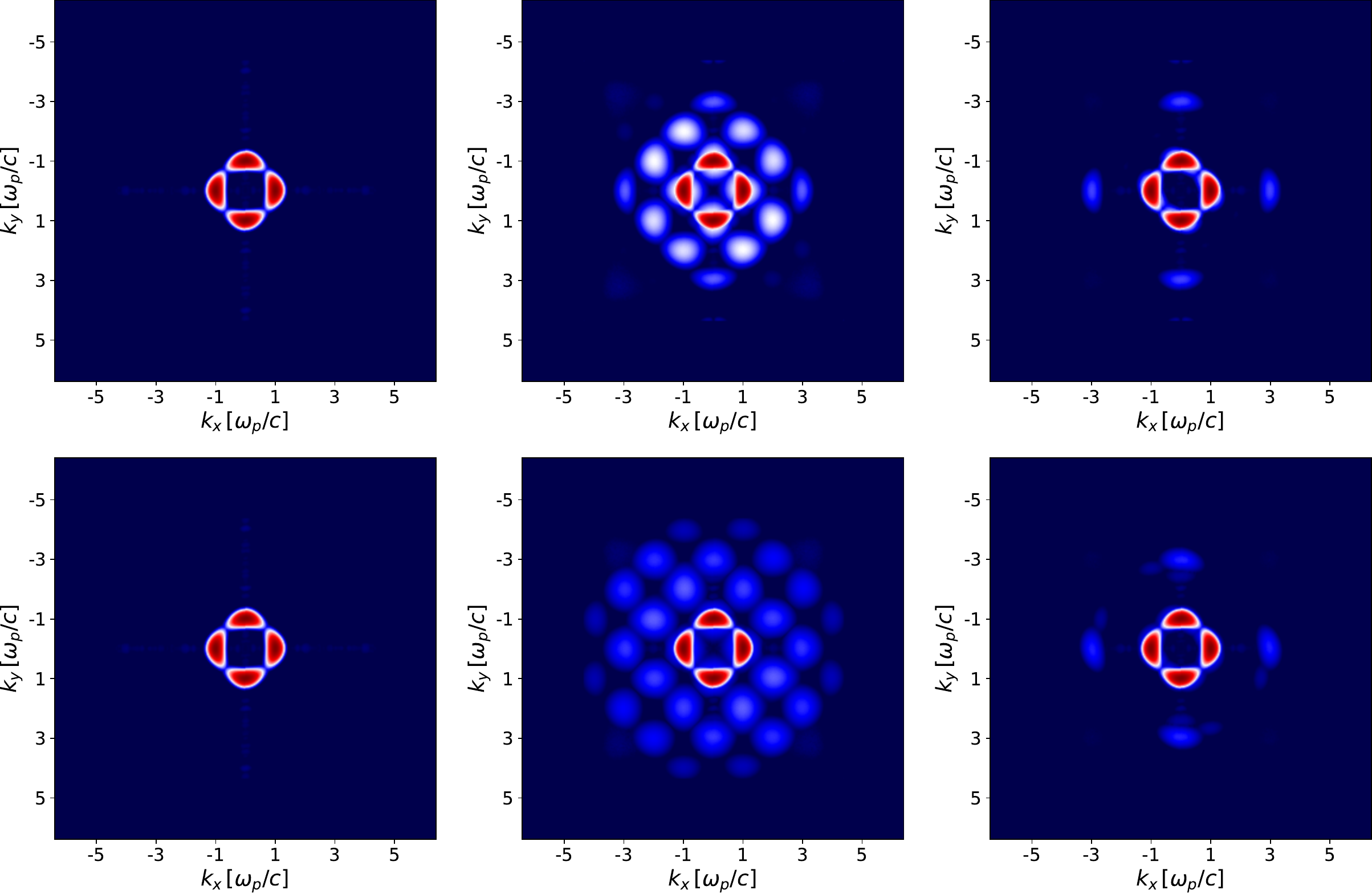}
	\caption[Harmonics in 2D for perpendicular pulses (four- and six-photon processes distinguished)]{Frequency space of perpendicularly colliding pulses with same polarization.
	The left plots show the initial state, those in the middle the overlap state, and the right ones the final state.
	\textit{Top}: only four-photon diagrams included. \textit{Bottom}: only six-photon diagrams included.}
	\label{fig:kspace_90_46}
\end{figure}

\begin{figure}
	\centering
	\includegraphics[width=\linewidth]{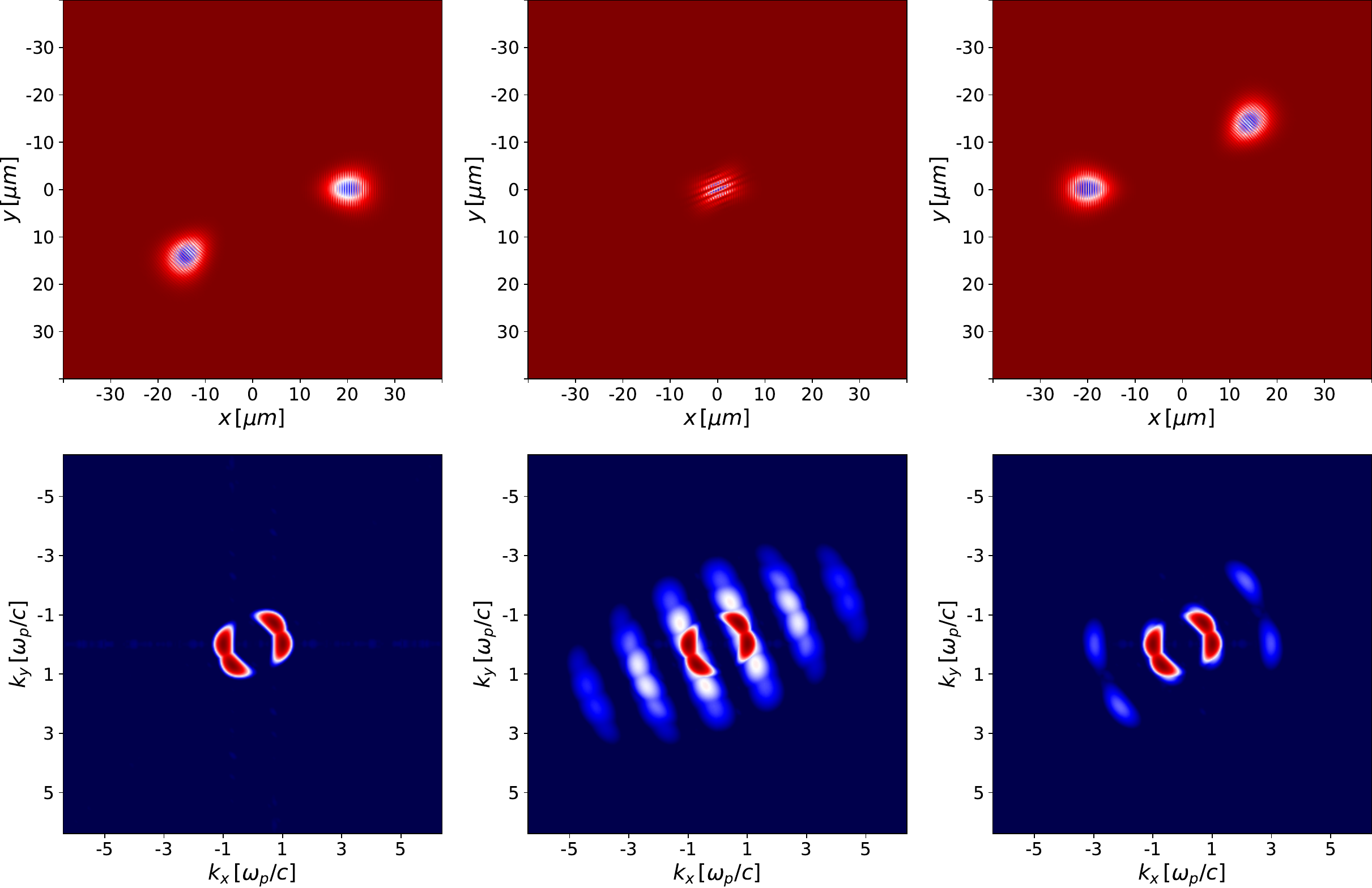}
	\caption[Harmonics in 2D for pulses at 135\textdegree]{Pulses with the same polarization colliding at an angle of 135\textdegree.
	The left plots show the initial state, those in the middle the overlap state, and the right ones the final state.
	\textit{Top}: position space. \textit{Bottom}: frequency space.}
	\label{fig:kspace_135}
\end{figure}

\begin{figure}
	\centering
	\includegraphics[width=\linewidth]{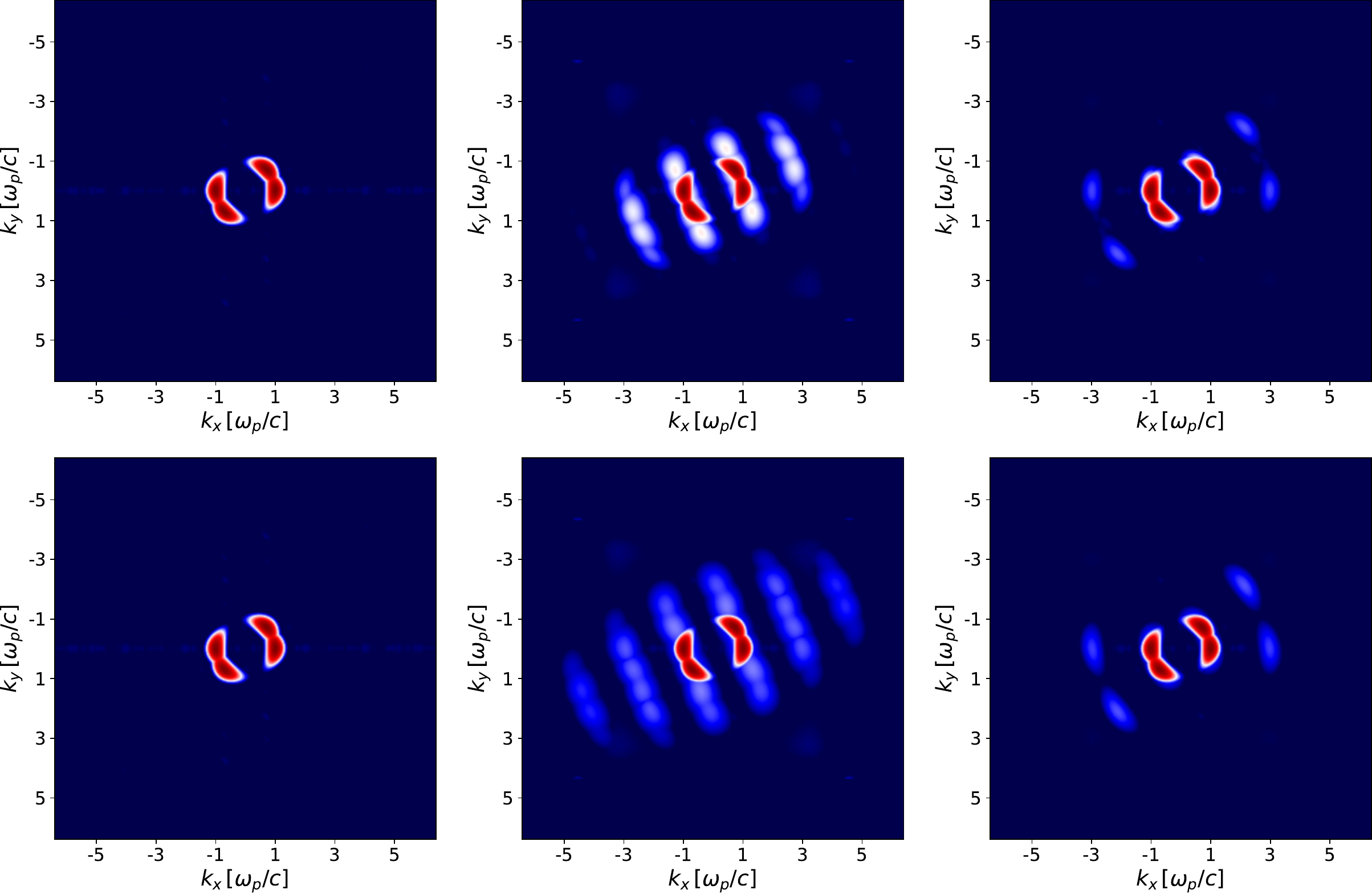}
	\caption[Harmonics in 2D for pulses at 135\textdegree \ (four- and six-photon processes distinguished)]{Frequency space of pulses with the same polarization colliding at an angle of 135\textdegree.
	The left plots show the initial state, those in the middle the overlap state, and the right ones the final state.
	\textit{Top}: only four-photon diagrams included. \textit{Bottom}: only six-photon diagrams included.}
	\label{fig:kspace_135_46}
\end{figure}

Going beyond coaxial pulses by varying the collision angle and thereby lifting the degeneracy of frequencies of the harmonics, geometry effects with rich spectra in 2D simulations can be observed, see Figures \ref{fig:kspace_90}--\ref{fig:kspace_135_46} for perpendicularly propagating and colliding pulses as well as for a collision angle of 135\textdegree.
Most signals vanish again in the asymptotic state.
These off-axis contributions occur due to the field spatio-temporal inhomogeneities \cite{Grismayeretal2021}.

The asymptotic harmonics that can be seen in the right frames of the frequency plots are on account of the self-interaction of the 2D Gaussian pulses \cite{NarozhnyFedotov2007,Grismayeretal2021}.
Time-resolving the processes reveals that these harmonics arise immediately as the dynamics begin, directly after the initial configuration shown at the lower left of the figure, and thus already before the pulses overlap.
This can be seen in the corresponding simulation videos in the \textit{Mendeley Data} repository\cite{Lindner2022}.
Both the four- and six-photon processes contribute to the asymptotic signals with $\vert \vec{k} \vert = \omega_p/c$ and $\vert \vec{k} \vert = 3 \omega_p/c$.

Six-photon processes contribute to all harmonics, in the overlap as well as in the asymptotic state.
In the overlap states various merging processes become directly visible.
A more precise analysis shows that the four-photon spectrum in the overlap state is even richer.
Some signals, however, are of the order of accumulating numerical errors.
Such errors appear, e.g., at the corners of the frequency plots for four-photon and six-photon processes, irrespective of the pulse alignment.

The signals of asymptotic harmonics generated by four-photon interactions are aligned on the initial propagation axes, while six-photon processes seem to generate a tiny off-axis twist.
Note that the initial symmetry of the two-pulse system is thereby conserved.

\begin{figure}
	\centering
\begin{subfigure}{\textwidth}
	\includegraphics[width=\linewidth]{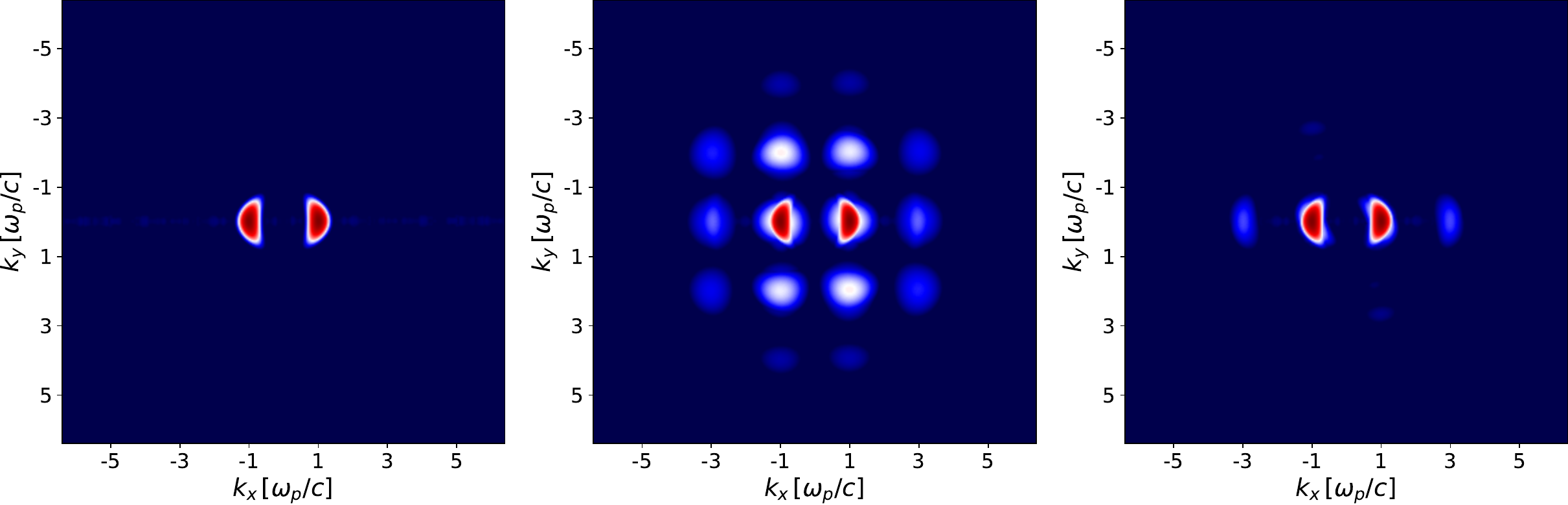}
\end{subfigure}

\begin{subfigure}{\textwidth}
	\includegraphics[width=\linewidth]{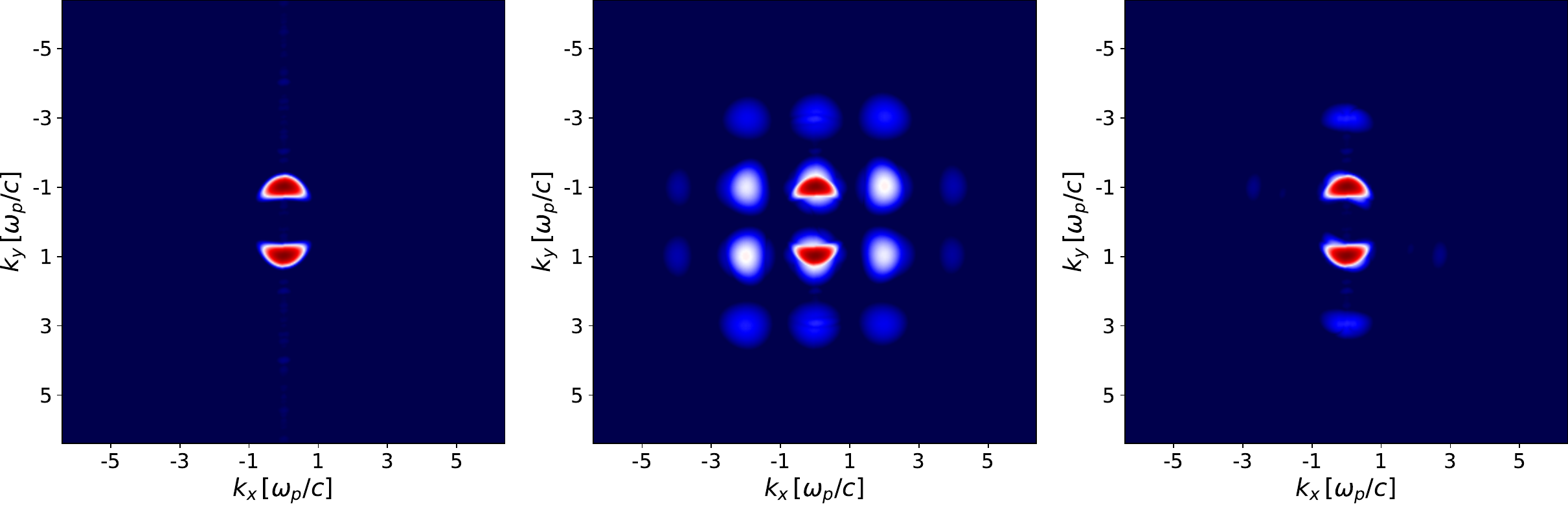}
\end{subfigure}
\caption[Harmonics in 2D for perpendicular pulses and orthogonal polarization]{Frequency space of perpendicularly colliding pulses with orthogonal relative polarization.
The left plots show the initial state, those in the middle the overlap state, and the right ones the final state.
\textit{Top}: the $E_z$ component is shown.
\textit{Bottom}: the $B_z$ component is shown.}
\label{fig:kspace_90_orthogonal_pol}
\end{figure}

\begin{figure}
	\centering
\begin{subfigure}{\textwidth}
	\includegraphics[width=\linewidth]{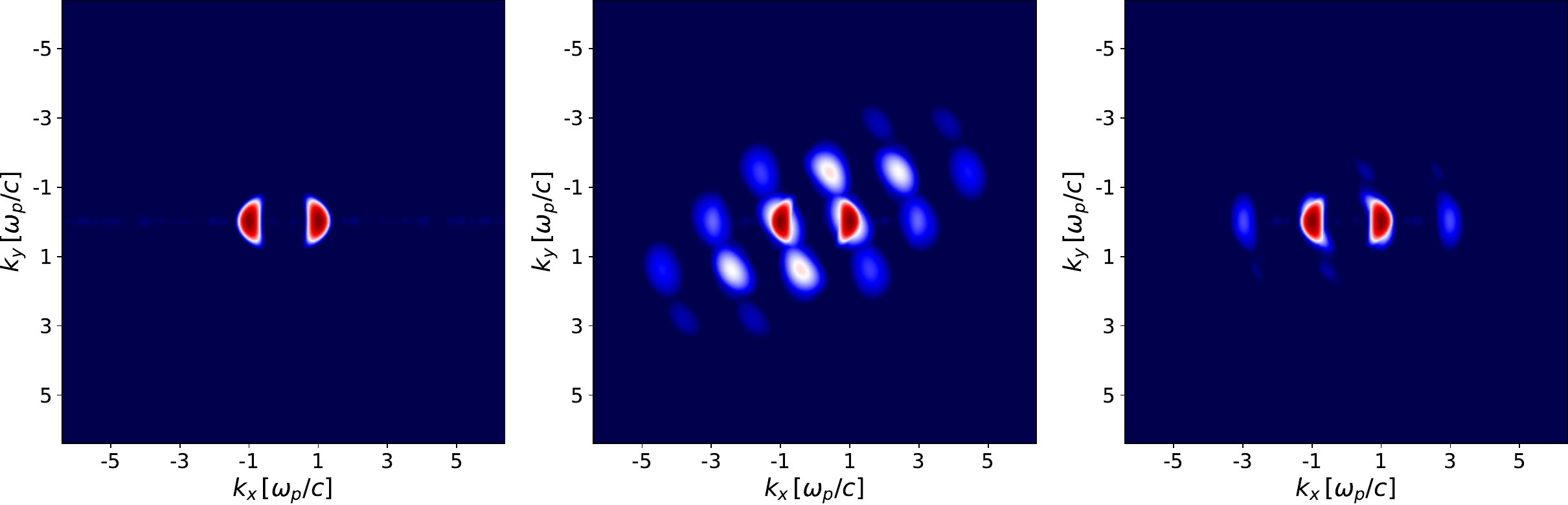}
\end{subfigure}%

\begin{subfigure}{\textwidth}
	\includegraphics[width=\linewidth]{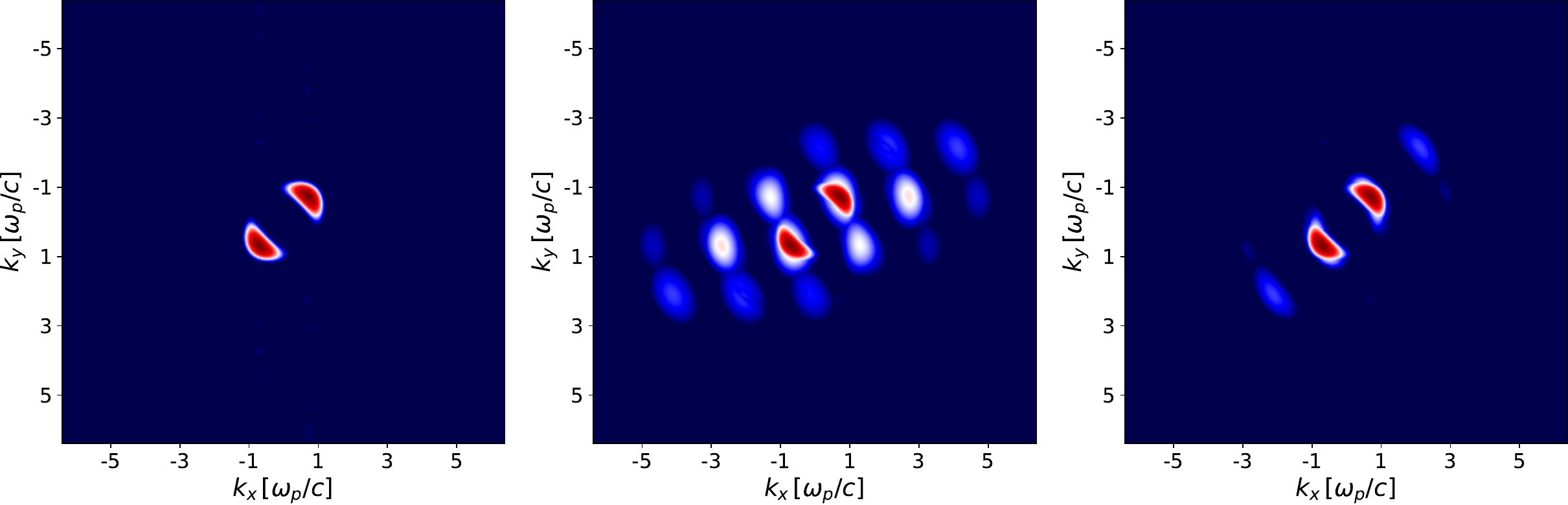}
\end{subfigure}
\caption[Harmonics in 2D for pulses at 135\textdegree \ and orthogonal polarization]{Frequency space of pulses colliding at an angle of 135\textdegree \ with orthogonal relative polarization.
The left plots show the initial state, those in the middle the overlap state, and the right ones the final state.
\textit{Top}: the $E_z$ component is shown.
\textit{Bottom}: the $B_z$ component is shown.}
\label{fig:kspace_135_orthogonal_pol}
\end{figure}

Giving the pulses different polarization directions, the harmonics can be tagged.
The frequency space of the simulations visualized in Figures \ref{fig:kspace_90_orthogonal_pol} and \ref{fig:kspace_135_orthogonal_pol} show again collision angles of 90\textdegree \ and 135\textdegree , but the pulse propagating from the left is now polarized along the $E_y$-direction.
Hence, the shown polarization components can be identified with one of the two pulses, and thus also the harmonics.

\begin{figure}
	\centering
	\begin{subfigure}{0.4\textwidth}
		\includegraphics[width=\linewidth]{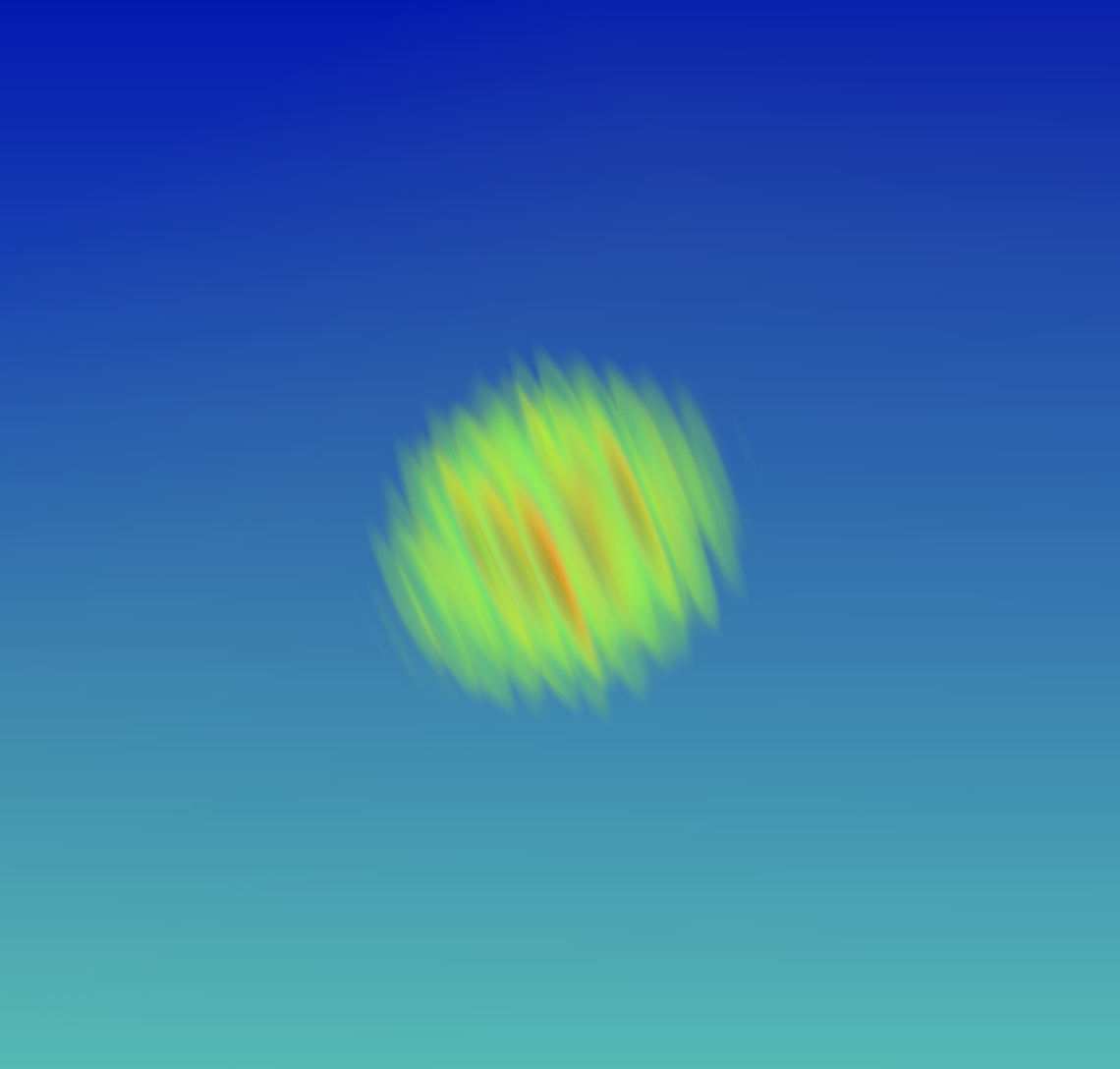}
	\end{subfigure}\qquad
	\begin{subfigure}{0.4\textwidth}
		\includegraphics[width=\linewidth]{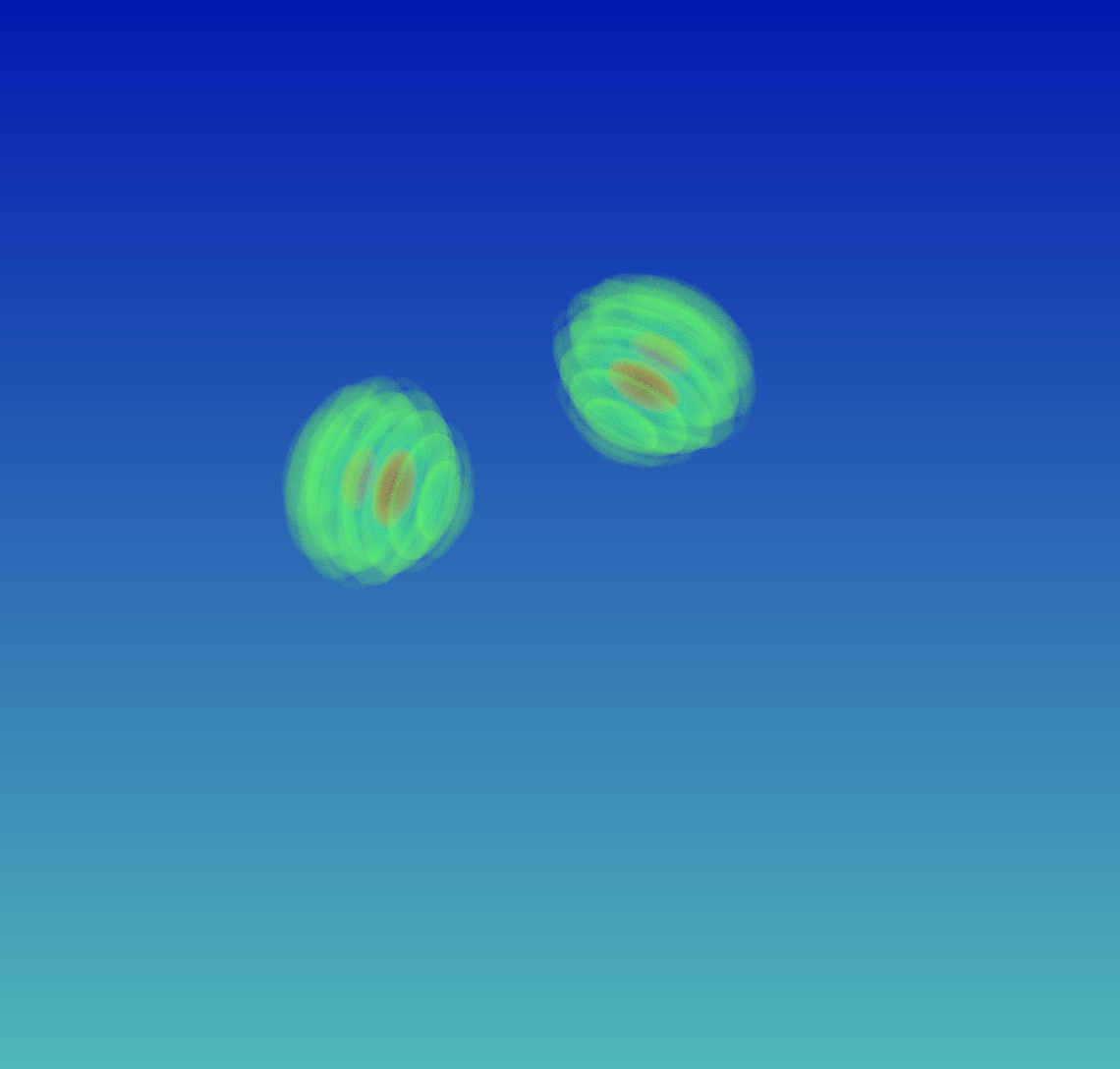}
	\end{subfigure}\\[0.8cm]
	\begin{subfigure}{0.4\textwidth}
		\includegraphics[width=\linewidth]{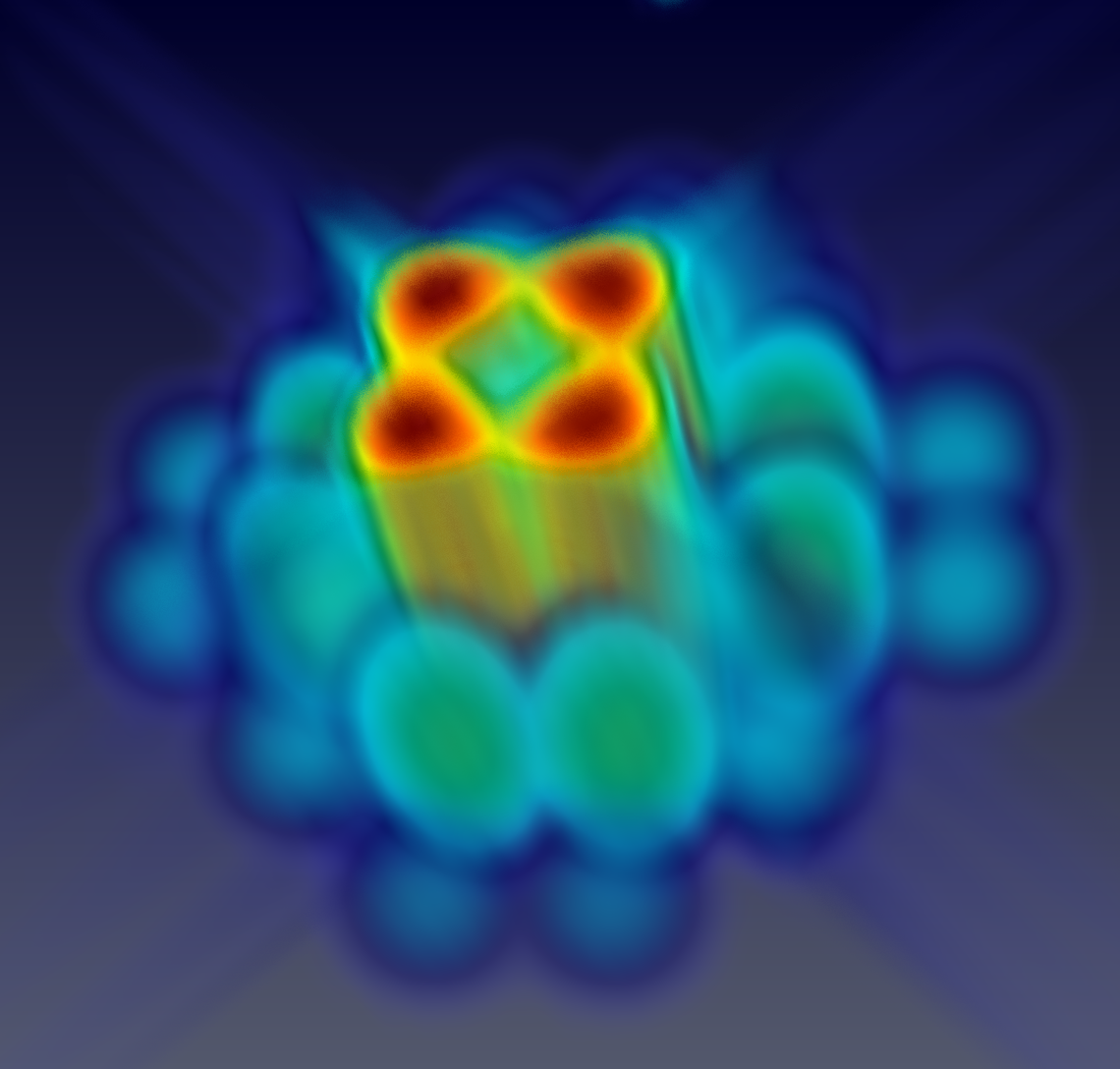}
	\end{subfigure}\qquad
	\begin{subfigure}{0.4\textwidth}
		\includegraphics[width=\linewidth]{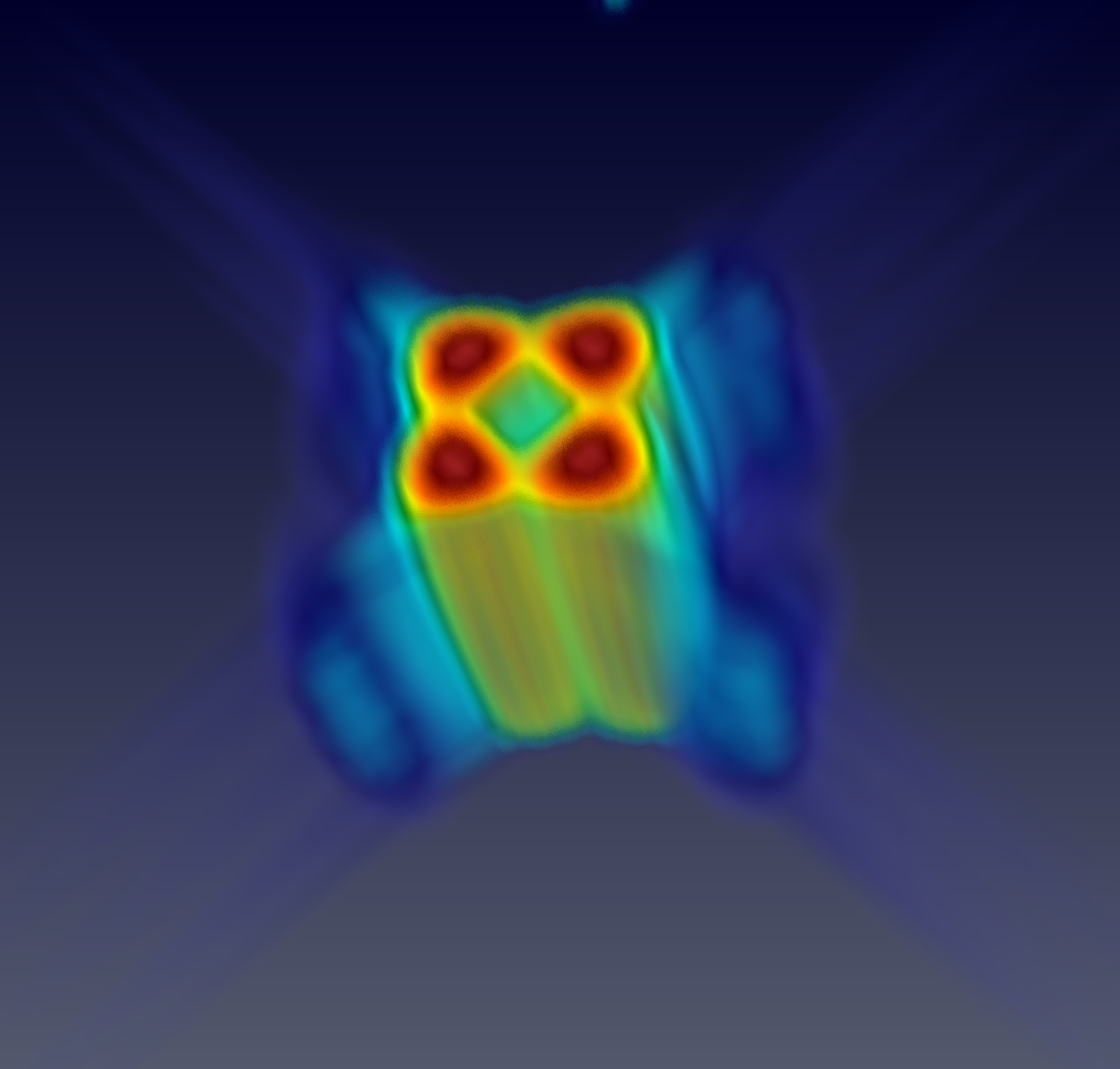}
	\end{subfigure}
	\caption[3D simulation of two perpendicularly colliding Gaussian pulses]{3D simulation of two perpendicularly colliding Gaussian pulses.
		The visualizations on the left show the overlap state, the visualizations on right the final state.
		\textit{Top}: position space. \textit{Bottom}: frequency space.}
	\label{fig:3d}
\end{figure}

\begin{figure}
	\centering
	\begin{subfigure}{0.47\textwidth}
		\includegraphics[width=\linewidth]{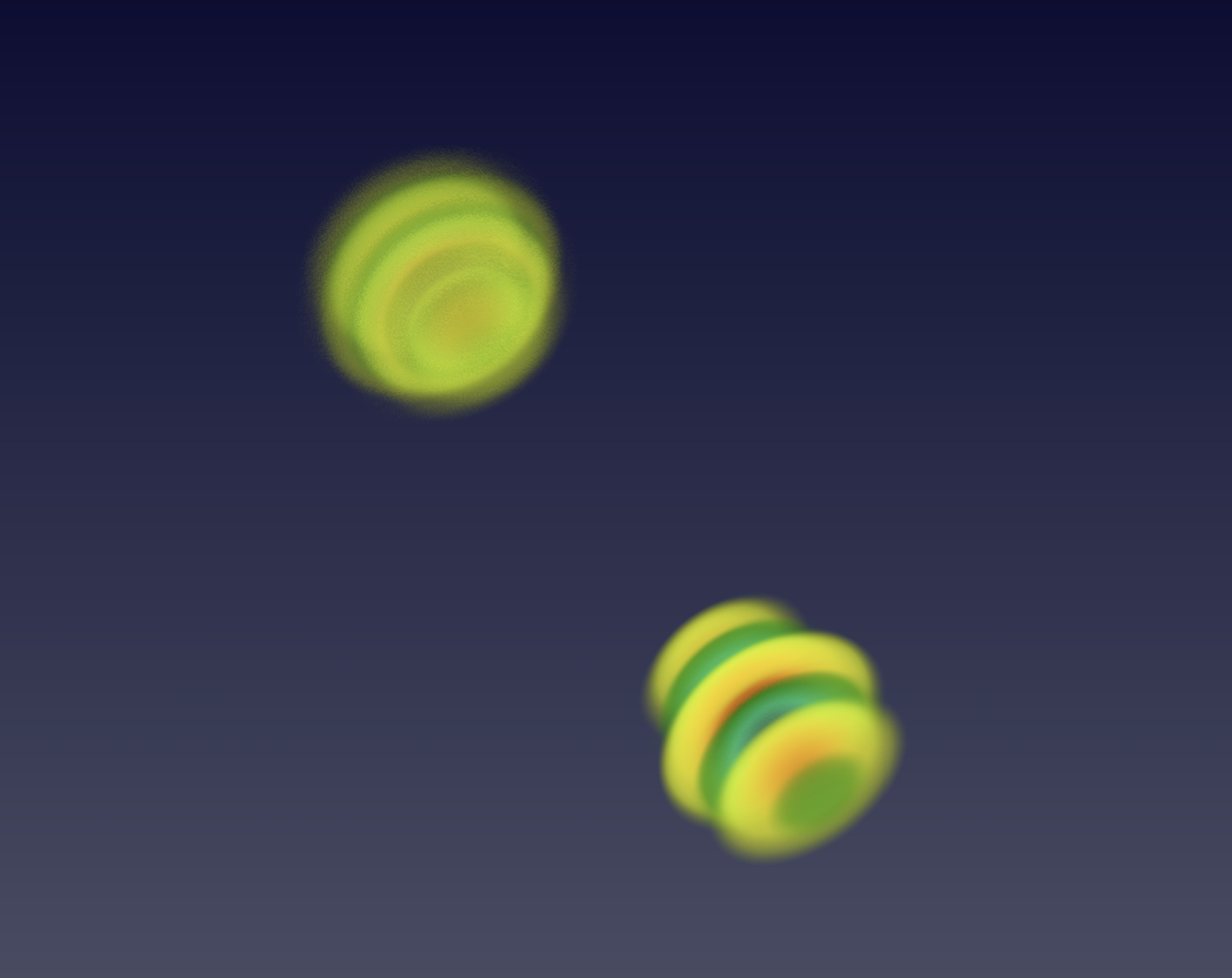}
	\end{subfigure}\qquad
	\begin{subfigure}{0.47\textwidth}
		\includegraphics[width=\linewidth]{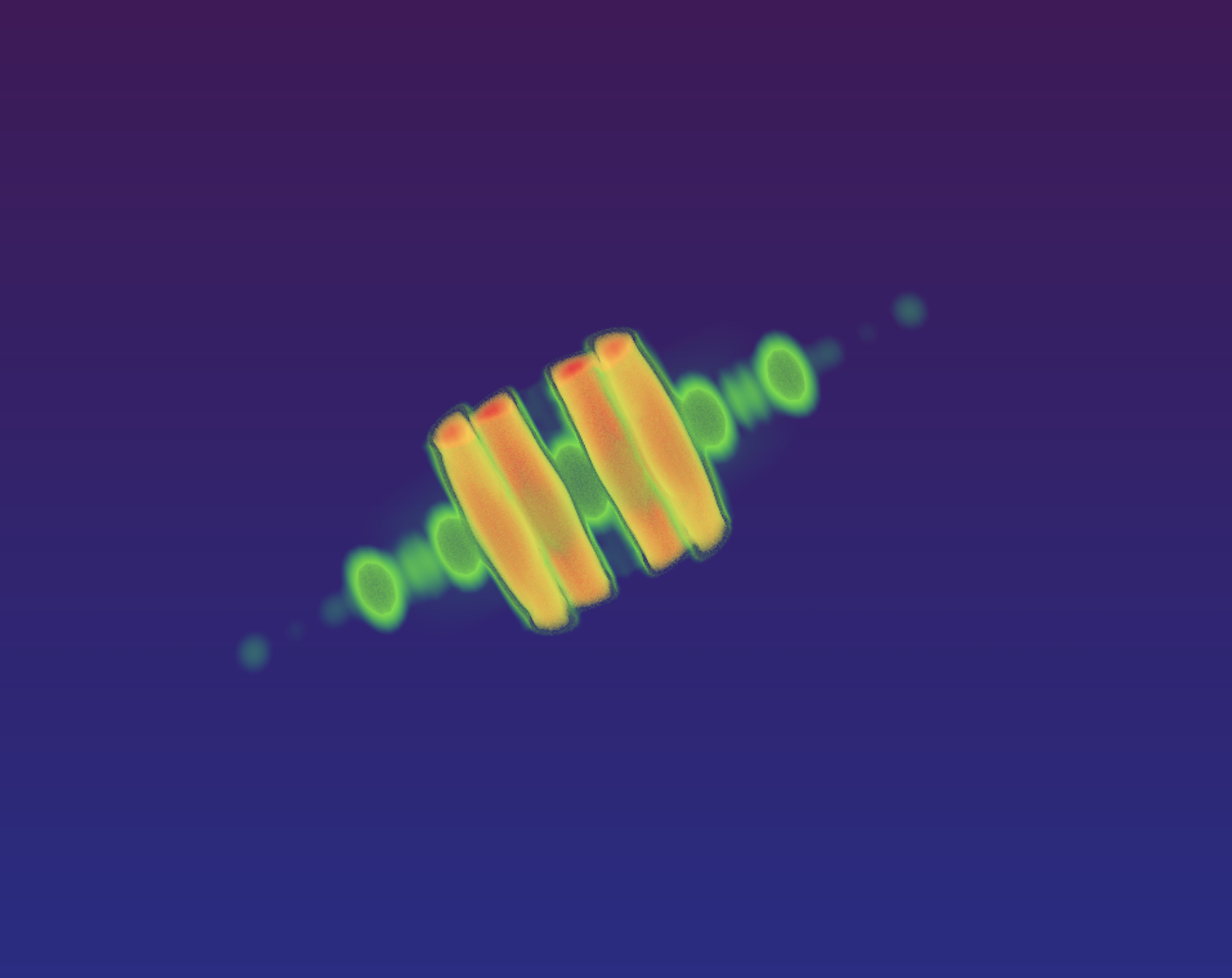}
	\end{subfigure}
	\caption[3D simulation of two coaxially colliding Gaussian pulses]{3D simulation of two coaxially colliding Gaussian pulses with different frequencies.
		\textit{Left:} initial pulse configuration.
		\textit{Right:} harmonics spectrum at the overlap position.}
	\label{fig:3d_coax}
\end{figure}

\begin{figure}[H]
	\centering
	\includegraphics[width=.8\linewidth]{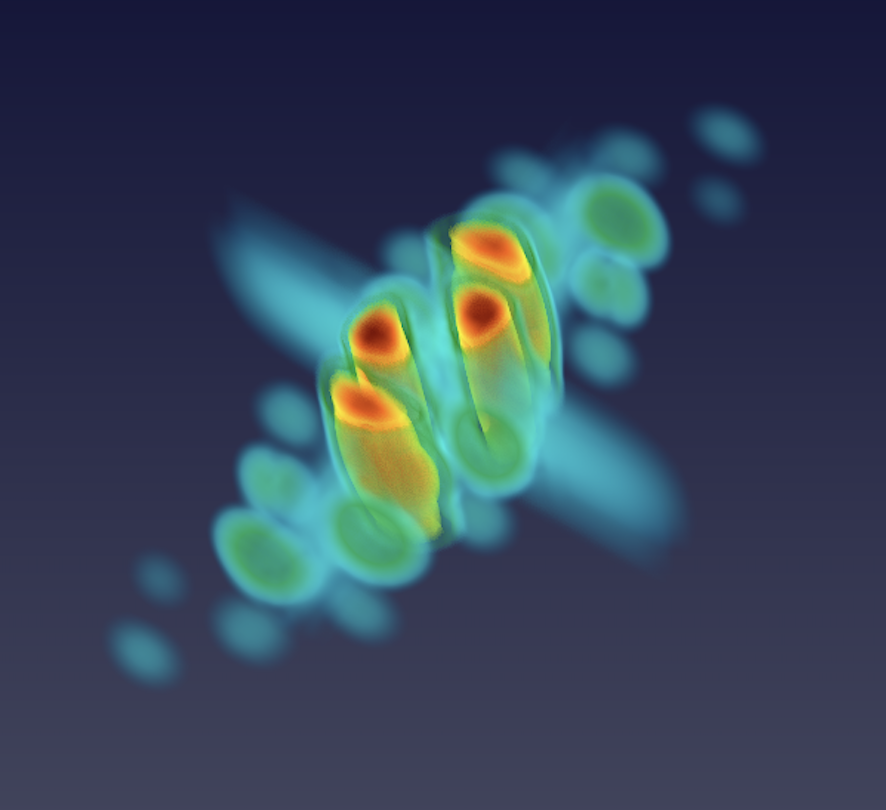}
	\\[0.5cm]
	\caption[Rich harmonics spectrum in 3D]{Rich harmonics spectrum in 3D. This spectrum is generated by the two Gaussian pulses of Figure \ref{fig:3d_coax} colliding at an angle of 135\textdegree.
		There are numerical remnants present on the coordinate axes, where the zero values in frequency space are located and small numerical errors accumulate.
		These can also be observed in Figure \ref{fig:3d}.}
	\label{fig:3d_varangle}
\end{figure}

Ultimately, in order to take into account all geometry effects, simulations have to be conducted in full three spatial dimensions.
A demonstration of configurations similar to those in 2D as shown in Figure \ref{fig:kspace_90}, in 3D yields results visualized in Figure \ref{fig:3d}.
Pulses with frequencies differing by a factor two colliding collinearly produce the results of Figure \ref{fig:3d_coax}.
By virtue of the differing frequencies expressly rich harmonics spectra can be observed.
The corresponding 1D case with a breakdown of the frequencies is demonstrated in the examples provided in the code repository \cite{Lindneretal2023a}.
Colliding the two pulses at an angle of 135\textdegree \ generates the harmonics shown in Figure \ref{fig:3d_varangle}.
It can be seen that the weakest generated signals have about the magnitude of numerical artifacts.

\vfill

\section{Conclusion and outlook\label{sec:Outlook}}

Within the limits of the Heisenberg--Euler weak-field approximation the numerical approach represents an efficient solver for the complete dynamical response of the nonlinear vacuum for complicated pulse setups in up to three dimensions.

Every simulation of the Heisenberg--Euler model incorporates the complete nonlinear physics in the weak-field approximation, whereas analytical calculations normally have to concentrate on single, isolated effects, leaving others aside, and hence often miss the complete picture of the interaction.
By taking into account the whole dynamics of the nonlinear vacuum, the solver captures in particular back-reactions to the radiation fields.
Withal, simulations permit to describe the temporal evolution of nonlinear vacuum processes, which is beyond the reach of many analytical approaches.
Simulations permit to time-resolve the nonlinear vacuum processes, which is beyond the reach of many analytical approaches.

The validity of the presented numerical scheme for solving the modified Maxwell equations in the Heisenberg--Euler weak-field expansion relies on two basic assumptions:
I) field strengths below the critical values $E_{\textrm{cr}} $ and $ B_{\textrm{cr}}$, and II) wavelengths larger than the Compton length of the electron.
The laser pulses are considered on a macroscopic level implying that the individual photons the pulses consist of are not resolved.

A validated solver of the Heisenberg--Euler dynamics is very useful for strong-field QED research going on at present.
The current \textit{C++} implementation of the numerical scheme discussed in this paper allows distinct simulations of the linear
Maxwell vacuum, the four-photon processes, the six-photon processes, and any combination of the latter processes.

Of paramount significance is the dispersion relation, lying at the heart of the algorithm, which ensures stability throughout the frequency spectrum and moreover creates an imaginary part that annihilates nonphysical modes.
Furthermore, the linear vacuum-like behavior of the dispersion relation for a large frequency range is an essential ingredient.

A good agreement with analytical results is achieved by making use of high discretization orders of the numerical scheme and high expansion orders of the effective Heisenberg--Euler model.
The computational cost scales strongly with the number of lattice points but weakly with the discretization orders of the scheme and the expansion orders of the Heisenberg--Euler model.
The impact of the discretization order on the computational cost, however, does become relevant in 3D with increasing \textit{MPI} communication required in order to transfer spatial derivative data.
Making use of high discretization orders and their favorable dispersion relations facilitates the use of comparably small lattices to accurately model the involved waves.
Nonetheless, for high-frequency pulses the required grid resolutions can become unfeasible.

There are facilities constructed with the purpose to detect vacuum birefringence \cite{Zavattinietal2012,Cadeneetal2014}.
The goal is to provide a tool for experimentalists to second their setups with simulation data.
While vacuum birefringence effects in 2D are simulated in a parallel project,
even the employed cluster computing system does not provide enough computing power to for such simulations in 3D as a consequence of the small probe wavelengths required to enhance the effect, c.f. Equation \eqref{eq:Pflip}.

Ideas are being developed in order to overcome the obstacle of extremely large 3D grids.
One promising, ongoing, and important project is on multi-scale simulation capability.
This research is intended to pave the way for a dynamical grid, adapting its resolution regionally on the fly and on demand in order to reduce the computational load, while at the same time maximizing the accuracy in important spatial regions.
An adaptive grid is particularly suited for the prominent low-frequency pump pulse and high-frequency probe pulse setup to detect vacuum nonlinearities, where the computational load can in principle be dramatically reduced.

\if false
The scope of the code is far-reaching.
The main purpose of the present paper is to introduce the numerical scheme and to prove its validity at the hand of cross-checked and further simulation results.
As a proven solver of the Heisenberg--Euler dynamics its usefulness in today's strong-field QED research is indisputable.
\fi
\if false
Not only the high-energy, low-intensity regime employed in particle accelerators, but also the low-energy, high-intensity regime might be a path towards new physics \cite{Gies2008}.
As the quantum fluctuations in the vacuum theoretically consist of all existing particles, it may also serve as a portal to new physics \cite{Karbsteinetal2019}.
\fi

\section*{Acknowledgments}

This work has been funded by the German Research Foundation (DFG) under Grant Nos. \href{https://gepris.dfg.de/gepris/projekt/416611371}{416611371}; \href{https://gepris.dfg.de/gepris/projekt/416607684}{416607684} within the Research Unit \href{https://gepris.dfg.de/gepris/projekt/392856280}{FOR2783}: Probing the Quantum Vacuum at the High-Intensity Frontier.

The authors thank the members of the Research Unit FOR 2783 for their support and fruitful discussions.
Special thanks deserve Felix Karbstein, Holger Gies, Carsten M\"uller, and Alina Golub.

The introductory work of Arnau Domenech and Hartmut Ruhl is acknowledged. In particular, the present paper has been inspired by Arnau's PhD project.
Hartmut Ruhl proposed the work, helped devising the numerical algorithm and setting up the code.

Parts of the computations have been performed on the KSC cluster computing system of the Arnold Sommerfeld Center for Theoretical Physics at the LMU Munich, hosted at the Leibniz-Rechenzentrum (LRZ) in Garching and funded by the German Research Foundation under Grant No. \href{https://gepris.dfg.de/gepris/projekt/409562408}{409562408}.

The hospitality of the Arnold Sommerfeld Center is acknowledged.

\section*{Data statement}
The simulation data produced for this work are archived on servers of the Arnold Sommerfeld Center (ASC) for Theoretical Physics in Munich, hosted by the Leibniz-Rechenzentrum (LRZ), in compliance with the regulations of the German Research Foundation (DFG).
The code is publicly available under the BSD 3-Clause License and maintained on an institutional \textit{GitLab} server \cite{Lindneretal2023a}. 
There is a \textit{Mendeley Data} repository containing extra and supplementary materials \cite{Lindner2022}.

\appendix

\section{Dispersion relations up to order thirteen}\label{app:stencils}

\begin{figure}
			\centering
	\begin{subfigure}{.75\textwidth}
		\centering
		\includegraphics[width=\linewidth]{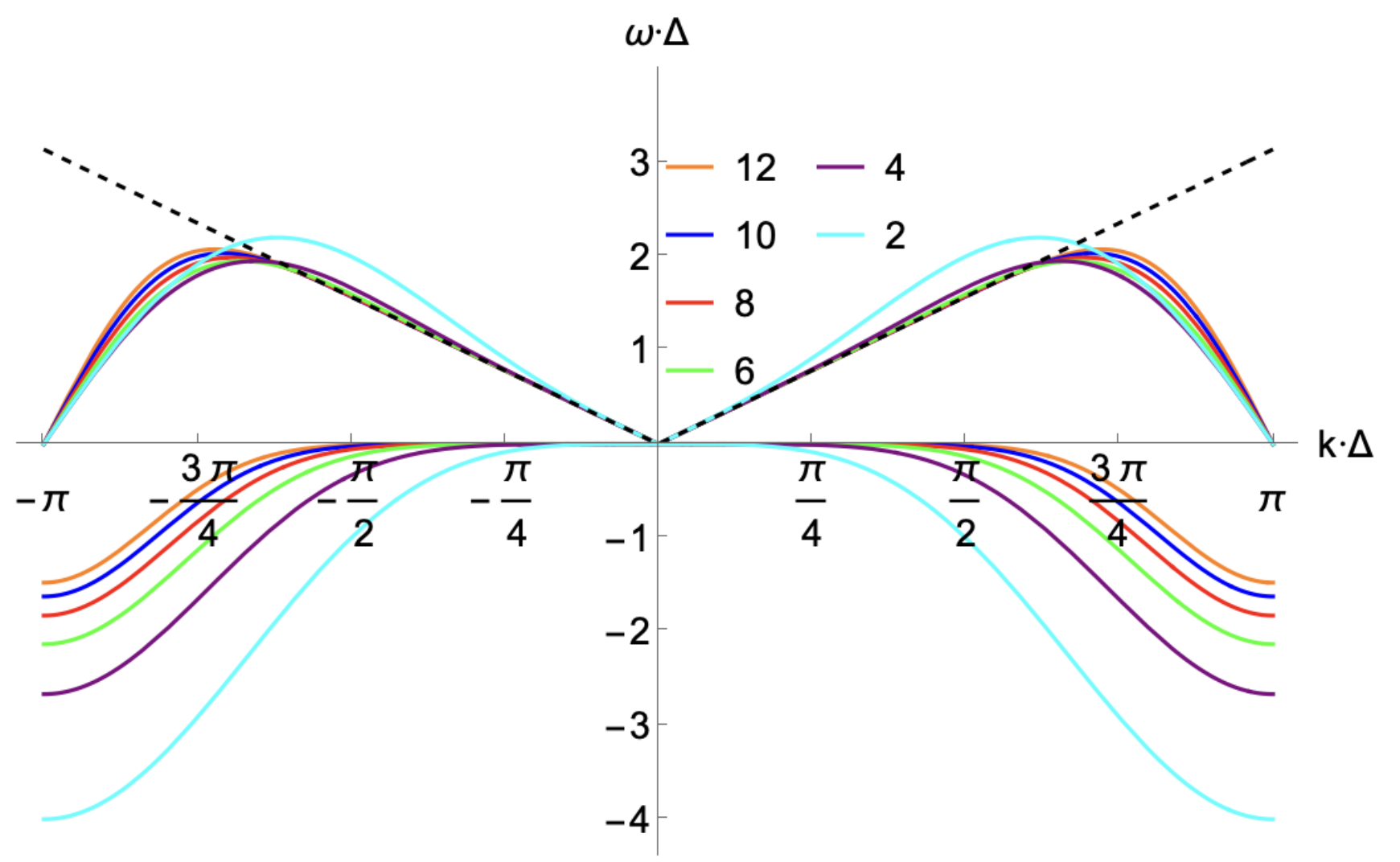}
	\end{subfigure}
	\\[0.7cm]
	\begin{subfigure}{.75\textwidth}
		\centering
		\includegraphics[width=\linewidth]{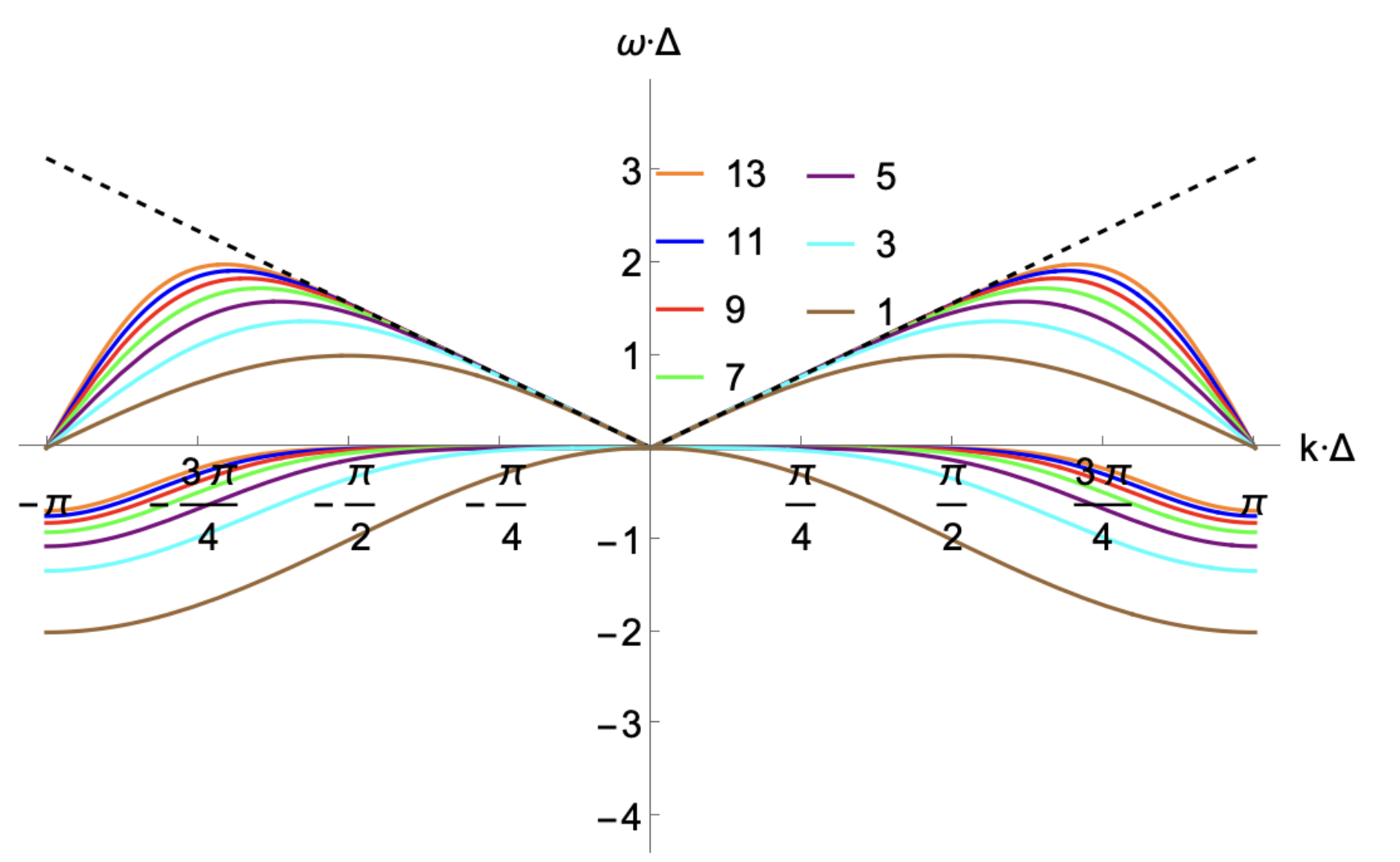}
	\end{subfigure}
	\\[0.3cm]
	\caption[Dispersion relations from order one to thirteen]{Dispersion relations up to order thirteen.
		Real and imaginary parts are shown in the same color and can be distinguished by their form.
		The black dotted line represents the real vacuum dispersion relation.
		\textit{Top}: even orders.
		\textit{Bottom}: uneven orders.
	}
	\label{fig:dispersions}
\end{figure}

For completeness, the elements of the minimally biased stencil matrices $S_\nu$ from order one to thirteen, listed in the form of the array components as in Equation \eqref{eq:stencil_components} for the fourth order, are given by
\begin{align*}
	s^1_f \big\rvert_{\nu=-1,0} &=\{-1,1\} \ ,
	\\
	s^2_f \big\rvert_{\nu=-2,-1,0} &=\left\{\frac{1}{2},-2,\frac{3}{2}\right\} \ ,
	\\
	s^3_f \big\rvert_{\nu=-2,...,1} &= \left\{\frac{1}{6},-1,\frac{1}{2},\frac{1}{3}\right\} \ ,
	\\
	s^4_f \big\rvert_{\nu=-3,...,1} &= \left\{-\frac{1}{12},\frac{1}{2},-\frac{3}{2},\frac{5}{6},\frac{1}{4}\right\} \ ,
	\\
	s^5_f  \big\rvert_{\nu=-3,...,2} &= \left\{-\frac{1}{30},\frac{1}{4},-1,\frac{1}{3},\frac{1}{2},-\frac{1}{20}\right\} \ ,
	\\
	s^6_f  \big\rvert_{\nu=-4,...,2} &= \left\{\frac{1}{60},-\frac{2}{15},\frac{1}{2},-\frac{4}{3},\frac{7}{12},\frac{2}{5},-\frac{1}{30}\right\} \ ,
	\\
	s^7_f  \big\rvert_{\nu=-4,...,3} &= \left\{\frac{1}{140},-\frac{1}{15},\frac{3}{10},-1,\frac{1}{4},\frac{3}{5},-\frac{1}{10},\frac{1}{105}\right\} \ ,
	\\
	s^8_f \big\rvert_{\nu=-5,...,3} &= \left\{-\frac{1}{280},\frac{1}{28},-\frac{1}{6},\frac{1}{2},-\frac{5}{4},\frac{9}{20},\frac{1}{2},-\frac{1}{14},\frac{1}{168}\right\} \ ,
	\\
	s^9_f \big\rvert_{\nu=-5,...,4} &= \left\{-\frac{1}{630},\frac{1}{56},-\frac{2}{21},\frac{1}{3},-1,\frac{1}{5},\frac{2}{3},-\frac{1}{7},\frac{1}{42},-\frac{1}{504}\right\} \ ,
	\\
	s^{10}_f \big\rvert_{\nu=-6,...,4} &= \left\{\frac{1}{1260},-\frac{1}{105},\frac{3}{56},-\frac{4}{21},\frac{1}{2},-\frac{6}{5},\frac{11}{30},\frac{4}{7},-\frac{3}{28},\frac{1}{63},-\frac{1}{840}\right\} \ ,
	\\
	s^{11}_f \big\rvert_{\nu=-6,...,5} &= \left\{\frac{1}{2772},-\frac{1}{210},\frac{5}{168},-\frac{5}{42},\frac{5}{14},-1,\frac{1}{6},\frac{5}{7},-\frac{5}{28},\frac{5}{126},-\frac{1}{168},\frac{1}{2310}\right\} \ ,
	\\
	s^{12}_f \big\rvert_{\nu=-7,...,5} &= \left\{-\frac{1}{5544},\frac{1}{396},-\frac{1}{60},\frac{5}{72},-\frac{5}{24},\frac{1}{2},-\frac{7}{6},\frac{13}{42},\frac{5}{8},-\frac{5}{36},\frac{1}{36},-\frac{1}{264},\frac{1}{3960}\right\} \ ,
	\\
	s^{13}_f \big\rvert_{\nu=-7,...,6} &= \left\{-\frac{1}{12012},\frac{1}{792},-\frac{1}{110},\frac{1}{24},-\frac{5}{36},\frac{3}{8},-1,\frac{1}{7},\frac{3}{4},-\frac{5}{24},\frac{1}{18},-\frac{1}{88},\frac{1}{660},-\frac{1}{10296}\right\} \ .
\end{align*}
Solutions to the dispersion relation for all given orders are visualized in Figure \ref{fig:dispersions}.

{\footnotesize
\bibliography{../refs}
}
\end{document}